\documentclass[preprint]{aastex6}

\usepackage{subfigure}
\usepackage{url}
\usepackage{graphicx,color}

\usepackage{soul}

\begin{document}

\accepted{\today}
\journalinfo{ApJ}

\shorttitle{}
\shortauthors{Avallone et al.}

\graphicspath{{diagrams/}}

\title{Critical magnetic field strengths for solar coronal plumes in quiet regions and coronal holes?}

\author{Ellis A. Avallone\altaffilmark{1}, Sanjiv K. Tiwari\altaffilmark{2}\(^{,}\)\altaffilmark{3}\(^{,}\)\altaffilmark{4}, Navdeep K. Panesar\altaffilmark{4}, Ronald L. Moore\altaffilmark{4}\(^{,}\)\altaffilmark{5}, Amy Winebarger\altaffilmark{4}}

\altaffiltext{1}{University of Washington, Seattle, WA 98195, USA}
\altaffiltext{2}{Lockheed Martin Solar and Astrophysics Laboratory, 3251 Hanover Street, Bldg. 252, Palo Alto, CA 94304, USA}
\altaffiltext{3}{Bay Area Environmental Research Institute, NASA Research Park, Moffett Field, CA 94035, USA}
\altaffiltext{4}{NASA Marshall Space Flight Center, Mail Code ST 13, Huntsville, AL 35812, USA}
\altaffiltext{5}{Center for Space and Aeronomic Research, The University of Alabama in Huntsville, Huntsville, AL 35805, USA}

\begin{abstract}
Coronal plumes are bright magnetic funnels found in quiet regions (QRs) and coronal holes (CHs). They extend high into the solar corona and last from hours to days. The heating processes of plumes involve dynamics of the magnetic field at their base, but the processes themselves remain mysterious. Recent observations suggest that plume heating is a consequence of magnetic flux cancellation and/or convergence at the plume base. These studies suggest that the base flux in plumes is of mixed polarity, either obvious or hidden in SDO/HMI data, but do not quantify it. To investigate the magnetic origins of plume heating, we select ten unipolar network flux concentrations, four in CHs, four in QRs, and two that do not form a plume, and track plume luminosity in SDO/AIA 171 \AA\, images along with the base flux in SDO/HMI magnetograms, over each flux concentration's lifetime. We find that plume heating is triggered when convergence of the base flux surpasses a field strength of $\sim$200-600 G. The luminosity of both QR and CH plumes respond similarly to the field in the plume base, suggesting that the two have a common formation mechanism. Our examples of non-plume-forming flux concentrations, reaching field strengths of 200 G for a similar number of pixels as for a couple of our plumes, suggest that a critical field might be necessary to form a plume but is not sufficient for it, thus, advocating for other mechanisms, e.g. flux cancellation due to hidden opposite-polarity field, at play.
\end{abstract}


\section{Introduction}
\begin{figure}
\includegraphics[trim=0cm 1.5cm 0cm 1.5cm,clip,width=\textwidth]{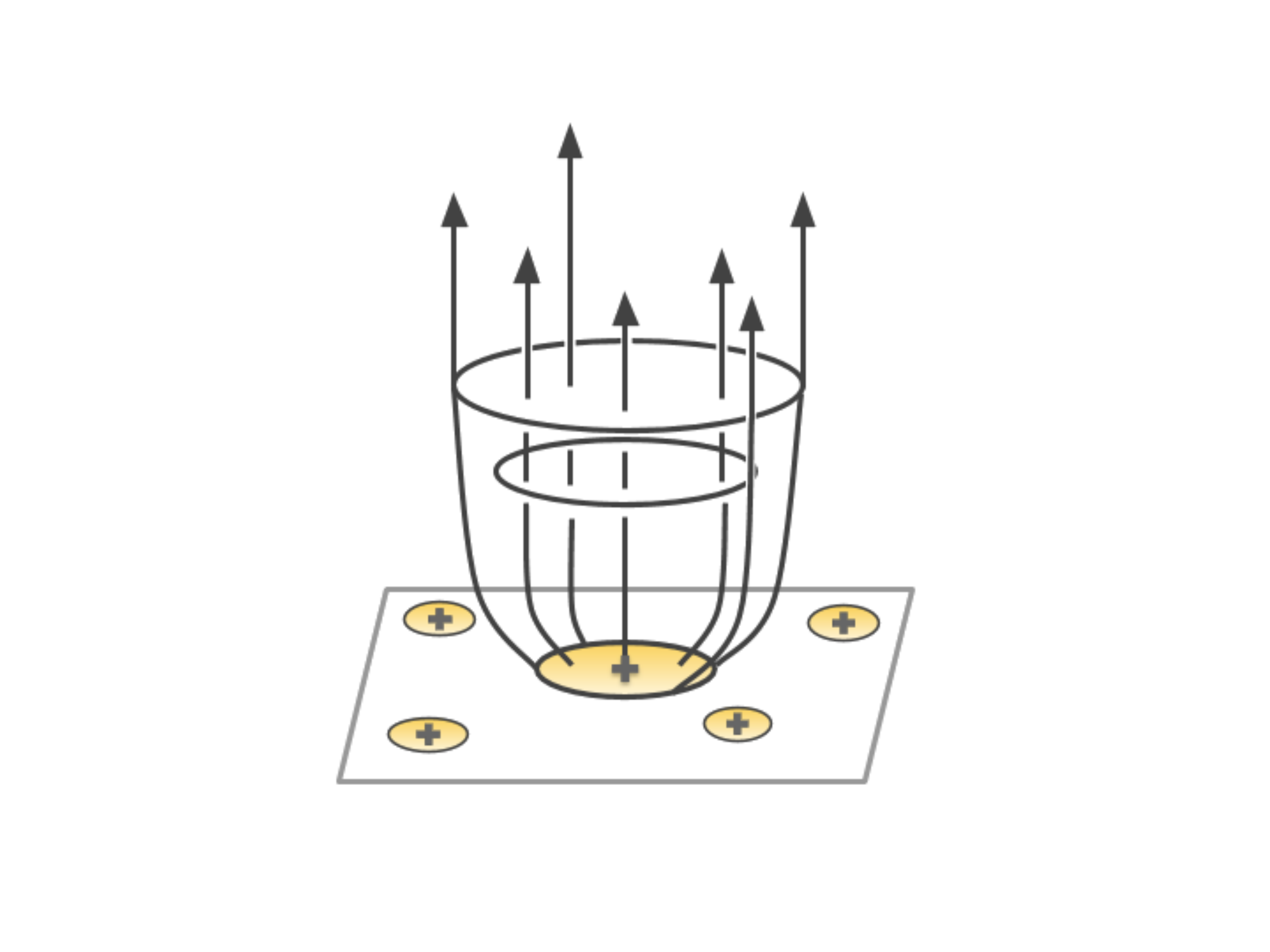}
\caption{Diagram showing the structure of a coronal plume with a dominant positive-polarity magnetic field. The large patch at the base of the plume represents the visibly-unipolar base flux, while the smaller patches represent the predominantly unipolar external magnetic field in either a quiet region or a coronal hole.
\label{plume_diagram}}
\end{figure}

Coronal plumes are bright, sporadic, fountain-like structures in the solar corona whose lifetimes range from a few hours to several days (for detailed reviews with references, see \citet{2011A&ARv..19...35W} and \citet{2015LRSP...12....7P}). Each plume is a magnetic funnel rooted in a strong patch of predominantly-unipolar magnetic flux, or base flux, surrounded by a predominantly-unipolar magnetic field, as shown by the sketch in Figure \ref{plume_diagram}. As seen in magnetograms from the Solar Dynamics Observatory (SDO) Helioseismic and Magnetic Imager (HMI), the base flux is visibly unipolar in most plumes (as seen in this work), but may contain visible minority-polarity flux in some cases \citep[e.g.][]{2014ApJ...787..118R}. 

Plumes were initially observed in white-light eclipse images as `rays' extending from the poles in polar coronal holes (PCHs). Examples of such observations include \citet{1950BAN....11..150V}, \citet{1958PASJ...10...49S}, \citet{1965PASJ...17....1S}, and \citet{1965ApJ...141..832H}. With subsequent extreme-ultraviolet (EUV) observations, it was found that the same structures were found at lower latitudes in equatorial coronal holes (ECHs) (e.g. \citet{1995ApJ...446L..51W} and \citet{2008SoPh..249...17W}). It was also found that plumes are best observed in AIA 171 \AA\, Fe IX emission, but they can also be observed in AIA 193 \AA\, Fe XII emission and occasionally in AIA 211 \AA\, Fe XIV emission. EUV spectroscopic observations also emerged. Such observations expanded our understanding of low-latitude plumes and further characterized coronal plume morphology \citep[e.g.][]{1999JGR...104.9753D}. 

These first observations assumed that plumes only existed in coronal hole (CH) regions. However, as described later, we find that non-CH/quiet Sun regions (QRs) also contain similar structures to on-disk CH plumes. In non-CH QRs these structures are most likely the feet of closed loops. \citet{2016ApJ...818..203W} speculated that plumes in both CHs and QRs have a similar formation mechanism, based on observations of plume-like 171 \AA\, emission outside CH regions, but the results were purely qualitative. In order to further investigate this idea, we include these plume-like structures in QRs in our sample and compare them with on-disk CH plumes to find similarities or differences between the two. Similar studies have been done with jets in CHs \citep{2018ApJ...853..189P} and QRs (see e.g. \citet{2017ApJ...844..131P} and \citet{2016SoPh..291.1129N}), but there has been no quantitative comparison between known plumes in CHs and plume-like structures in QRs.

Multiple observations have been presented regarding how plumes form and disappear, primarily with off-limb plumes in PCHs. The dynamics and flow patterns in on-disk CH plumes have also recently been studied by \citet{2011ApJ...736..130T}, \citet{2014ApJ...794..109F}, and \citet{2015ApJ...807...71P}. With on-disk observations of the Sun, we have been able to further probe the magnetic processes which govern plume evolution through coordinated EUV and magnetogram observations. 

The primary mechanism that has been the forefront of investigations of plume evolution since the emergence of coordinated EUV and magnetogram observations is magnetic reconnection due to mixed-polarity, both hidden and visible, in the base flux. The role of mixed polarity in plume formation was proposed in \citet{1994ApJ...435L.153W}, who showed that strong additional heating due to magnetic reconnection in the base could account for plume formation. This idea was first investigated in \citet{1997ApJ...484L..75W} and \citet{1998ApJ...501L.145W}, inspired by early Skylab and EIT observations showing bright "plume haze" forming rapidly above some (but not all) decaying bright points within PCH. This suggested that plumes resulted from the interaction between minority-polarity flux and adjacent unipolar flux concentrations. Therefore, it is difficult to explain such observations on the basis of unipolar flux concentrations alone. Although magnetic reconnection due to mixed polarity is still considered a likely mechanism for plume formation, others have concluded that plumes are purely unipolar and contain no minority-polarity flux (see e.g. \citet{1968SoPh....3..321N} and \citet{1997SoPh..175..393D}).

In order to further understand the processes which govern plume evolution, we base our work on two previous studies of on-disk coronal plumes: \citet{2014ApJ...787..118R} and \citet{2016ApJ...818..203W}. \citet{2014ApJ...787..118R} observed that plume emergence follows the appearance of transient bright points, or `jetlets', and inferred that plume heating is triggered and sustained by magnetic flux cancellation in the plume base. \citet{2016ApJ...818..203W} observed that plume formation and disappearance is a result of convergence and divergence of the base flux. Both papers suggest some dependence on mixed polarity in the base flux, and state that their visibly unipolar plumes may have underlying mixed polarity that can't be resolved by current instruments. 

\citet{2014ApJ...787..118R} and \citet{2016ApJ...818..203W} present purely qualitative observations. Both papers do not provide any quantitative measurements of the base flux or intensity in any EUV wavelengths, or consider whether other factors such as a critical magnetic field strength are necessary for plume production. To further investigate the magnetic origins of plume heating and determine whether or not a critical magnetic field strength is required for plume formation, we track plume luminosity and base magnetic flux over the lifetimes of eight coronal plumes and two non-plume-forming flux concentrations using SDO/Atmospheric Imaging Assembly (AIA) 171 \AA\, Fe IX images \citep{2012SoPh..275...17L} and SDO/HMI line-of-sight magnetograms \citep{2012SoPh..275..207S}.

\section{Data and Methods}
SDO/AIA provides high-resolution, high-cadence, full-disk images of the Sun in seven EUV band passes, detecting temperatures ranging from $\sim$4500 K to $\sim$10 MK. AIA takes images of the Sun at a 12 second cadence with a spatial resolution of 1.5" \citep{2012SoPh..275...17L}. This high spatial and temporal resolution allows us to track minute details across the solar disk ranging from the upper chromosphere to the corona. 

SDO/HMI provides high-cadence, full-disk line-of-sight (LOS) magnetograms. HMI obtains magnetograms at a 45 second cadence with a spatial resolution of 1" \citep{2012SoPh..275..207S}. Simultaneous observations of AIA and HMI on SDO make it possible to observe how the upper chromosphere and corona respond to transient magnetic activity in the photosphere. 

To perform our analysis, we used SDO/AIA 171 \AA\, Fe IX images and SDO/HMI LOS magnetograms. While plumes can be observed in AIA 171 \AA, AIA 193 \AA, and AIA 211 \AA\, emission, we chose AIA 171 \AA\, emission for our measurements (at $\sim$0.6-0.8 MK), as it shows the largest extent of the plume structure. We selected eight isolated, apparently unipolar plumes using JHelioviewer software \citep{2017arXiv170507628M}, and chose four plumes in CHs and four plumes in QRs. We focused on finding plumes close to the center of the solar disk to avoid projection effects (particularly in magnetograms, e.g. \citet{2016ApJ...833L..31F}) and avoided selecting plumes on the CH/QR boundary to ensure that no factors based on region would confuse our results. SDO/AIA 171 \AA\, data was used to locate plumes, and SDO/AIA 211 \AA\, data was used to confirm whether each plume was in a QR or a CH. We also selected two isolated, apparently unipolar network flux concentrations that do not form a plume, but still behave similarly to regions that do form plumes. We used SDO/HMI data to locate these concentrations, selecting them using the same selection criteria as for plume-forming regions, and used SDO/AIA 211 \AA\, data to confirm whether or not each concentration was in a QR or CH.

AIA 171 \AA\, level-1 data and HMI magnetogram data were then acquired using the Joint Science Operations Center \footnote{\url{http://jsoc.stanford.edu}}, and downloaded at a three minute cadence over the lifetime of each plume, beginning one hour before formation and ending one hour after disappearance to obtain a good baseline level for both plume luminosity and base flux.  

AIA and HMI data were then imported into SolarSoft IDL, and derotated using the drot.pro routine. AIA images were divided by exposure time in order to obtain uniform luminosity measurements. Both AIA and HMI data were then converted into maps in SolarSoft IDL and imported into Python. Python scripts were written and used to calculate luminosity in AIA 171 \AA\, and base magnetic flux over each plume's lifetime. Both of these measurements were taken for each plume by first visually selecting regions of interest in AIA images and HMI magnetograms which contained the plume throughout its entire lifetime. The region of interest for AIA 171 \AA\, data was selected to overlap with the region of interest in HMI magnetogram data as well as possible while still including the part of the plume that extends beyond the base. The region of interest for HMI magnetogram data was selected to include all flux patches which contribute to the plume's evolution and avoid any extraneous flux patches. Any flux patches which cross the edge of the region of interest boundary are considered non-contributing, or their contribution is negligible. Once the region of interest was selected, plume luminosity in AIA 171 \AA\, was measured by adding up all the pixel values in that region of interest. Base magnetic flux was computed in a similar manner. However we only added up pixel values above a given value for magnetic field strength, or threshold magnetic field value. 

\subsection{Critical Magnetic Field Strength Calculation}
\begin{figure}[ht!]
\centering
\plottwo{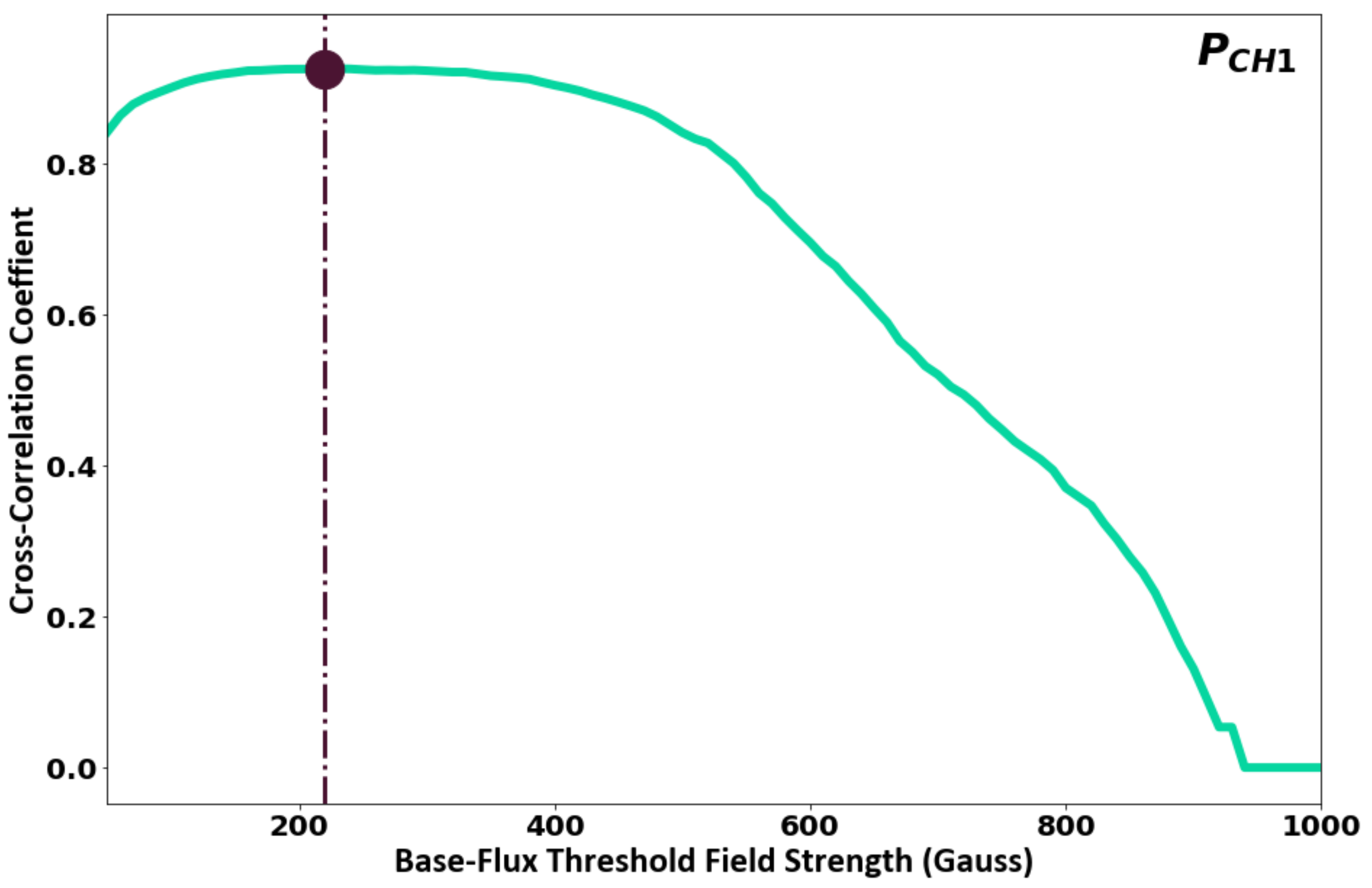}{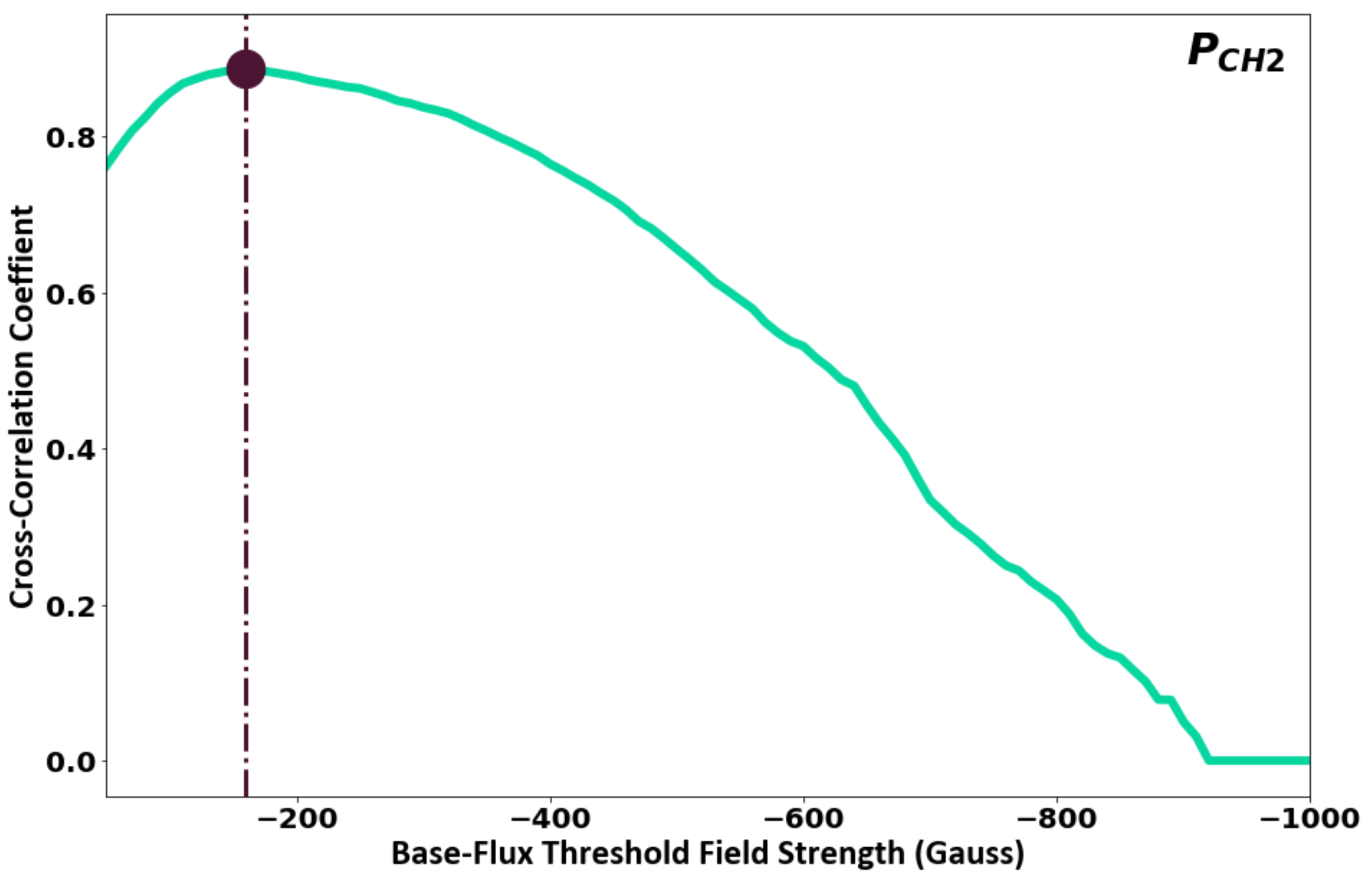}
\plottwo{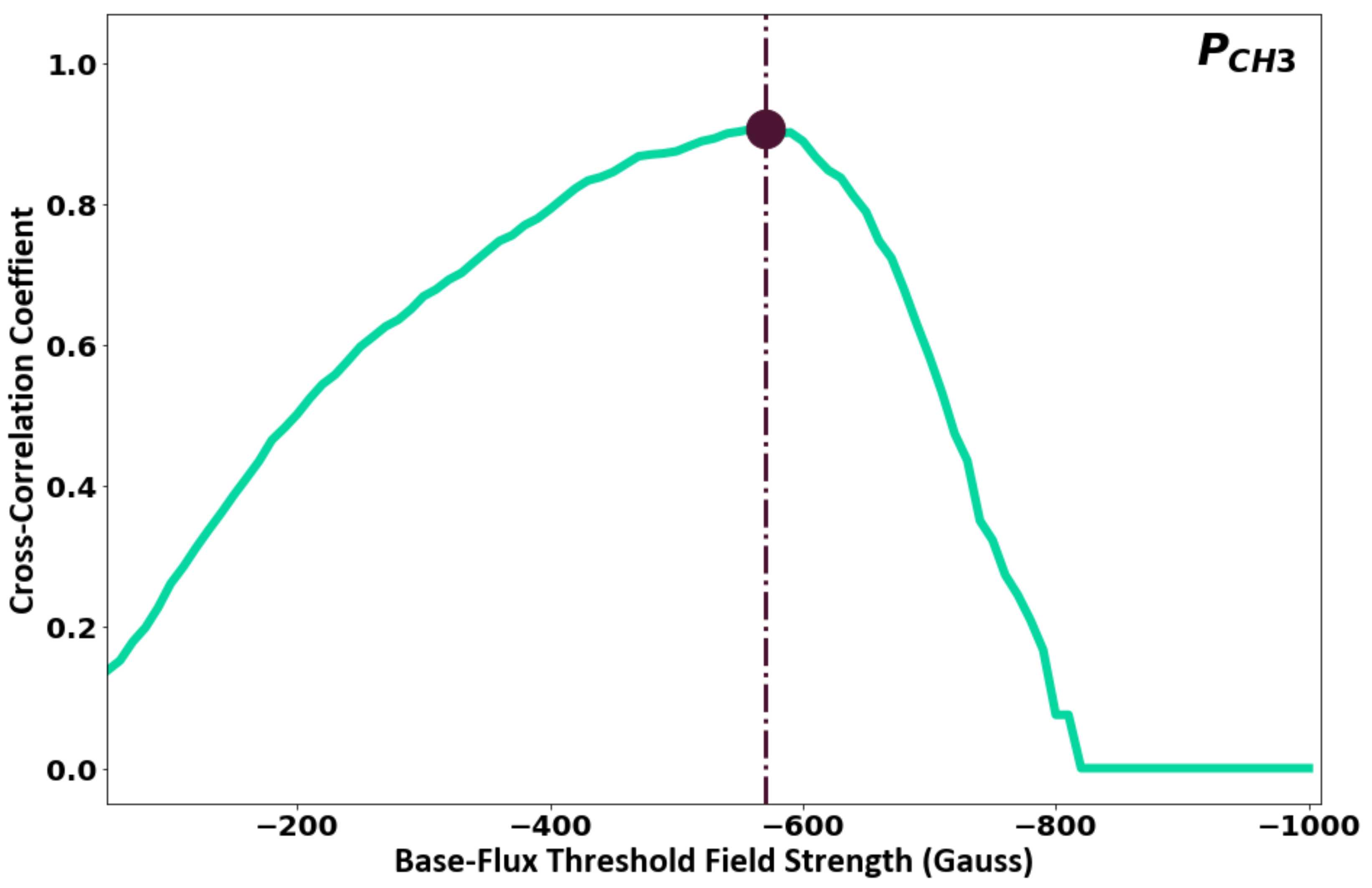}{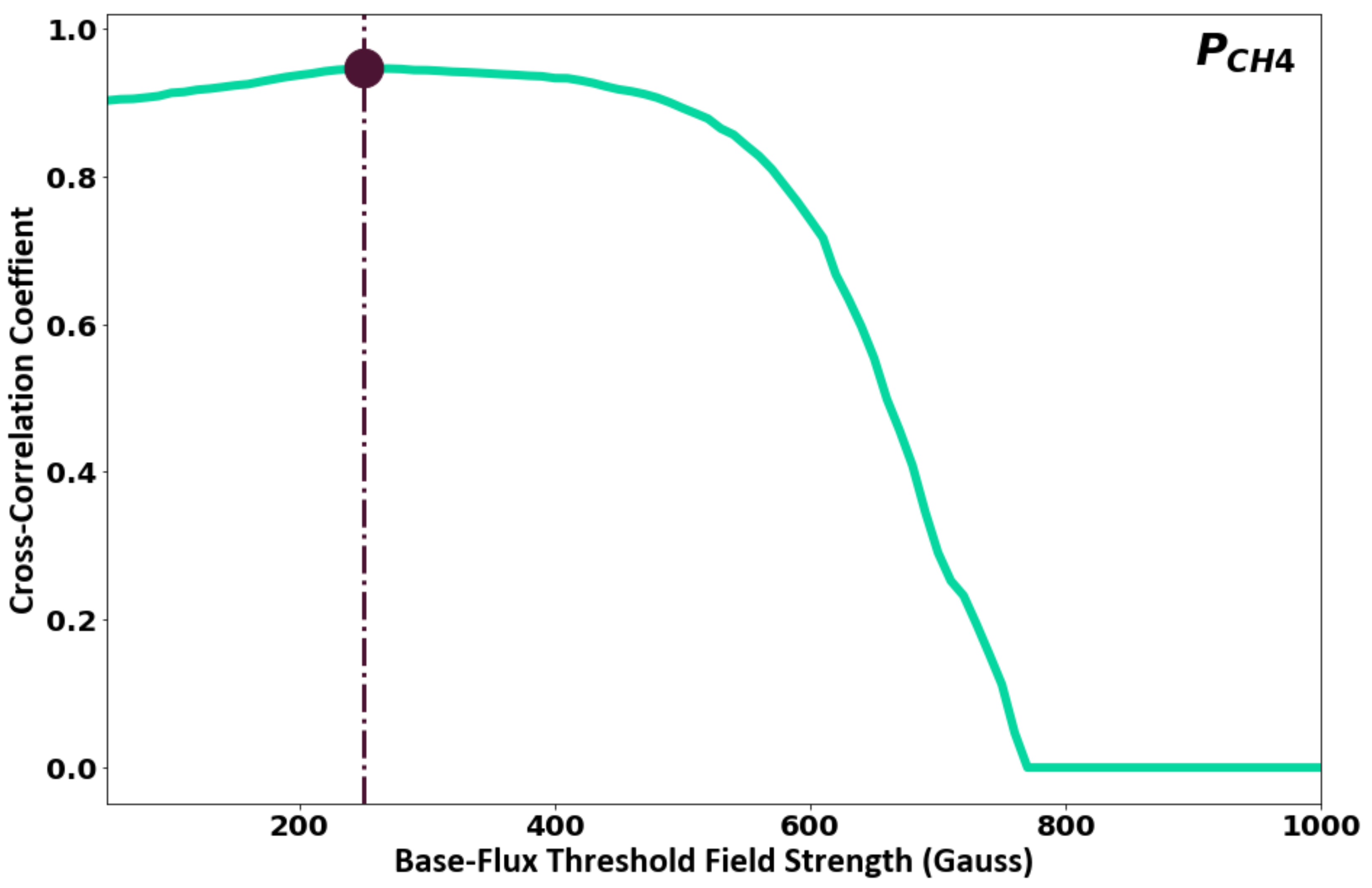}
\plottwo{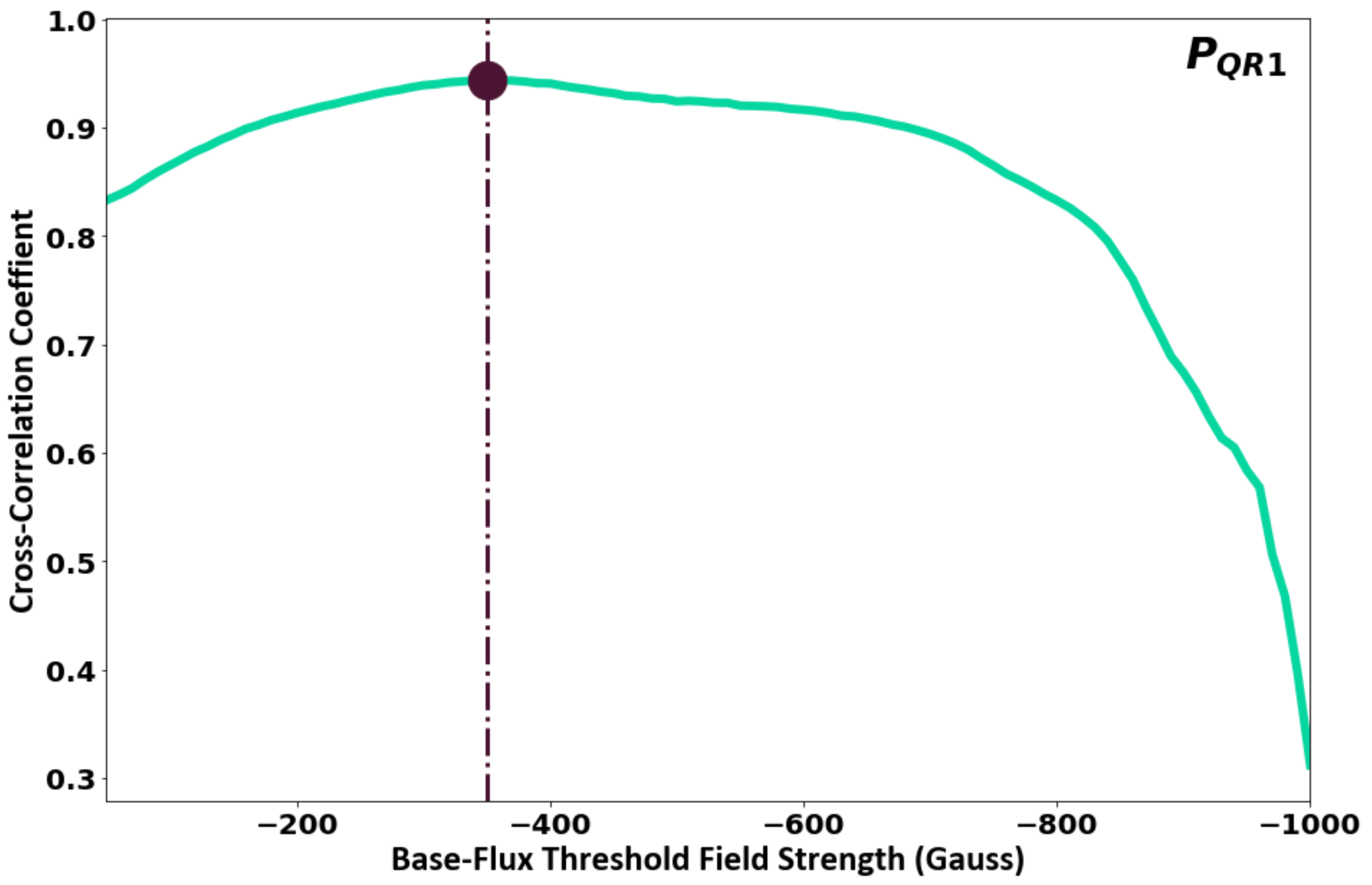}{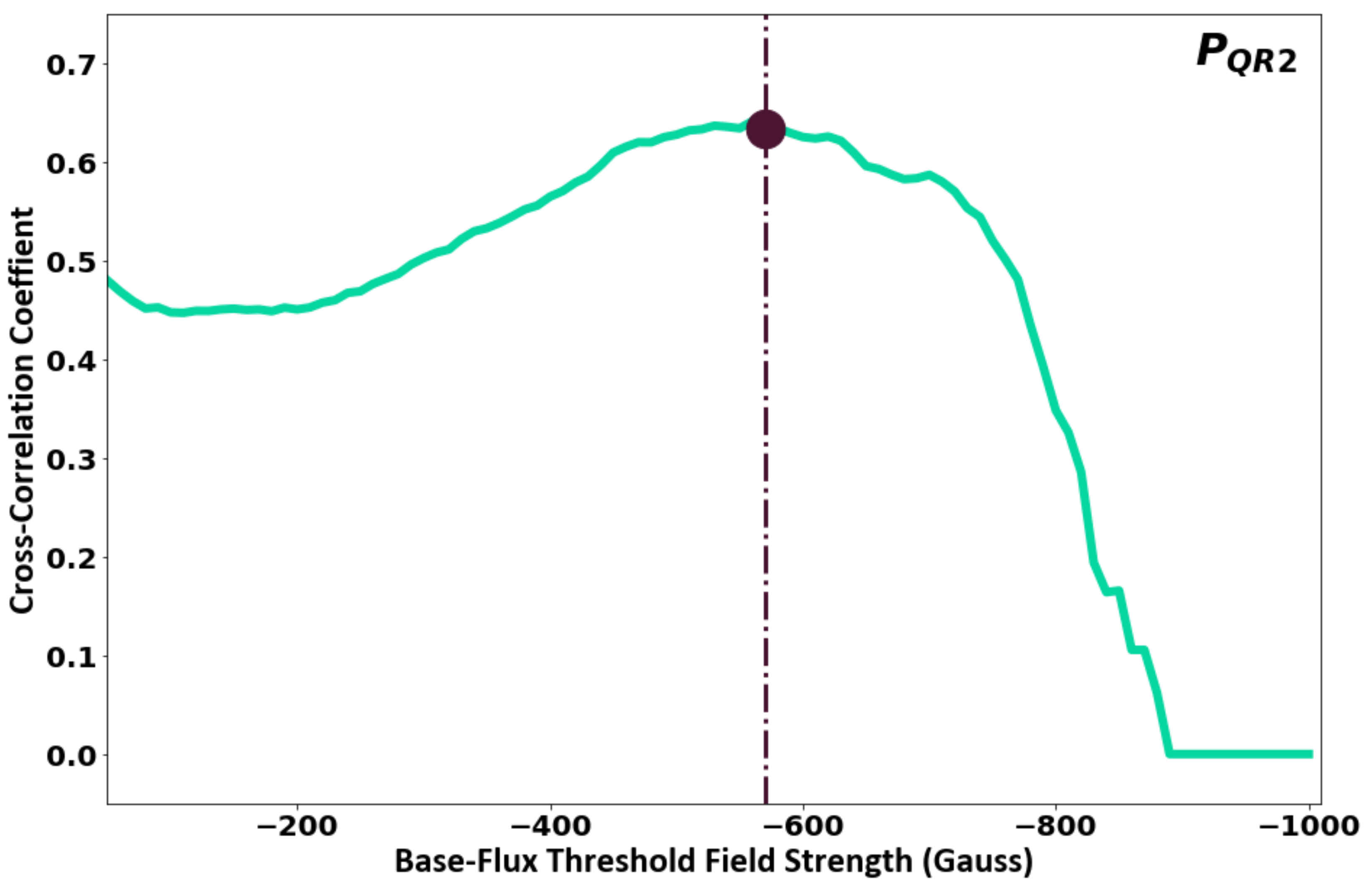}
\plottwo{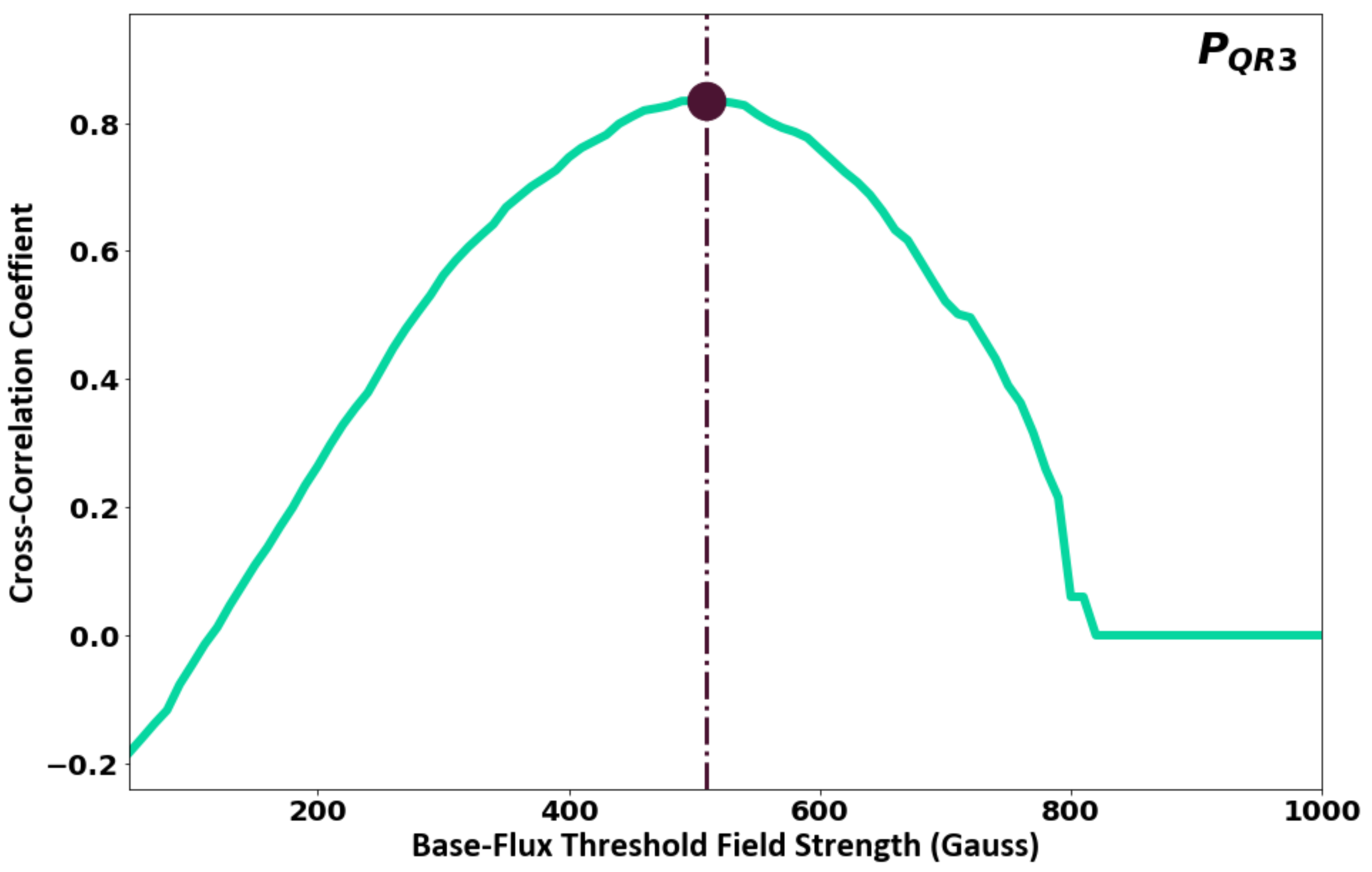}{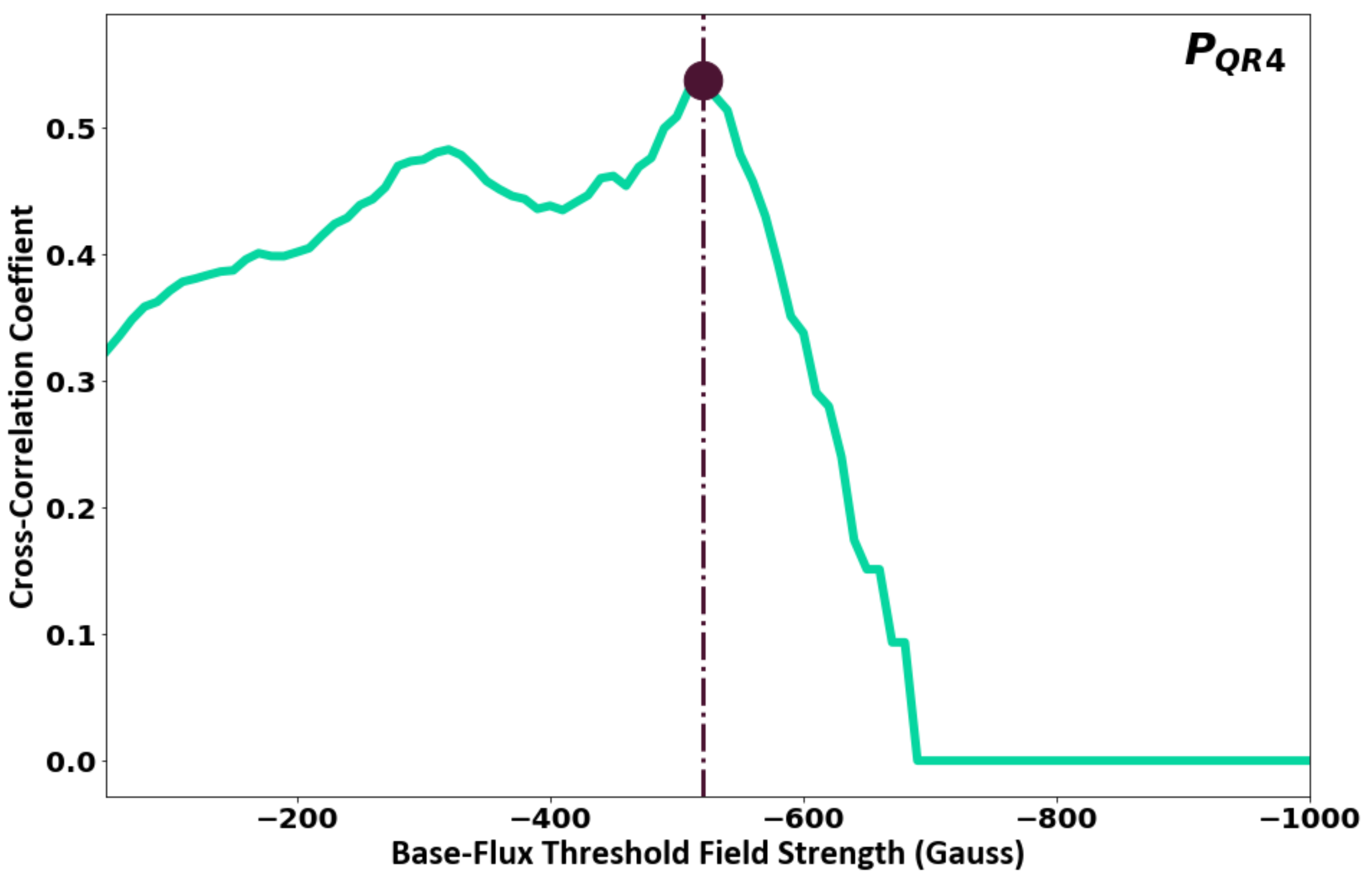}
\caption{Plots of cross-correlation coefficient (for the correlation of the plume’s AIA 171 \AA\, luminosity time profile with the plume-base HMI LOS magnetic flux time profile) versus the base-flux threshold field strength, from which we obtain the critical magnetic field strength for each example plume.  Each plot’s label (in the upper right corner) denotes the plume’s region type and the chronological order of the plume among our example plumes for that type of region (see Table \ref{t1}). The plots with negative x-axis values are for plumes that have negative-polarity flux. The dot in each plot marks the critical base-flux threshold field strength, i.e., the threshold field strength that gives the largest correlation coefficient, or the strongest correlation between the plume’s luminosity time profile and it’s base-flux time profile. Further details about each example can be found in Tables \ref{t1}, \ref{t2}, and \ref{t3}.
\label{fig:corr_coeff}}
\end{figure}

To determine the critical magnetic field strength for the formation of each plume, we selected potential minimum absolute pixel values, our threshold magnetic field strength values, and used those thresholds in the magnetic flux calculation over the plume's lifetime. We then calculated a correlation coefficient between the computed base flux profile and the AIA 171 \AA\, luminosity profile. Prior to computing correlation coefficients, we first smooth both profiles by a boxcar smoothing factor of 50, and use the smoothed plume luminosity and base flux measurements in our correlation coefficient calculation. We first determine correlation coefficients for thresholds ranging from 100 Gauss (G) to 1000 G. Based on \citet{2016ApJ...818..203W}'s observation that flux convergence and divergence directly correspond to plume formation and disappearance, we take the critical magnetic field strength to be the threshold value that yields the maximum correlation coefficient value, i.e. the correlation coefficient that is closest to one. The correlation coefficient plots used to determine the critical magnetic field strength for each plume are shown in Figure \ref{fig:corr_coeff}, where the dot indicates the critical magnetic field strength. 

\section{Results}
In Table \ref{t1} we summarize our results for all eight plume examples and two non-plume forming regions and introduce the labels we will use throughout the remainder of this paper. Our examples are organized and labeled based on region and start time with respect to other plumes in that region. Presented results include the label used to refer to the example throughout the remainder of the paper, location on the solar disk, duration, critical magnetic field strength, peak luminosity, and peak flux at the critical magnetic field strength threshold.

Based on our entire sample, we find that plumes require a critical magnetic field strength of between $\sim$ 200-600 G to become noticeably bright in AIA 171 \AA. The base magnetic flux plots used to find the critical magnetic field strengths were on average 70\%\, correlated with their corresponding AIA 171 \AA\, luminosity plots. Those plumes with lower correlation coefficients were found to have peaks in their flux plots that led or lagged the peaks in their luminosity plots. 

Although plumes in QRs and CHs are surrounded by different magnetic networks, and may or may not be the footprints of loop structures, we find that this has no effect on their behavior. Based on observing similar trends in duration and flux and luminosity measurements, we find that there is no difference between plumes in QRs and CHs.

Tables \ref{t2} and \ref{t3} show a more detailed comparison of base magnetic flux values for all plumes, with Table \ref{t2} showing peak base flux values for varying thresholds and Table \ref{t3} showing the number of pixels used to determine the base flux value. 

\begin{table}[ht!]
\centering
\begin{tabular}{cccccccc}
\hline
Label & \shortstack{Region,\\ Polarity} & \shortstack{Peak Location \\ (arcsec)} & \shortstack{Start Time \\ (UT)} & \shortstack{Duration \\ (hours)} & \shortstack{Critical Field \\ B$_{critical}$} (G)& \shortstack{Peak Luminosity \\ (DN/s)} & \shortstack{Peak Flux at \\ B = B\(_{critical}\) (Mx)} \\ \hline
\(P_{CH1}\) & Coronal Hole, + & (43, 258) & 5-Jul-2011 18:30 & 25.5 & 290 G & \(2.28\times 10^{6}\) & \(9.21\times 10^{19}\) \\
\(P_{CH2}\) & Coronal Hole, - & (-102,256) & 8-Sep-2011 16:00 & 24 & 160 G & \(1.20\times 10^{6}\) & \(1.04\times 10^{20}\) \\
\(P_{CH3}\) & Coronal Hole, - & (-302,223) & 5-Oct-2013 15:00 & 16 & 560 G & \(6.03\times 10^{5}\) & \(1.09\times 10^{19}\) \\
\(P_{CH4}\) & Coronal Hole, + & (-64,496) & 12-May-2016 12:30 & 19.5 & 250 G & \(7.05\times 10^{5}\) & \(4.17\times 10^{19}\) \\
\(P_{QR1}\) & Quiet Sun, - & (194,-294) & 1-Feb-2015 2:00 & 25.5 & 350 G & \(3.04\times 10^{6}\) & \(1.08\times 10^{20}\) \\
\(P_{QR2}\) & Quiet Sun, - & (185,-391) & 4-Dec-2015 7:00 & 14.5 & 570 G & \(1.73\times 10^{6}\) & \(1.56\times 10^{19}\) \\
\(P_{QR3}\) & Quiet Sun, + & (-178,302) & 23-Dec-2016 20:00 & 14 & 510 G & \(1.19\times 10^{6}\) & \(1.59\times 10^{19}\) \\
\(P_{QR4}\) & Quiet Sun, - & (-305,236) & 12-May-2017 13:00 & 13 & 520 G & \(1.55\times 10^{6}\) & \(7.04\times 10^{18}\) \\
\(P_{no1}\) & Coronal Hole, + & (75, 307) & 5-Jul-2011 18:30 & 25.5 & N/A & \(5.58\times 10^{5}\) & N/A \\
\(P_{no2}\) & Quiet Sun, - & (-69,-153) & 5-Aug-2015 0:00 & 16.5 & N/A & \(3.08\times 10^{5}\) & N/A \\ \hline
\end{tabular}
\caption{Location, duration, critical magnetic field strength, peak luminosity, and peak flux at the critical magnetic field strength threshold for all eight plume examples and two non-plume forming examples. We can see that there is no significant difference between coronal hole plumes and quiet Sun plumes with respect to the response of the luminosity in the corona to the behavior of the magnetic field at the base.}
\label{t1}
\end{table}

\begin{table}[ht!]
\centering
\begin{tabular}{cccccc}
\hline
Label & \shortstack{Peak Flux at \\ T = 200 G (Mx)} & \shortstack{Peak Flux at \\ T = 300 G (Mx)} & \shortstack{Peak Flux at \\ T = 400 G (Mx)} & \shortstack{Peak Flux at \\ T = 500 G (Mx)} & \shortstack{Peak Flux at \\ T = 600 G (Mx)} \\ \hline
\(P_{CH1}\) & \(1.17\times 10^{20}\) & \(8.99\times 10^{19}\) & \(6.44\times 10^{19}\) & \(4.34\times 10^{19}\) & \(2.83\times 10^{19}\) \\
\(P_{CH2}\) & \(9.76\times 10^{19}\) & \(7.63\times 10^{19}\) & \(6.05\times 10^{19}\) & \(3.83\times 10^{19}\) & \(2.39\times 10^{19}\) \\
\(P_{CH3}\) & \(5.44\times 10^{19}\) & \(3.89\times 10^{19}\) & \(2.62\times 10^{19}\) & \(1.82\times 10^{19}\) & \(8.59\times 10^{18}\) \\
\(P_{CH4}\) & \(4.59\times 10^{19}\) & \(3.71\times 10^{19}\) & \(2.94\times 10^{19}\) & \(2.22\times 10^{19}\) & \(1.53\times 10^{19}\) \\
\(P_{QR1}\) & \(1.41\times 10^{20}\) & \(1.17\times 10^{20}\) & \(9.73\times 10^{19}\) & \(7.86\times 10^{19}\) & \(5.90\times 10^{19}\) \\
\(P_{QR2}\) & \(3.83\times 10^{19}\) & \(3.02\times 10^{19}\) & \(2.31\times 10^{19}\) & \(1.83\times 10^{19}\) & \(1.40\times 10^{19}\) \\
\(P_{QR3}\) & \(4.96\times 10^{19}\) & \(3.43\times 10^{19}\) & \(2.42\times 10^{19}\) & \(1.66\times 10^{19}\) & \(1.22\times 10^{19}\) \\
\(P_{QR4}\) & \(3.09\times 10^{19}\) & \(2.15\times 10^{19}\) & \(1.38\times 10^{19}\) & \(8.37\times 10^{18}\) & \(3.39\times 10^{18}\) \\
\(P_{no1}\) & \(6.46\times 10^{19}\) & \(4.33\times 10^{19}\) & \(2.70\times 10^{19}\) & \(1.60\times 10^{19}\) & \(8.18\times 10^{18}\) \\
\(P_{no2}\) & \(4.65\times 10^{19}\) & \(3.32\times 10^{19}\) & \(1.95\times 10^{19}\) & \(1.34\times 10^{19}\) & \(8.19\times 10^{18}\) \\ \hline
\end{tabular}
\caption{Peak fluxes at threshold values of 200 G, 300 G, 400 G, 500 G, and 600 G for all eight plume examples and two non-plume forming examples. }
\label{t2}
\end{table}

\begin{table}[ht!]
\centering
\begin{tabular}{cccccc}
\hline
Label & \shortstack{Number of Pixels \\ for T = 200 G} & \shortstack{Number of Pixels \\ for T = 300 G} & \shortstack{Number of Pixels \\ for T = 400 G} & \shortstack{Number of Pixels \\ for T = 500 G} & \shortstack{Number of Pixels \\ for T = 600 G} \\ \hline
\(P_{CH1}\) & 252 & 160 & 96 & 55 & 32 \\
\(P_{CH2}\) & 193 & 123 & 88 & 49 & 25 \\
\(P_{CH3}\) & 121 & 72 & 41 & 25 & 10 \\
\(P_{CH4}\) & 91 & 64 & 41 & 28 & 18 \\
\(P_{QR1}\) & 266 & 169 & 129 & 90 & 61 \\
\(P_{QR2}\) & 77 & 53 & 32 & 22 & 15 \\
\(P_{QR3}\) & 119 & 61 & 34 & 21 & 14 \\
\(P_{QR4}\) & 73 & 38 & 22 & 11 & 4 \\
\(P_{no1}\) & 154 & 80 & 42 & 21 & 10 \\
\(P_{no2}\) & 106 & 61 & 30 & 17 & 9 \\
\hline
\end{tabular}
\caption{The number of pixels used to compute the base magnetic flux at various thresholds for all eight plume examples and two non-plume forming examples. The number of pixels is an indication of the size of the base of the plume at peak convergence for a certain threshold value.}
\label{t3}
\end{table}

We can see that the peak flux values at varying thresholds are within an order of magnitude of each other for both QR and CH examples. A similar number of pixels above the given threshold field strength is also found in all examples, indicating no specific difference between plume-forming and non plume-forming flux concentrations. While we expected non-plume forming regions to exhibit lower peak flux values over a smaller number of pixels, the similarity in flux to plume-forming patches could arise from a difference in magnetic field strength over different concentrations of strong-field pixels, i.e. perhaps plume-forming regions have a higher concentration of strong-field pixels over a smaller area, while non-plume forming regions have strong-field pixels distributed over a larger area.

To present further details about our results, we focus on three examples: one plume in a CH, one plume in a QR, and one flux concentration which does not produce a plume. The two plume-forming examples were chosen as our highlighted examples because of how well the flux and luminosity plots correlate, while the non-plume forming example was chosen because of how poorly the flux and luminosity plots correlate. 

For the two plume-forming examples, we show an AIA 211 \AA\, image of the plume's location with respect to the solar disk and frames of the plume in AIA 171 \AA\, and HMI LOS magnetograms over its entire lifetime at the pre-plume stage, brightening stage, peak brightness, dimming stage, and post-plume stage. The magnetograms in these frames are saturated at \(\pm\)100 G. We also provide the following plots: a plot showing the normalized luminosity of the plume over its entire lifetime along with three normalized flux plots which have threshold values above, below, and at the critical magnetic field strength value, and a plot showing the normalized luminosity with the best-fitting normalized flux plot alone. For our non-plume forming example, we show frames of the plume in AIA 171 \AA\, and HMI LOS magnetograms over its entire lifetime at the pre-flux-convergence stage, flux-convergence stage, peak-flux convergence, flux-divergence stage, and post-flux-divergence stage. The magnetograms in these frames are also saturated at \(\pm\)100 G. We also provide plots of luminosity in AIA 171 \AA\, over the flux patch's lifetime along with the magnetic flux at thresholds of 100 G, 200 G, and 300 G over the patch's lifetime and a plot showing the magnetic flux at a threshold of 200 G alone in order to clearly show base flux convergence and divergence. As supplementary material, we also provide movies of AIA 171 \AA\, image data and HMI LOS magnetogram data for all our examples. The movies are named using the same labels given in Table \ref{t1}. The aforementioned figures for the remaining seven examples not presented in this section can be found in the appendix ordered by region and observation date.

\subsection{Coronal Hole Plume}
\begin{figure}[ht!]
\gridline{\fig{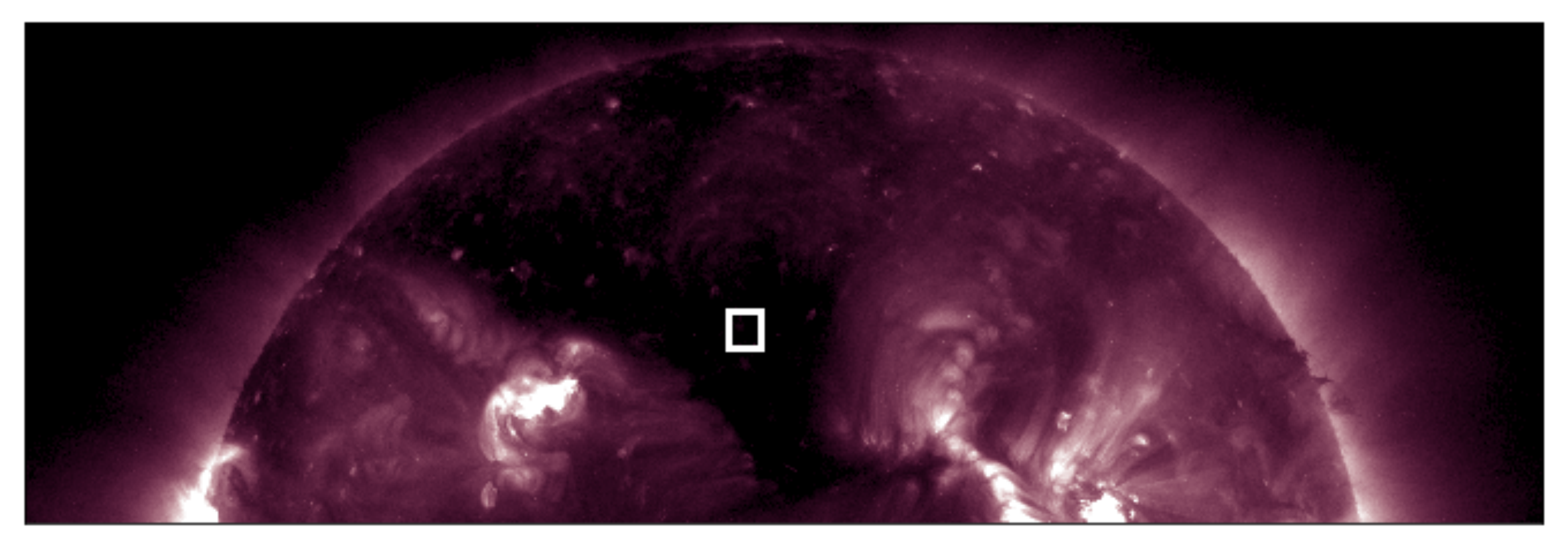}{0.92\textwidth}{(a)}}
\gridline{\fig{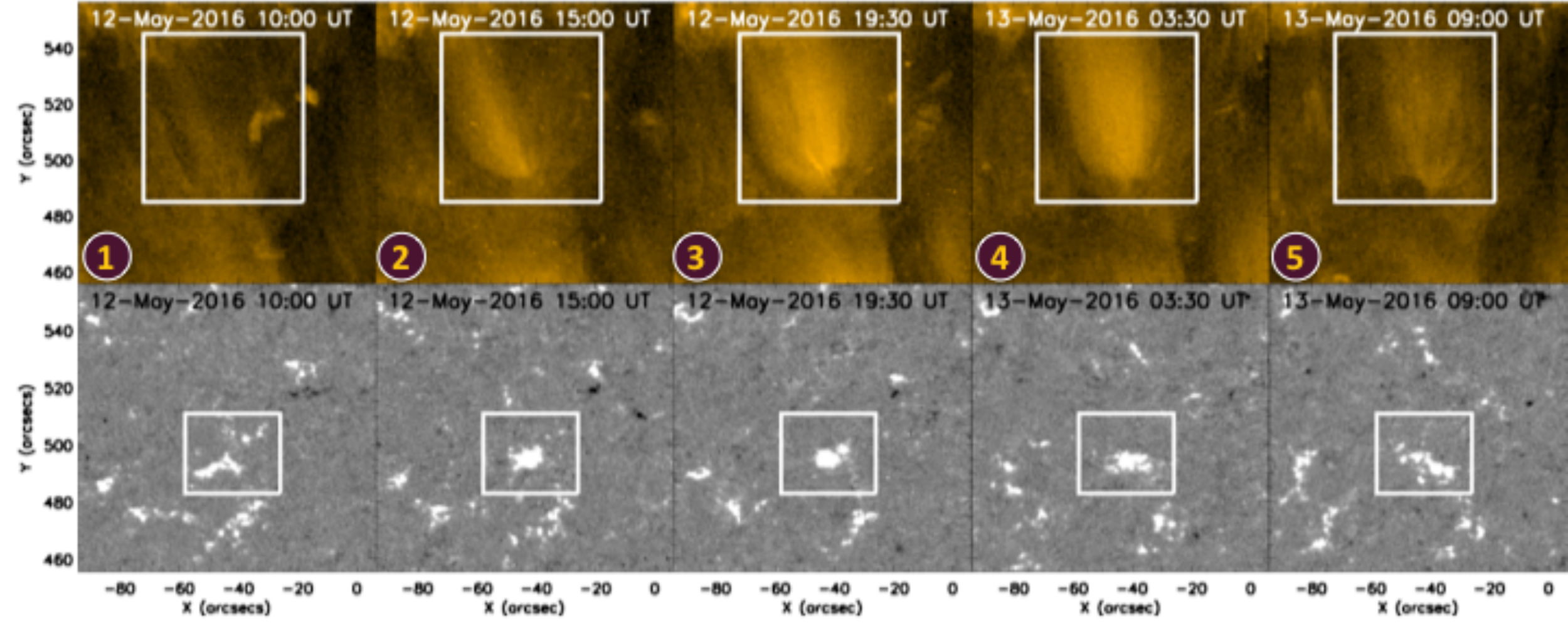}{0.99\textwidth}{(b)}}
\gridline{\fig{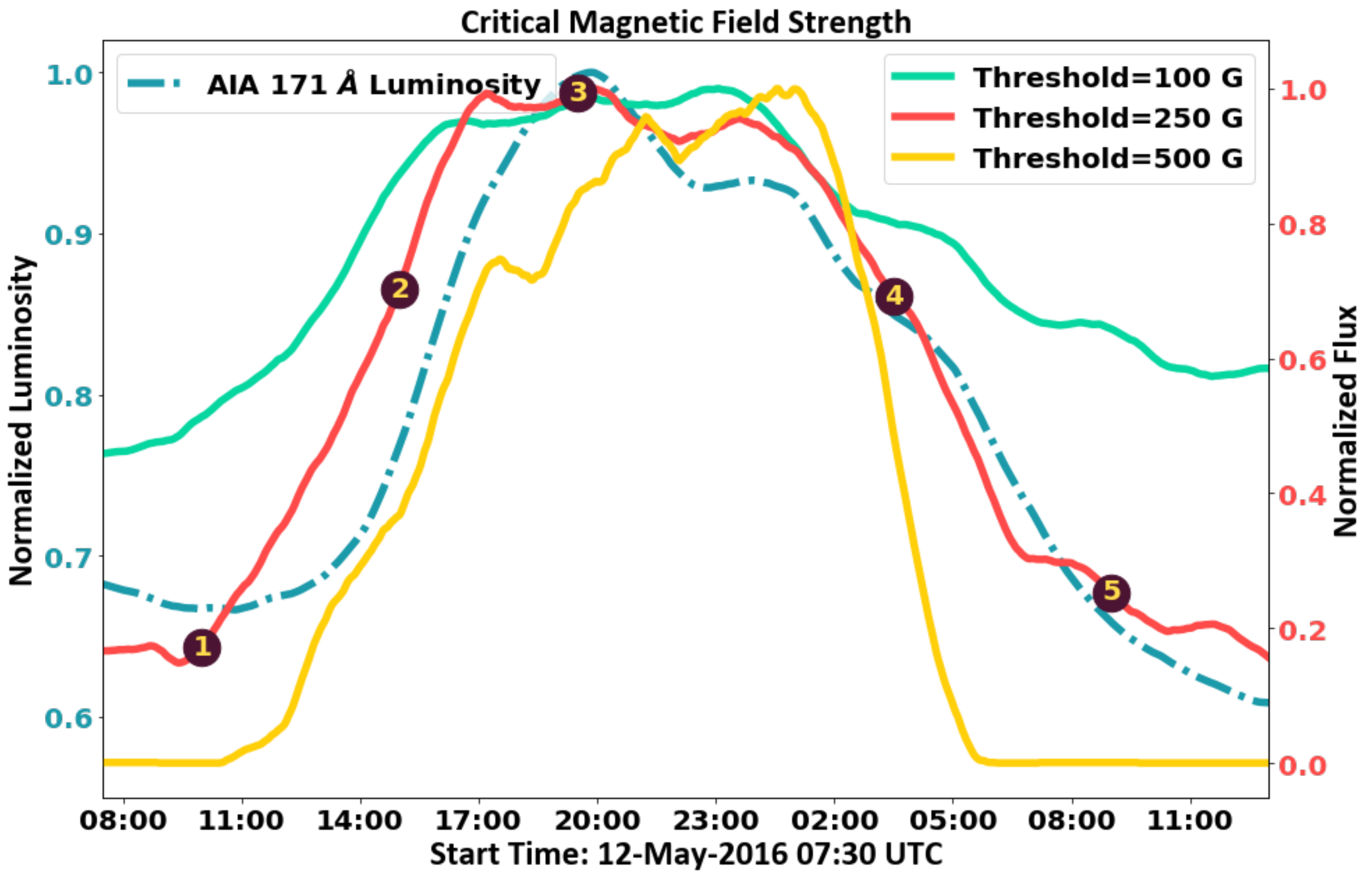}{0.49\textwidth}{(c)}
            \fig{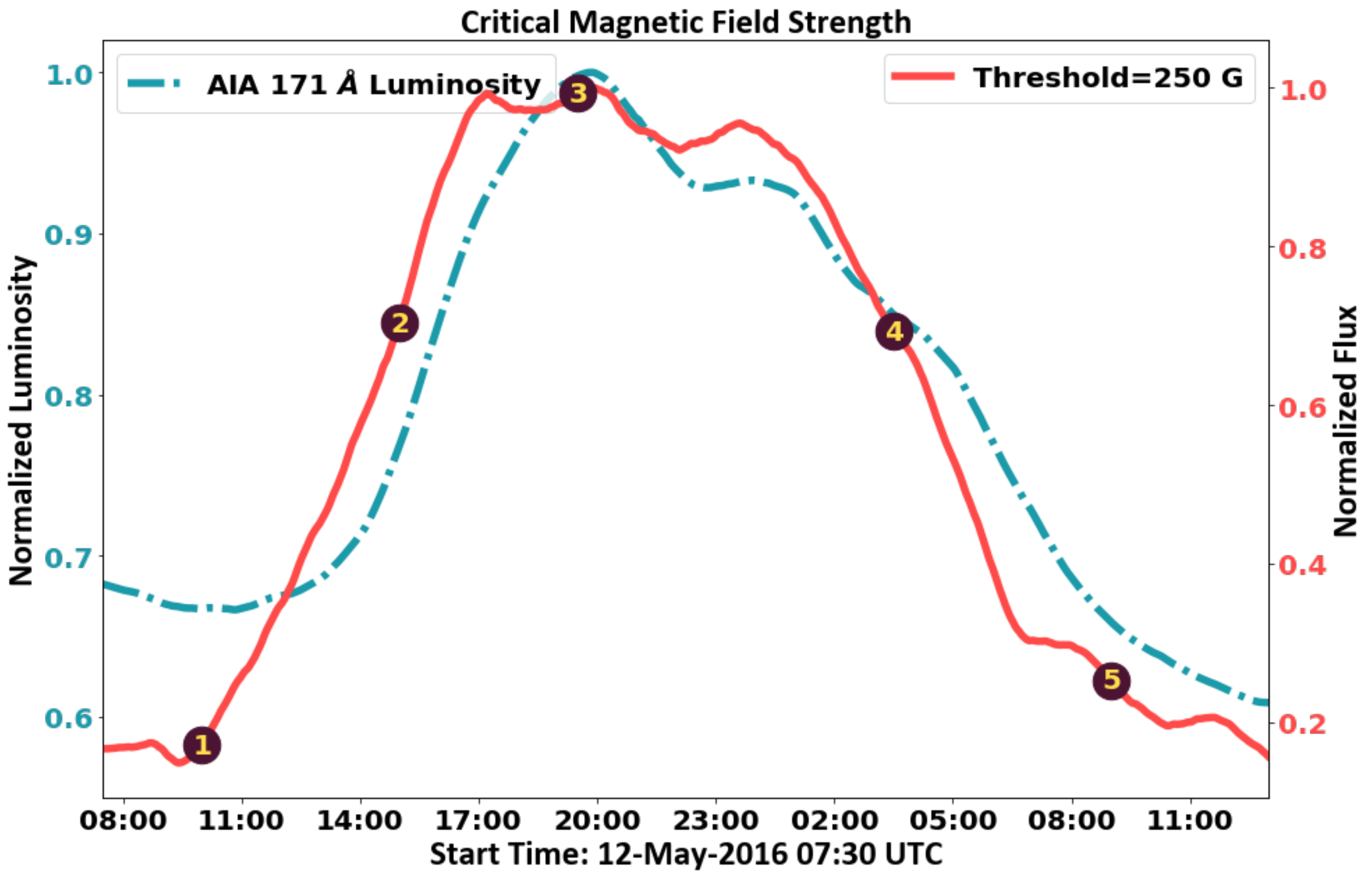}{0.49\textwidth}{(d)} \vspace{-0.2cm}}
\caption{\(P_{CH4}\), which occurred from May 12, 2016 to May 13, 2016. (a) shows an AIA 211 \AA\, image of the Northern half of the solar disk at 19:30 UTC, when the plume is at peak intensity (best seen in AIA 171 \AA, displayed in the upper row of panel (b)). The white box in (a) corresponds to the region in which AIA 171 \AA\, measurements were taken. We can see that \(P_{CH4}\) is inside a PCH. (b) shows AIA 171 \AA\, images and HMI magnetograms of \(P_{CH4}\) throughout its lifetime. HMI magnetograms are saturated at \(\pm\)100 G. (c) shows \(P_{CH4}\)'s normalized luminosity over its lifetime along with the normalized flux over its lifetime at three different thresholds: 100 G, 250 G, and 500 G. (d) shows the normalized luminosity and normalized flux over its lifetime at its critical magnetic field strength threshold, 250 G. The dots on (c) and (d) correspond to the frame locations in (b). A movie of AIA 171 \AA\ images and HMI magnetograms, CH4{\_}movie.mp4, is available, beginning at 12-May-2016 07:30 UTC and ending at 13-May-2016 13:00 UTC with a 47 second duration.
\label{CH4}}
\end{figure}

We first present a coronal hole plume example, \(P_{CH4}\). This plume, which occurred from May 12, 2016 to May 13, 2016, was located in a PCH in the northern hemisphere. \(P_{CH4}\) has the similar hazy, funnel-like appearance of other observed plumes. Figure \ref{CH4}a shows an AIA 211 \AA\, image showing the plume's location relative to the solar disk. Figure \ref{CH4}b shows AIA 171 \AA\, images and HMI magnetograms \(P_{CH4}\) over its lifetime. Figure \ref{CH4}c and \ref{CH4}d show measurements of normalized luminosity and flux. Other properties of this plume can be found in Tables \ref{t1}, \ref{t2}, and \ref{t3}.

Figure \ref{CH4}c shows the normalized luminosity curve with three different normalized flux plots, which have thresholds at 100 G, 250 G, and 500 G. We can see that for thresholds both below 250 G and far above 250 G, the flux plot peaks after the luminosity plot. The best-fitting flux plot with a threshold of 250 G is 94\% correlated with the normalized luminosity plot. We will only discuss the luminosity plot and the best fitting flux plot in Figure \ref{CH4}d for the remainder of this section. To supplement our discussion of Figure \ref{CH4}d, we will also address the data in CH4{\_}movie.mp4. 

Our data for \(P_{CH4}\) begins on May 12, 2016 at 07:30 UTC. We begin to notice an increase in brightness from the background CH region at 14:30 UTC, which directly follows increased flux convergence at 13:30 UTC. We can see in Figure \ref{CH4}d that the start of flux convergence leads the start of plume brightening by about 2 hours. The plume then sharply increases in brightness from 15:00 UTC to 18:00 UTC, directly following the sharp increase in flux. The plume reaches peak intensity at 19:42 UTC, slightly before the peak in flux at 19:56 UTC. During this increase in brightness we see the distinct funnel shape start to form, and see the base flux become more condensed. From 16:30 UTC to 19:30 UTC, we see the plume plasma orient itself primarily along the field lines at the edge of the flux patch, leaving a region in the center of the flux patch free of plume material. We observe this same behavior in \(P_{CH2}\), \(P_{QR3}\), and \(P_{QR4}\), which can be seen in the respective movies of these plumes.  

About the peak in flux and luminosity, we see the flux patch condense into two distinct patches, one of which is much larger than the other, at $\sim$18:35 UTC. At around the same time, we see the plume cover the entire flux patch, leaving no regions visibly free of plume material, for the remainder of its lifetime. These two patches are present until $\sim$21:00 UTC when they coalesce back into a single base flux patch. The plume steadily diminishes until a secondary peak in the luminosity and flux plots emerges from $\sim$00:30 UTC until $\sim$1:00 UTC. This secondary peak corresponds to a few transient bright points which emerge near the base of the plume and last for several minutes. The base flux begins to substantially disperse at $\sim$03:00 UTC, which corresponds to a decrease in plume intensity in AIA 171 \AA\, data. Although this plume's base flux indicates the presence of a single polarity well, 03:00 UTC is one of a few instances where we do observe flux cancellation. We see minority-polarity flux get swept in to the majority polarity base on the left side of the flux patch. We can see some brightening happen in the corresponding AIA 171 \AA\, image near the left side of the plume, but that brightening is quite diffuse. The plume continues to diminish, reaching the initial background coronal hole brightness around 08:00 UTC on May 13, 2016. During the dimming stage, we see some bright points, small regions of increased intensity that last for a few minutes, in the field of view of the movie. However, these are extraneous to the plume and are not included in the box used for AIA 171 \AA\, measurements. 

\subsection{Quiet Sun Plume}
\begin{figure}[ht!]
\gridline{\fig{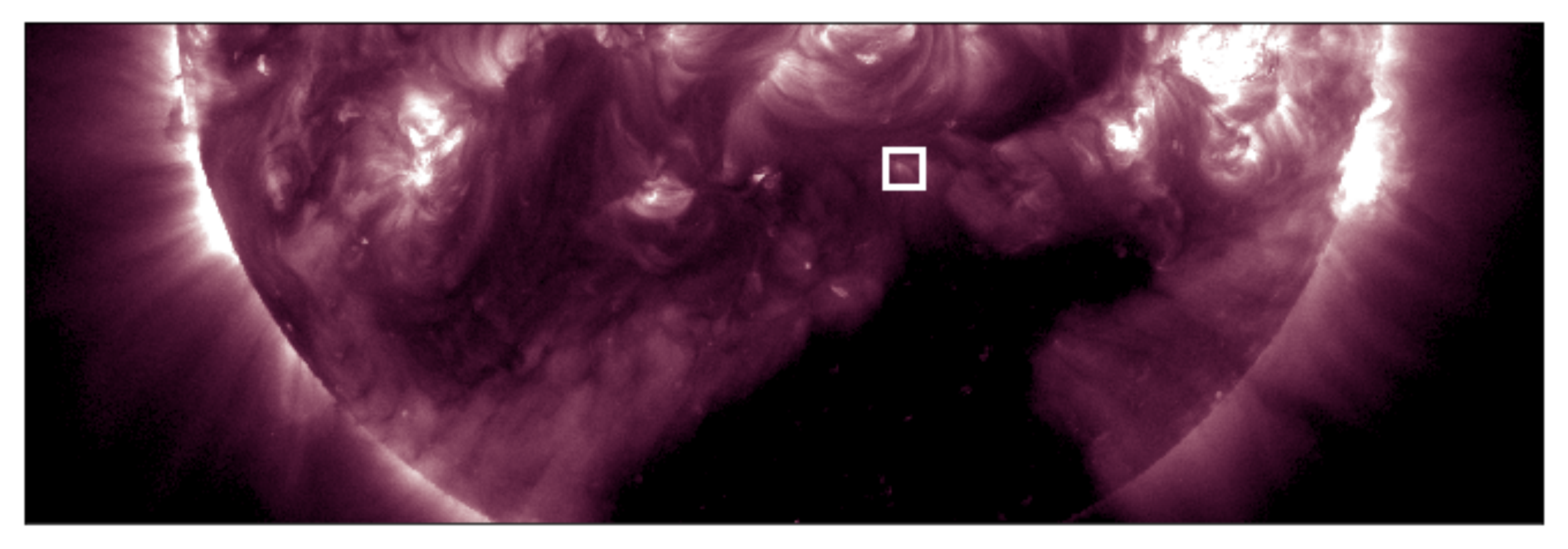}{0.92\textwidth}{(a)}}
\gridline{\fig{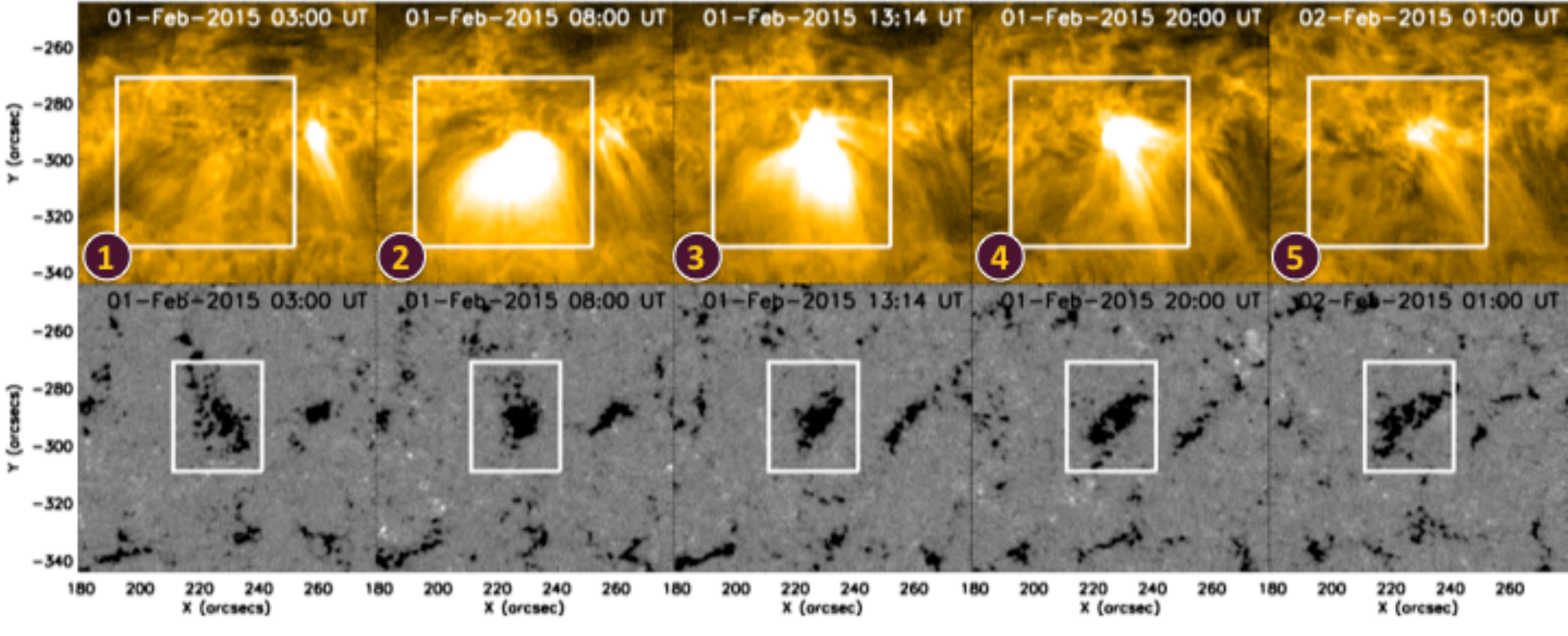}{0.99\textwidth}{(b)}}
\gridline{\fig{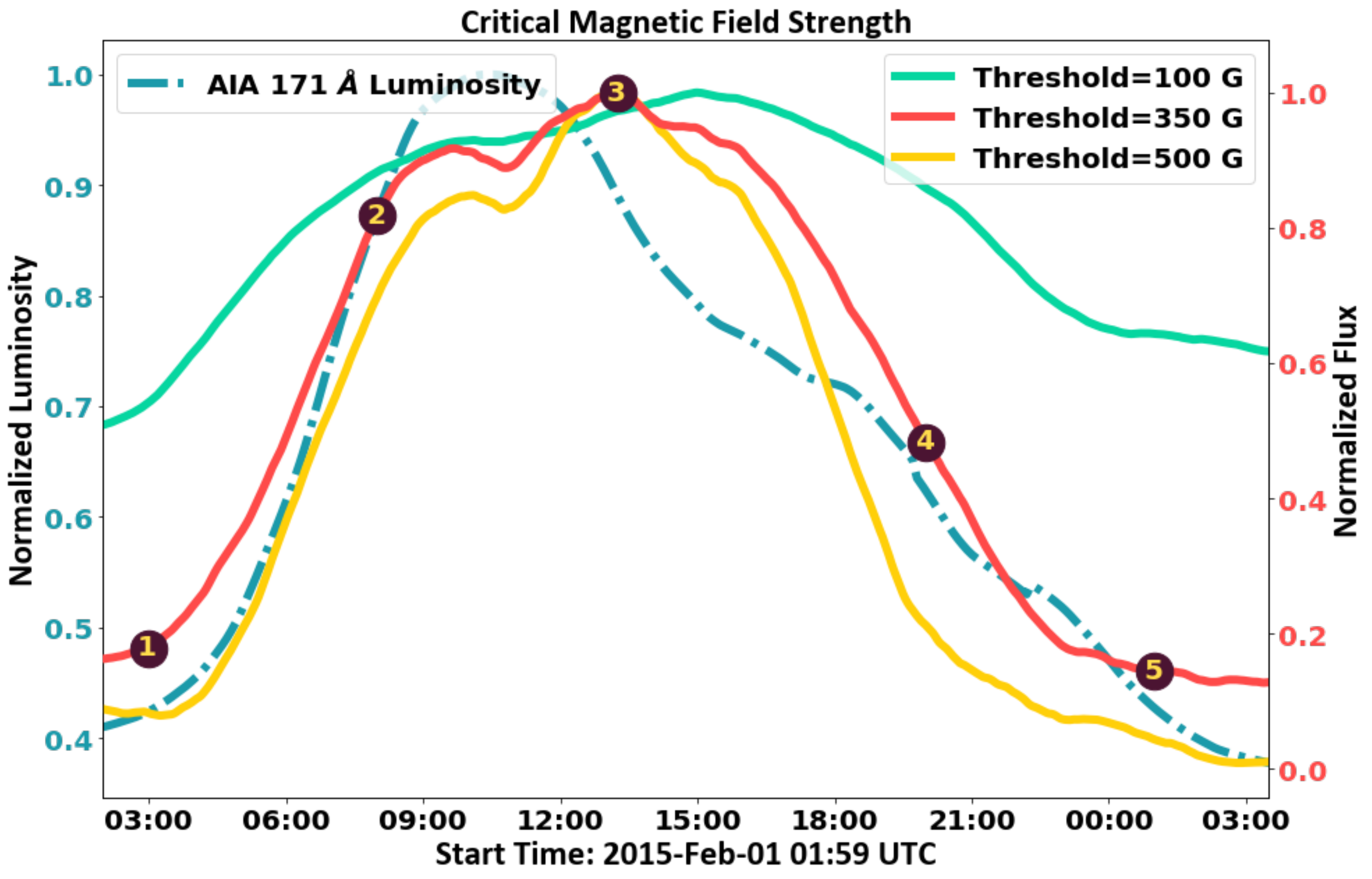}{0.49\textwidth}{(c)}
            \fig{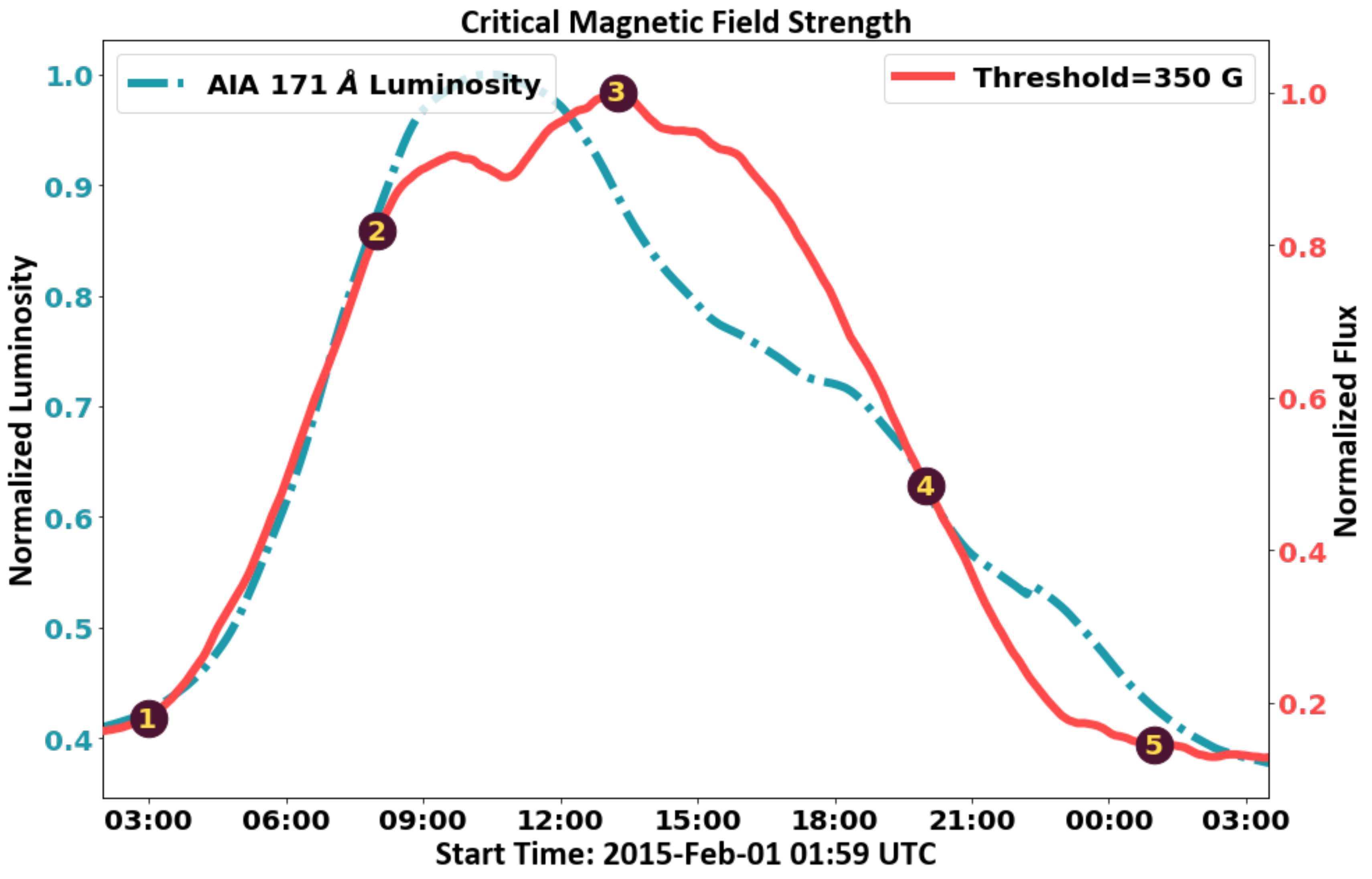}{0.49\textwidth}{(d)}}
\caption{\(P_{QR1}\), which occurred on February 1, 2015. (a) shows an AIA 211 \AA\, image of the Southern half the solar disk at 10:30 UTC, when the plume is at peak intensity (best seen in AIA 171 \AA, displayed in the upper row of panel (b)). The white box in (a) corresponds to the region in which AIA 171 \AA\, measurements were taken. We can see that \(P_{QR1}\) is in a non-CH QR. (b) shows AIA 171 \AA\, images and HMI magnetograms of \(P_{QR1}\) throughout its lifetime. HMI magnetograms are saturated at \(\pm\)100 G. (c) shows \(P_{QR1}\)'s normalized luminosity over its lifetime along with the normalized flux over its lifetime at three different thresholds: 100 G, 350 G, and 500 G. (d) shows the normalized luminosity and normalized flux over its lifetime at the critical magnetic field strength threshold, 350 G. The dots on (c) and (d) correspond to the frame locations in (b). A movie of AIA 171 \AA\ images and HMI magnetograms, QR1{\_}movie.mp4, is available, beginning at 1-Feb-2015 02:00 UTC and ending at 2-Feb-2015 03:30 UTC with a 41 second duration.
\label{QR1}}
\end{figure}

We now present a QR plume example, \(P_{QR1}\). This plume, which occurred from February 1, 2015 to February 2, 2015, was located in a QR in the southeast quadrant of the solar disk. Figure \ref{QR1}a shows an AIA 211 \AA\, image showing the plume's location relative to the solar disk. Figure \ref{QR1}b shows AIA 171 \AA\, images and HMI magnetograms of \(P_{QR1}\) over its lifetime. Figures \ref{QR1}c and \ref{QR1}d show measurements of normalized luminosity and flux. Further properties of this plume can be found in Tables \ref{t1}, \ref{t2}, and \ref{t3}.

Figure 4c shows the normalized luminosity curve with three different normalized flux plots, which have thresholds at 100 G, 350 G, and 500 G. We can see that for all thresholds both below 350 G and above 350 G, the flux plot peaks after the luminosity plot. The best-fitting flux plot with a threshold of 350 G is 94\% correlated with the normalized luminosity plot. We will only discuss the luminosity plot and the best fitting flux plot in Figure \ref{QR1}d for the remainder of this section. To supplement our discussion of Figure \ref{QR1}d, we will also address the data in QR1{\_}movie.mp4. 

Our data for \(P_{QR1}\) begins on February 1, 2015 at 02:00 UTC. We notice the plume starts around 04:00 UTC at the same time the base flux starts to converge. The plume's brightness steadily increases until it reaches peak luminosity at 10:30 UTC at the same time as a secondary peak in the base flux. The flux peak follows the peak in luminosity, reaching its maximum value at 13:14 UTC. Unlike \(P_{CH4}\), throughout the brightening stage, the base of the plume in AIA 171 \AA\, is concentrated near the center of the base flux patch and covers most of the area. However, the plume does frequently break up into smaller structures throughout its lifetime, both before and after the peaks in luminosity and flux. One instance where we see the base flux break up is slightly before the peak brightness at 10:18 UTC. We see a much smaller plume-like structure form at the top of \(P_{QR1}\). This structure forms above a smaller flux patch that converges at the top of the base flux patch but is still a part of the main base flux patch. This structure is present in AIA 171 \AA\, images until 13:30 UTC, when it coalesces back into the main plume.

After peak brightness, the plume diminishes until the patch reaches the background quiet region brightness at around 02:00 UTC on February 2, 2015. While this patch also shows the presence of a single polarity well, we see some minute flux cancellation during the dimming stage. This flux cancellation occurs from $\sim$19:00 UTC to $\sim$20:00 UTC, but no noticeable brightening in AIA 171 \AA\, results from the flux cancellation. During the dimming period, we also see more fluctuations in the plume structure. From 15:00 UTC to 17:30 UTC, we see a few fragments extending near the top of the field of view from the base of the plume. These fragments eventually condense back into the main plume structure. At 18:21 UTC, we see the plume break up again into three distinct plume-like structures. At the same time the base flux patch becomes more dispersed, but still appears as one cohesive structure. These three structures condense back into a single point at 19:45 UTC before disappearing entirely on February 2, 2015 at 03:00 UTC. 

\subsection{Non-Plume Forming Region}
\begin{figure}[ht!]
\gridline{\fig{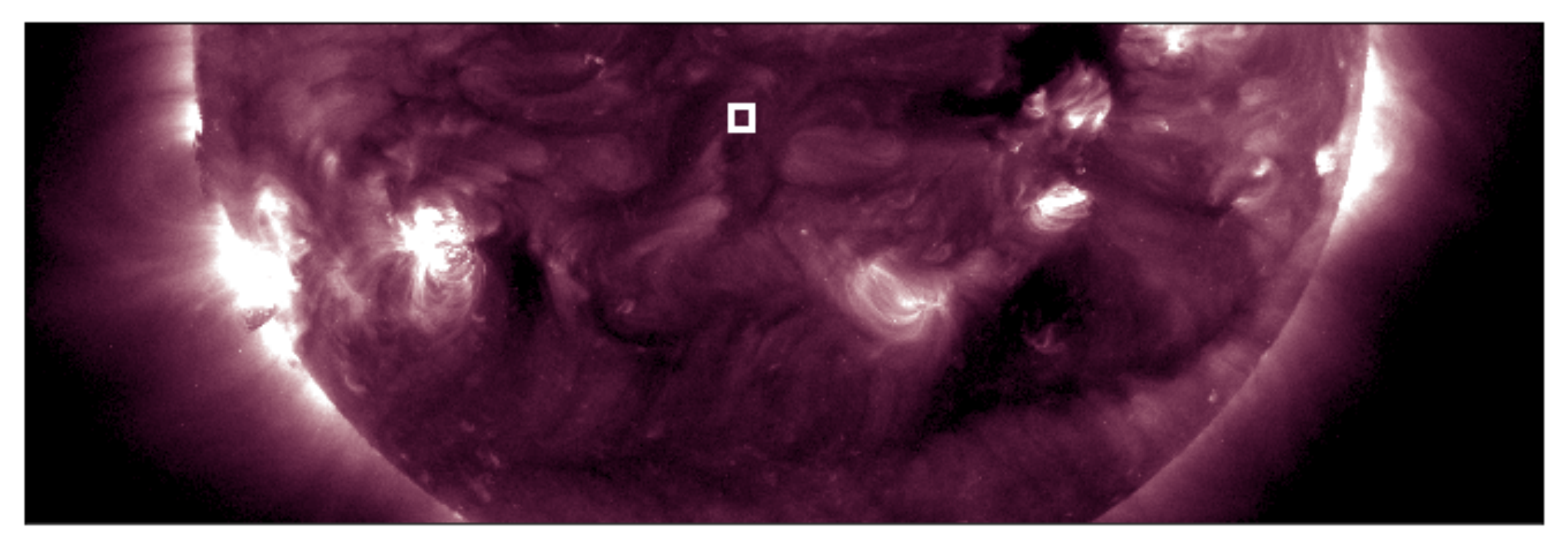}{0.92\textwidth}{(a)}}
\gridline{\fig{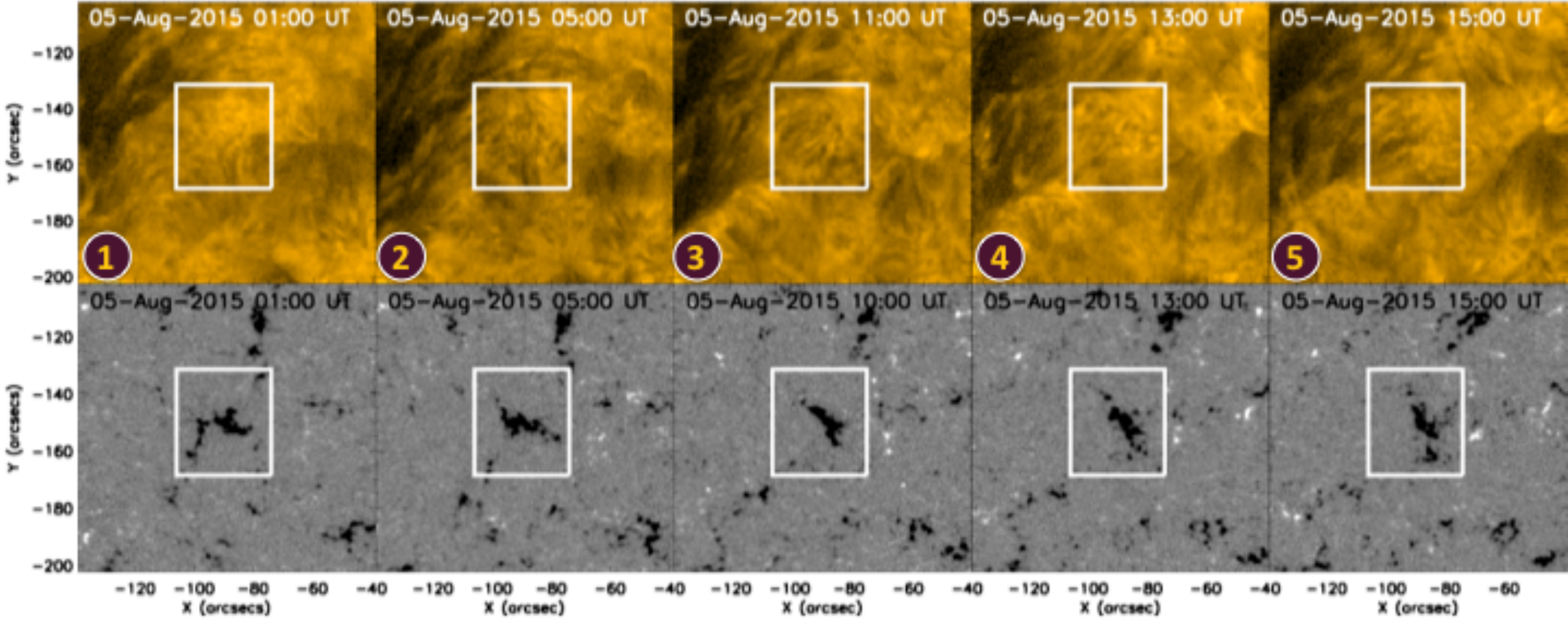}{0.99\textwidth}{(b)}}
\gridline{\fig{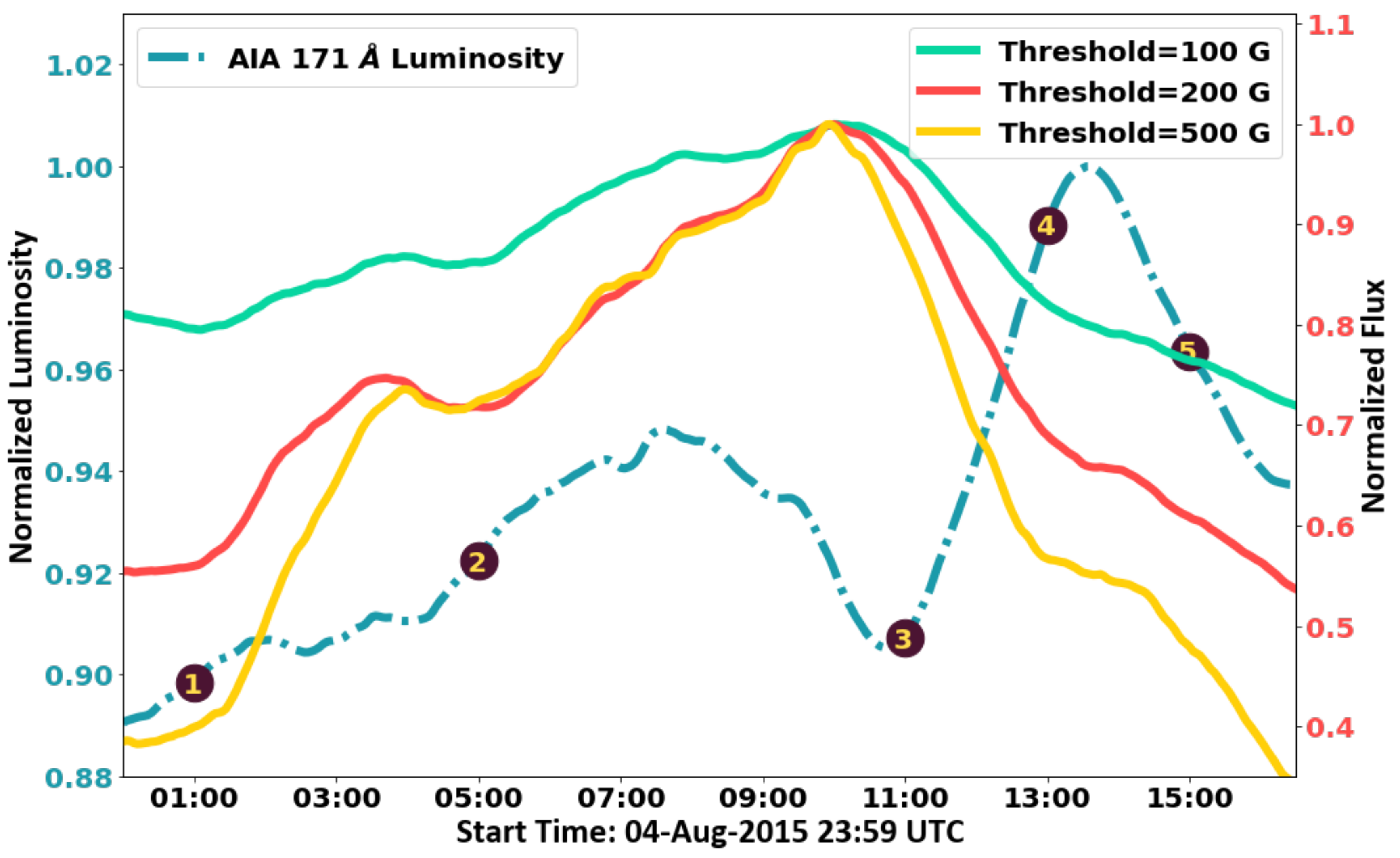}{0.49\textwidth}{(c)}
            \fig{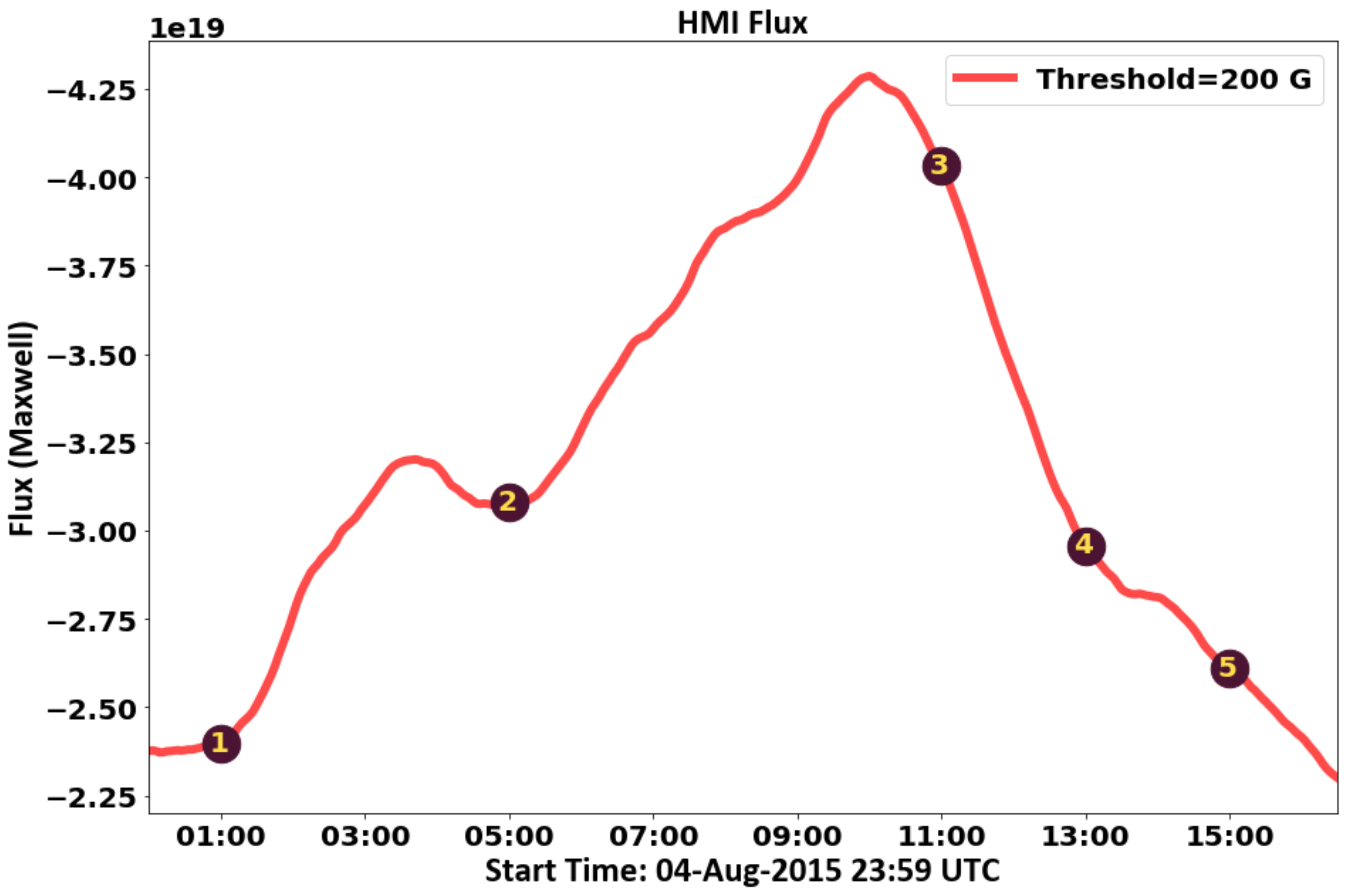}{0.49\textwidth}{(d)}}
\caption{\(P_{no2}\), a photospheric magnetic flux concentration which does not form a plume. This event occured on August 5, 2015 in a QR near the center of the solar disk. (a) shows an AIA 211 \AA\, image of half the solar disk at 11:00 UTC, when the flux patch is at peak convergence. The white box in (a) corresponds to the region in which AIA 171 \AA\, measurements were taken, and we can see that \(P_{no2}\) is in a QR. (b) shows AIA 171 \AA\, images and HMI magnetograms of \(P_{no2}\) throughout its lifetime. HMI magnetograms are saturated at \(\pm\)100 G. We can see that the flux plot (d) follows the same trend as the patches that form plumes. However, the luminosity of the region in AIA 171 \AA, shown in (c) with flux plots at thresholds of 100 G, 200 G, and 300 G, does not follow the plume trend exhibited by the flux plots or get as bright as flux concentrations which form plumes. The dots on (c) and (d) correspond to the frame locations in (b). A movie of AIA 171 \AA\ images and HMI magnetograms, no2{\_}movie.mp4, is available, beginning at 5-Aug-2015 00:00 UTC and ending at 5-Aug-2015 16:27 UTC with a 26 second duration.
\label{no2}}
\end{figure}

Finally, we present our non-plume forming photospheric magnetic flux concentration. This flux concentration, which occurred on August 5, 2015, was located in a QR near the center of the solar disk. Figure \ref{no2}a shows an AIA 211 \AA\, image showing the plume's location relative to the solar disk. Figure \ref{no2}b shows AIA 171 \AA\, images and HMI magnetograms \(P_{no2}\) over its lifetime. Figures \ref{no2}c and \ref{no2}d show measurements of normalized luminosity and flux. Further properties of this flux concentration can be found in Tables \ref{t1}, \ref{t2}, and \ref{t3}. 

Figure \ref{no2}c shows the normalized luminosity curve with three different normalized flux plots, which have thresholds at 100 G, 200 G, and 300 G. We can see that for all threshold values, the flux plot does not follow the same shape as the luminosity plot. Throughout \(P_{no2}\)'s lifetime, we do see some small bright points occasionally emerge in the AIA 171 \AA\, data, but these are extraneous to the flux patch, and do not arise as a result of the patch's behavior or substantially affect our measurements in AIA 171 \AA. Figure \ref{no2}d shows the base flux over the patch's lifetime at a threshold of 200 G alone, which will allow us to more easily discern fluctuations in the plot in our analysis. We will discuss only Figures \ref{no2}c and \ref{no2}d for the remainder of this section. To supplement our discussion of Figures \ref{no2}c and \ref{no2}d, we will also address the data in no2{\_}movie.mp4. 

Our data for \(P_{no2}\) begins on August 5, 2015 at 00:00 UTC. We notice the flux concentration start to converge at 00:30 UTC, converging more substantially at 02:00 UTC. However, as the flux concentration converges, we do not notice any brightening in AIA 171 \AA\, images. The flux concentration continues to converge until it reaches peak convergence at 10:00 UTC. After the peak, the flux concentration sharply diverges, reaching the background flux level at ~16:00 UTC. 

Throughout \(P_{no2}\)'s lifetime, the flux concentration primarily exists as one cohesive structure, with few instances where the concentration breaks up into smaller structures. However, we do see several instances where smaller regions of the same polarity get swept into the primary flux concentration. One such instance occurs from 02:30 UTC to 04:30 UTC, where we see a small area of flux converge slightly above and to the left of the primary flux concentration and then get swept into the rest of the flux concentration.

\section{Discussion and Conclusions}
Based on AIA and HMI observations of a sample of eight apparently unipolar on-disk coronal plumes, we find that plume formation requires enough convergence of the magnetic flux at the plume's base to surpass a base LOS magnetic field strength of $\sim$ 200-600 G. Formation and disappearance of plumes is observed to be directly linked to flux convergence and divergence, supporting the findings of \citet{2016ApJ...818..203W}. However, we see flux patches converge and diverge all over the Sun that do not form plumes, evoking this investigation of whether a sufficient flux convergence leading to a strong enough base magnetic field, or critical magnetic field strength, in a sufficient number of pixels is necessary for plume formation. 

Our observations suggest that the EUV visibility of apparently-unipolar coronal plumes depends on the amount and strength of magnetic field at their base along with base flux convergence and divergence. However, through observing two flux concentrations that do not form plumes, along with our eight plume-forming examples, we find that these concentrations exhibit similar peak flux values and number of pixels above the magnetic field threshold values. Thus, we conclude that although the convergence of flux to surpass a certain magnetic field strength is necessary, it is evidently not sufficient to trigger plume heating. This implies that there are additional mechanism(s) that trigger and/or sustain plume heating. 

One such mechanism could be small-scale magnetic flux cancellation at the plume-base, leading to the formation of numerous jetlets. Although the plumes in this investigation don't exhibit any obvious mixed polarity, and through our random sampling of plumes we find that plumes with apparently unipolar base flux are common whereas plumes with obvious mixed polarity are rare, we cannot rule out the possibility that there is hidden mixed polarity in the plume base flux that is unresolved by HMI, and that hidden mixed polarity may be a factor that contributes to plume heating. This idea is further supported by \citet{2016ApJ...818..203W}, who invariably found loop-like structures in the base of plumes in 171 \AA, which indicate the presence of mixed-polarity flux in visibly unipolar plumes.

A caveat to considering flux cancellation with hidden opposite-polarity field is that such a small amount of flux may not be sufficient to sustain a plume for its lifetime of days, unless a continuous generation of such small-scale minority-polarity flux is taking place. A flux cancellation rate of the order of \(10^{18}\) Mx/h, on average, is observed for jets in QRs and CHs (\citet{2016ApJ...832L...7P} and \citet{2018ApJ...853..189P}). If plumes are a result of numerous small coronal jets (jetlets), a much smaller amount of flux cancellation would be needed. The fact that jet activity stops after the minority polarity flux disappears in some example coronal jets in \citet{2016ApJ...832L...7P} and \citet{2018ApJ...853..189P}, further challenges the idea of plumes being formed by flux cancellation. Future generation solar telescopes like DKIST \citep{2016AN....337.1064T}, and advanced magnetohydrodynamic simulations, should clarify the issue of small-scale hidden mixed polarity field and its role in driving plume heating. 

We do observe some flux emergence and cancellation which produce transient bright points like those seen in \citet{2014ApJ...787..118R} in a few plumes, but none of these last longer than a few minutes. While there is the possibility that mixed polarity contributes to triggering and sustaining plume heating, the aforementioned short events, with obvious patches of opposite-polarity field, do not seem to be a significant contributor to plume heating. This also casts doubt on flux cancellation by hidden magnetic polarity being responsible for plume heating.

The other possible mechanism for plume heating could be the heating by waves. The recently proposed mechanism of turbulence build up and dissipation \citep{2018ApJ...854...32Z} could be an alternative possibility. In any case (whether heating is from mixed-polarity, waves, or turbulence) magnetoconvection drives the heating in plumes, similar to that seen for active regions \citep[e.g.,][]{2017ApJ...843L..20T}.

It has traditionally been thought that plumes are only present in CH regions. However, we observed similar structures in QRs and found that they have a similar structure and similar durations, peak luminosities, and peak magnetic fluxes as plumes in CHs. Thus, despite the fact that QR plumes are most likely the base of large-scale closed loops, unlike the plumes in CHs, which are apparently the base of open field structures, plumes in the two regions are still similar in their coronal emission and magnetic properties and likely have the same heating mechanism. \citet{2016ApJ...818..203W} surmised that if magnetic reconnection due to the presence of mixed polarity is the primary mechanism for plume heating, converging supergranular flows would drive reconnection between long, closed loops and small bipoles to form and sustain heating in QR plumes. 

There is a noticeable difference in the peak times of the magnetic flux and the luminosity in all QR plumes. This behavior is unexpected, and part of this could probably be attributed to the fact that, unlike CH regions, QR plumes have more minority-polarity flux (as a part of the closed field magnetic carpet: \citet{1998ASPC..154..345T}) thus making the isolation of base flux more difficult. 

The filling factor of QR and CH magnetic flux may be far from one (this may or may not be true for strong field regions e.g., sunspots: see for example \citet{2005ApJ...621..498K}), as considered in our calculations. However because we take only pixels into account above certain values ($\sim$200 - 600 G) of magnetic field, the pixels with a large amount of noise (and thus a smaller signal and smaller filling factors) are automatically avoided.       

We plan to extend this work to include more examples in order to further investigate plume production. Since we used the 171 \AA\, channel from AIA which shows Fe IX emission at about 0.6-0.8 MK, plumes forming at much cooler temperatures are not seen. We plan to include transition region emission lines using the Interface Region Imaging Spectrograph (IRIS) \citep{2014SoPh..289.2733D} to follow up on whether or not a cooler plume is formed with the convergence of magnetic flux, and whether or not this leads to a weaker critical magnetic field strength than that found for the plumes seen in AIA 171 \AA. We will use observations from IRIS to closely observe jet eruptions in the chromosphere and transition region in plume bases and flow patterns throughout plume lifetimes. We will also measure the abundance and strength of the magnetic field in these plumes using SDO/HMI and/or Hinode/SOT \citep{2008SoPh..249..167T}, since our range of $\sim$200-600 G may not apply to all plumes found in CHs and QRs.

\acknowledgments
Ellis Avallone would like to thank the University of Alabama in Huntsville and NASA Marshall Space Flight Center for hosting the REU program where this research was conducted. She would also like to thank Sophie Hourihane for her illustration in Figure \ref{plume_diagram}. This research is supported by the National Science Foundation under grant No. AGS-1460767. Sanjiv K. Tiwari gratefully acknowledges support by NASA contract NNG09FA40C (IRIS).

\bibliographystyle{aasjournal}
\bibliography{refs}

\appendix
\begin{figure}[ht!]
\centering
\gridline{\fig{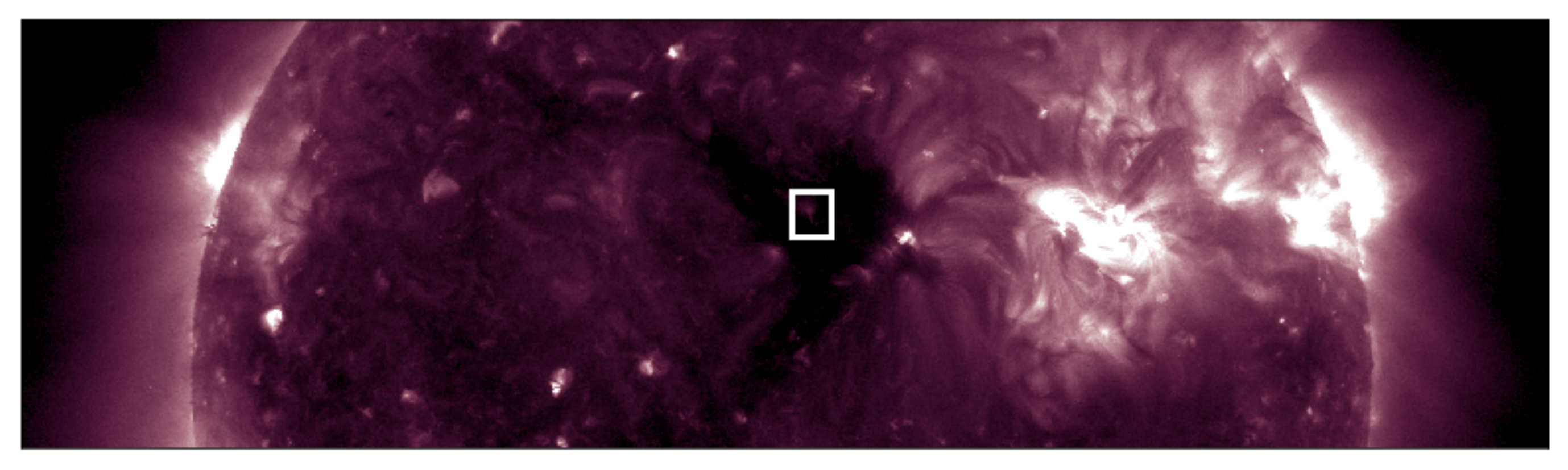}{0.92\textwidth}{(a)}}
\gridline{\fig{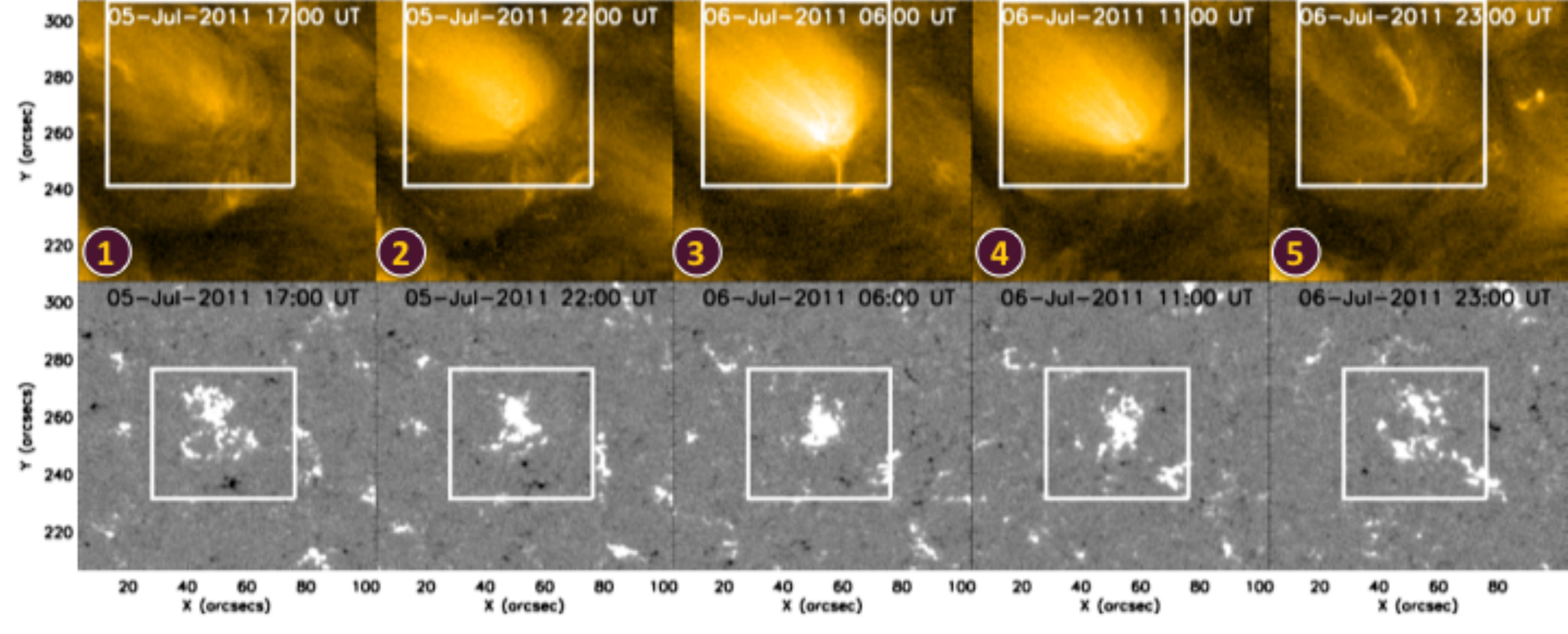}{0.99\textwidth}{(b)}}
\gridline{\fig{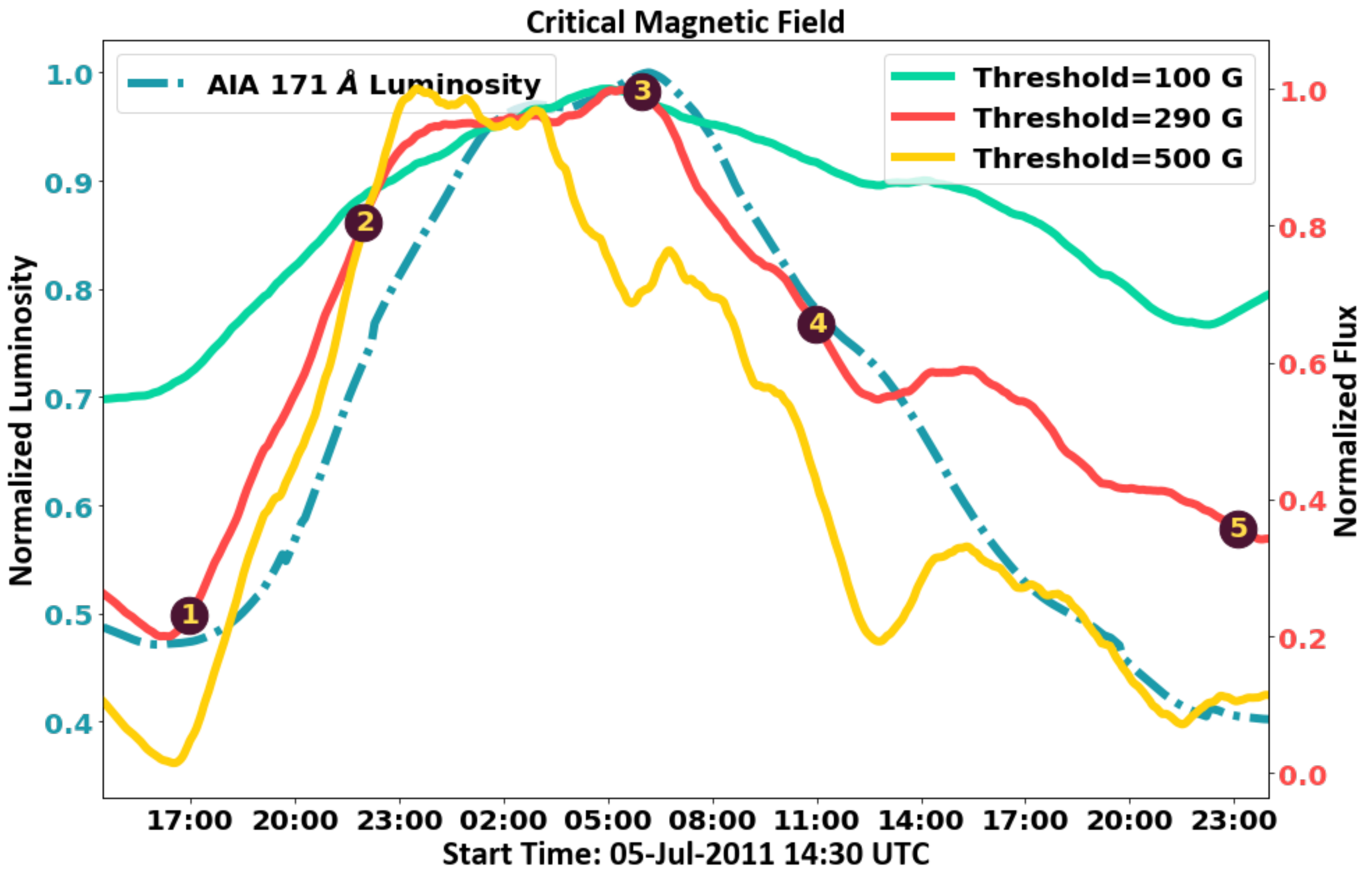}{0.49\textwidth}{(c)}
            \fig{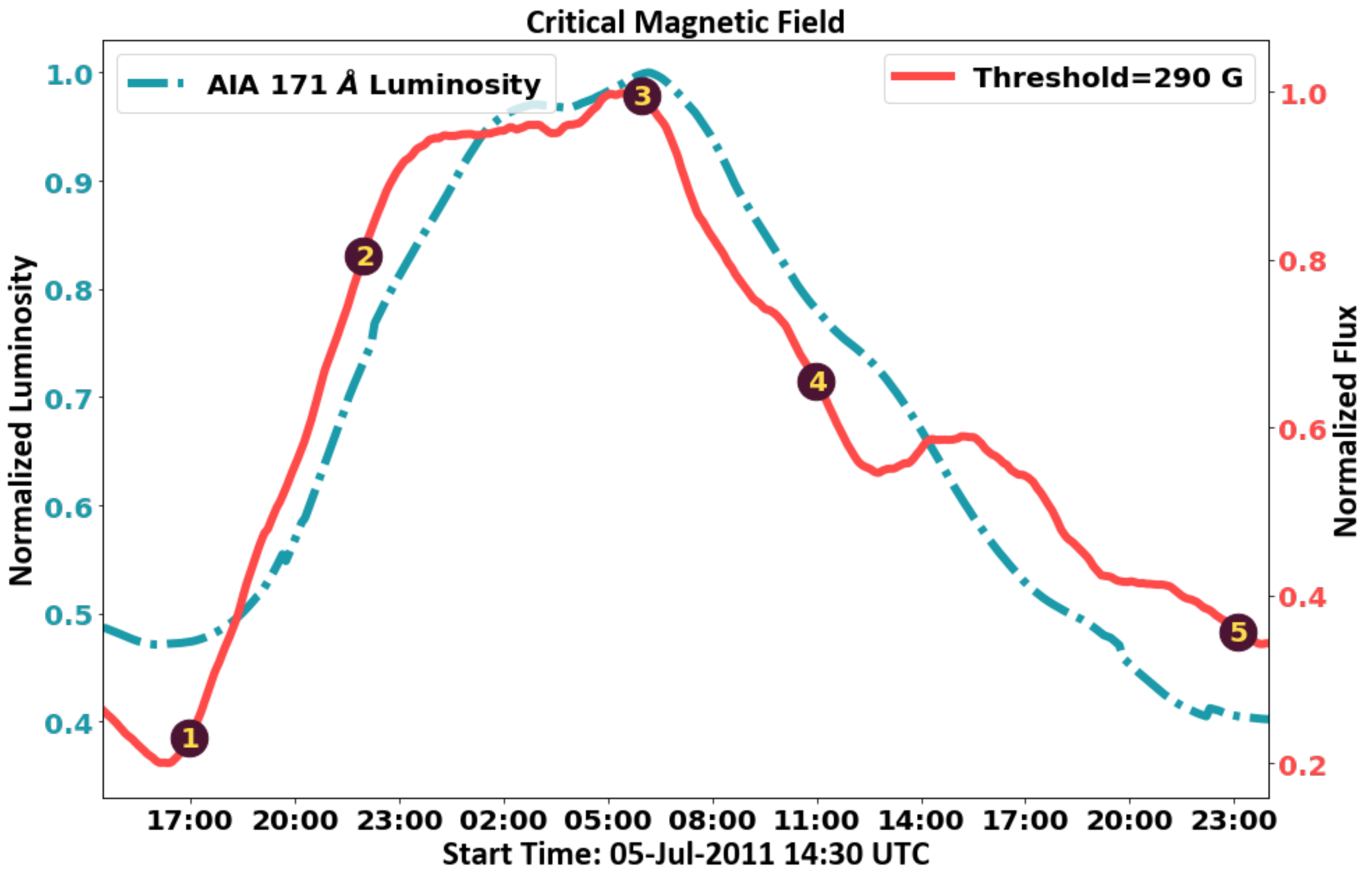}{0.49\textwidth}{(d)}}
\caption{\(P_{CH1}\), which occurred on July 5, 2011. (a) shows an AIA 211 \AA\, image of half the solar disk at 06:00 UTC on July 6, when the plume is at peak intensity (best seen in AIA 171 \AA, displayed in the upper row of panel (b)). The white box in (a) corresponds to the region in which AIA 171 \AA\, measurements were taken. We can see that \(P_{CH1}\) is in an ECH. (b) shows AIA 171 \AA\, images and HMI magnetograms of \(P_{CH1}\) throughout its lifetime. HMI magnetograms are saturated at \(\pm\)100 G. (c) shows this plume's normalized luminosity over its lifetime along with the normalized flux over its lifetime at three different thresholds: 100 G, 290 G, and 500 G. (d) shows the normalized luminosity and normalized flux over its lifetime at the critical magnetic field strength threshold, 290 G. The dots on (c) and (d) correspond to the frame locations in (b). The flux and luminosity data peak at approximately the same time, within one hour of each other. A movie of AIA 171 \AA\ images and HMI magnetograms, CH1{\_}movie.mp4, is available, beginning at 5-Jul-2011 14:30 UTC and ending at 6-Jul-2011 23:54 UTC with a 53 second duration.
\label{CH1}}
\end{figure}

\begin{figure}[ht!]
\gridline{\fig{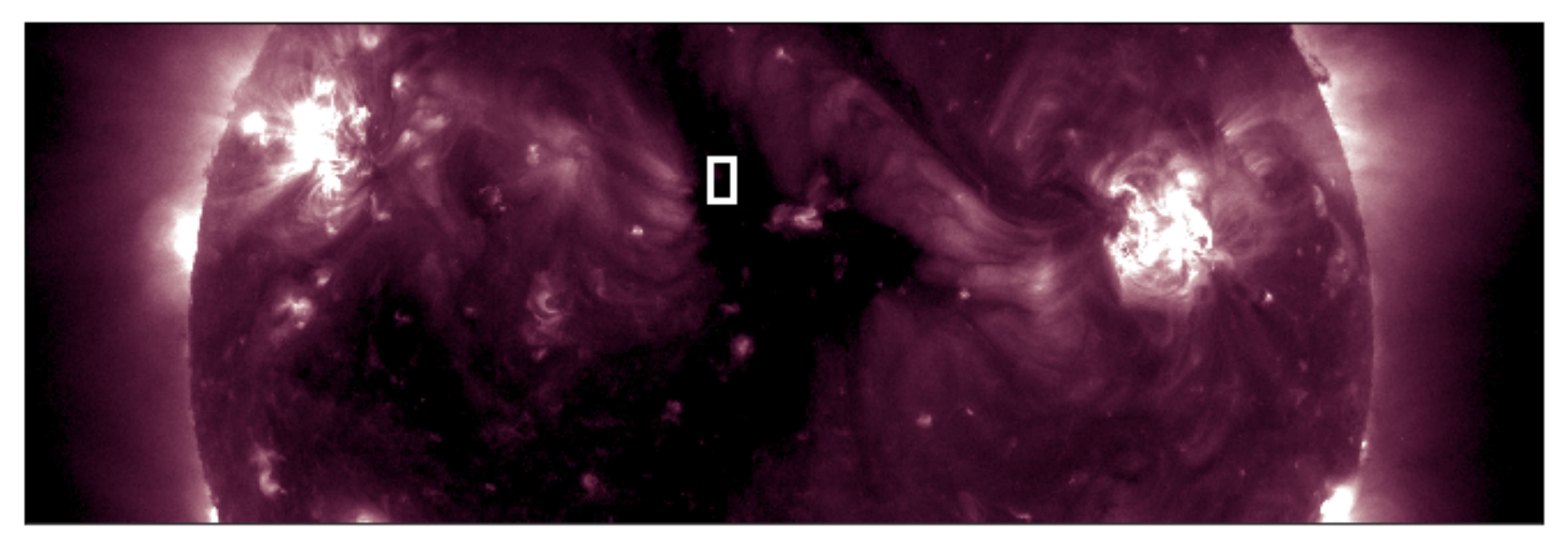}{0.92\textwidth}{(a)}}
\gridline{\fig{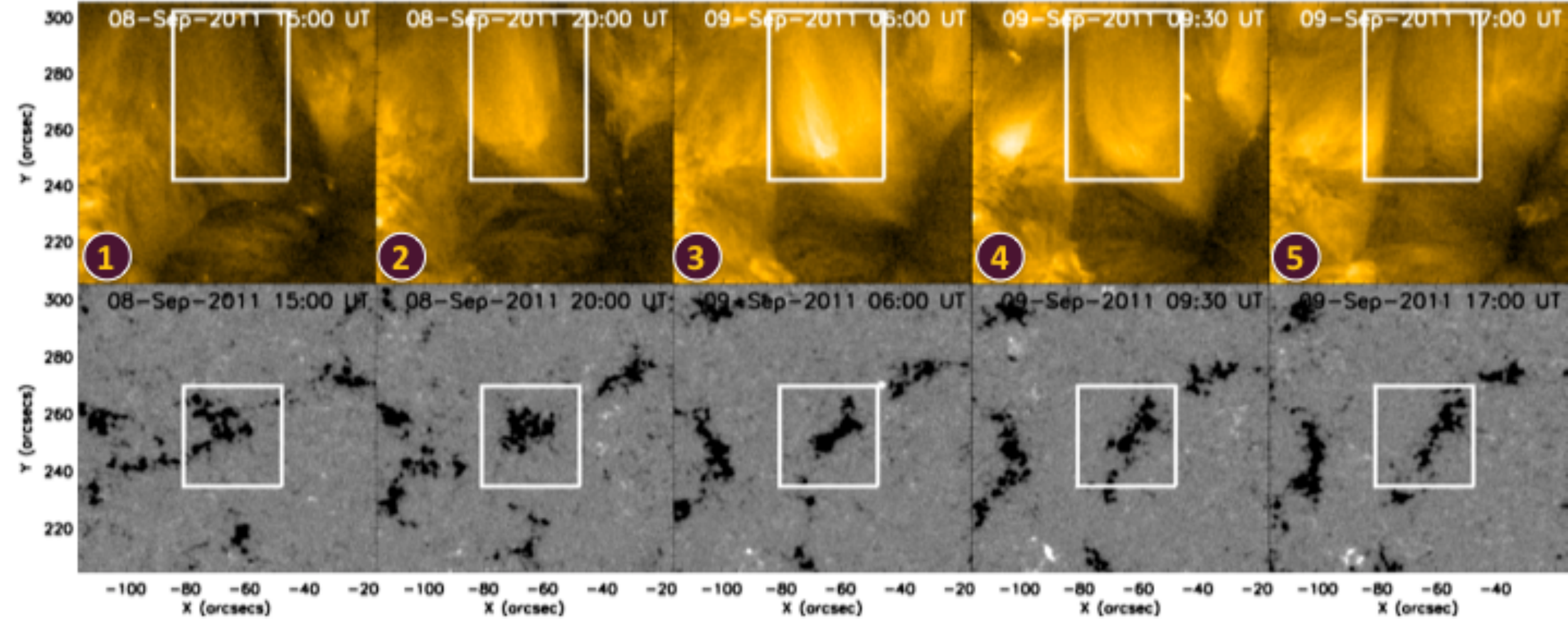}{0.99\textwidth}{(b)}}
\gridline{\fig{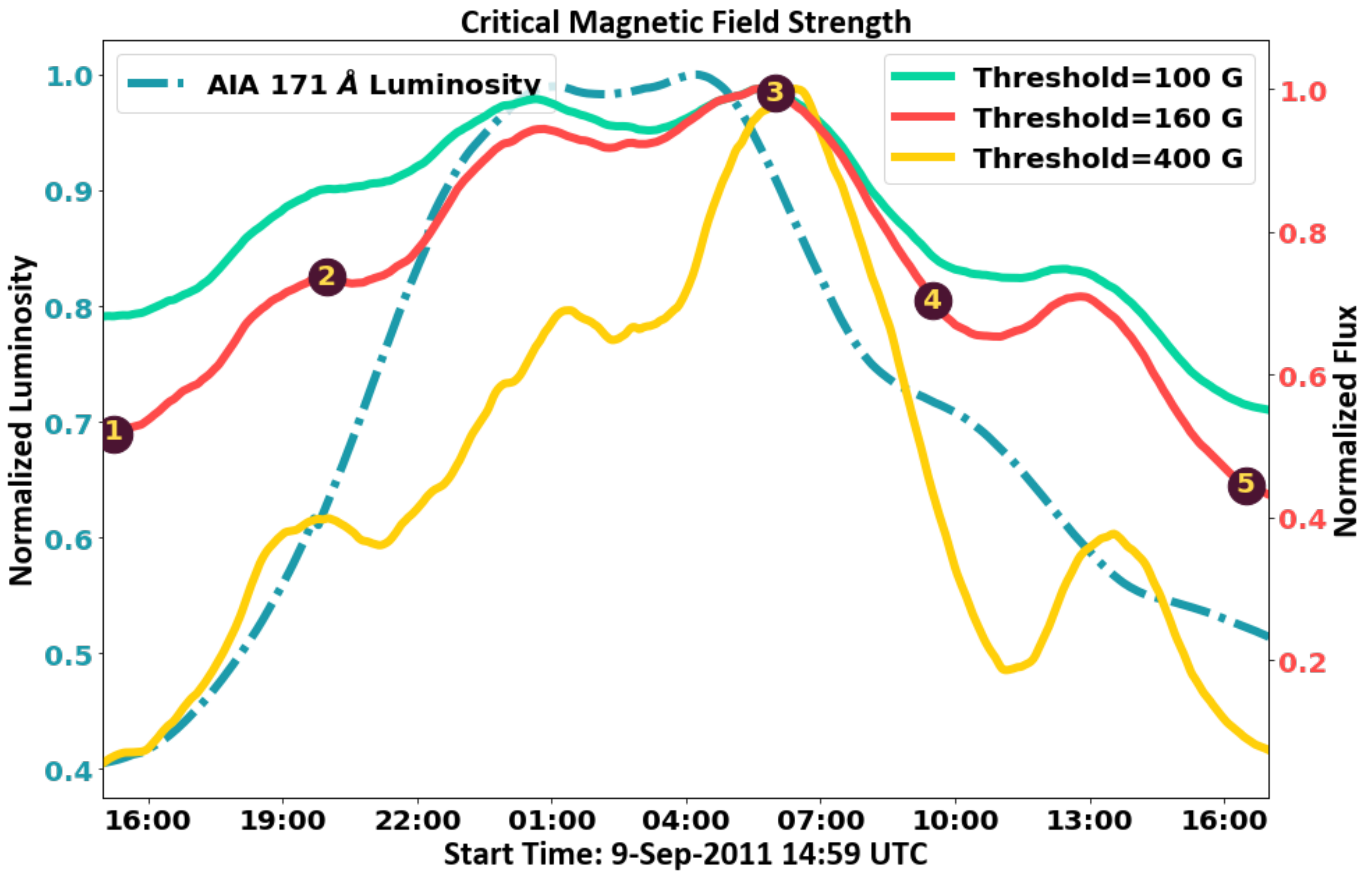}{0.49\textwidth}{(c)}
            \fig{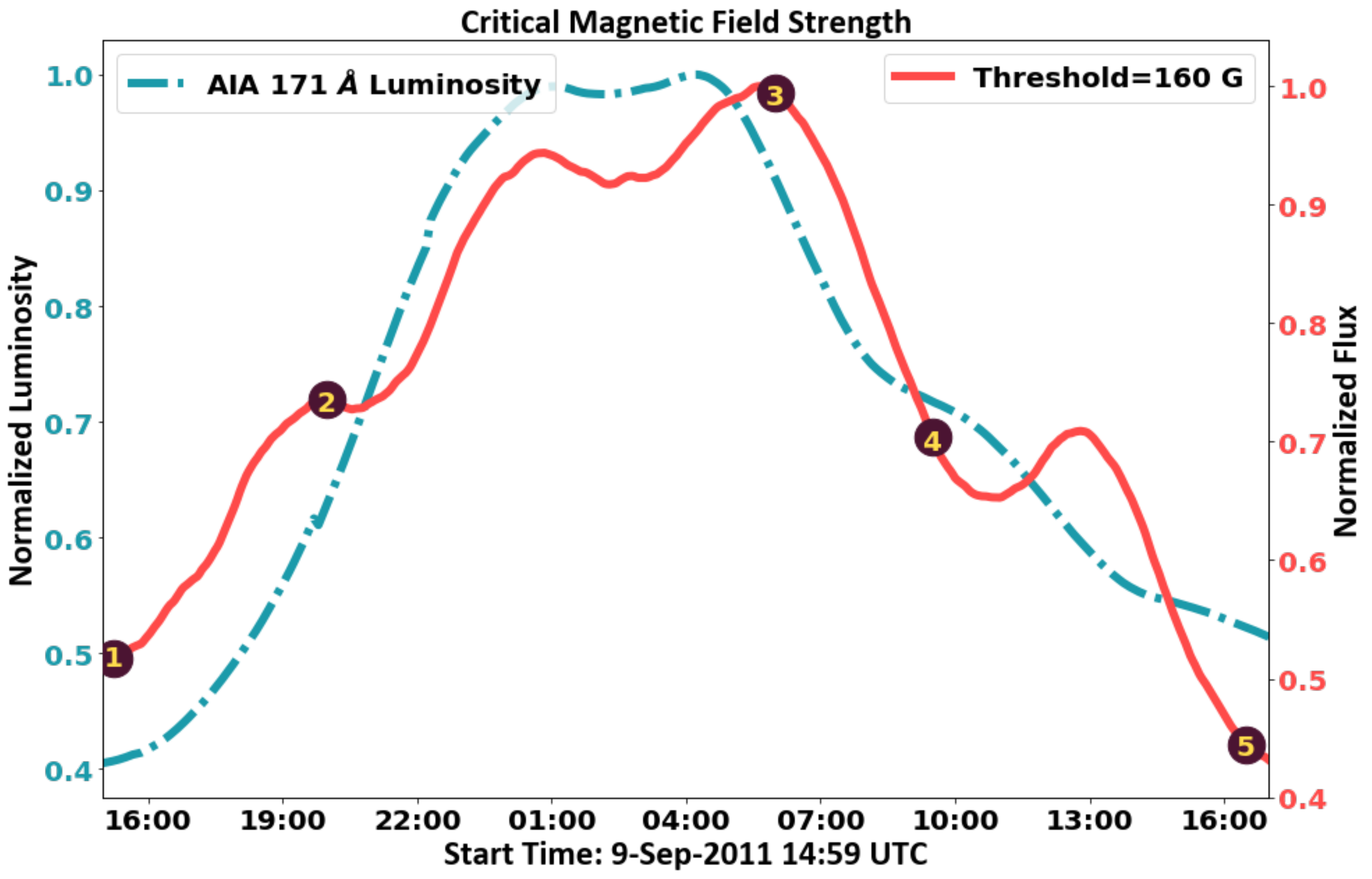}{0.49\textwidth}{(d)}}
\caption{\(P_{CH2}\), which occurred on September 8, 2011. (a) shows an AIA 211 \AA\, image of half the solar disk at 00:00 UTC on September 9, when the plume is at peak intensity (best seen in AIA 171 \AA, displayed in the upper row of panel (b)). The white box in (a) corresponds to the region in which AIA 171 \AA\, measurements were taken. We can see that \(P_{CH2}\) is in an ECH. (b) shows AIA 171 \AA\, images and HMI magnetograms of \(P_{CH2}\) throughout its lifetime. HMI magnetograms are saturated at \(\pm\)100 G. (c) shows this plume's normalized luminosity over its lifetime along with the normalized flux over its lifetime at three different thresholds: 100 G, 160 G, and 400 G. (d) shows the normalized luminosity and normalized flux over its lifetime at the critical magnetic field strength threshold, 160 G. The dots on (c) and (d) correspond to the frame locations in (b).The flux and luminosity data peak within two hours of each other, with flux lagging luminosity. A movie of AIA 171 \AA\ images and HMI magnetograms, CH2{\_}movie.mp4, is available, beginning at 8-Sep-2011 15:00 UTC and ending at 9-Sep-2011 17:00 UTC with a 42 second duration. 
\label{CH2}}
\end{figure}

\begin{figure}[ht!]
\gridline{\fig{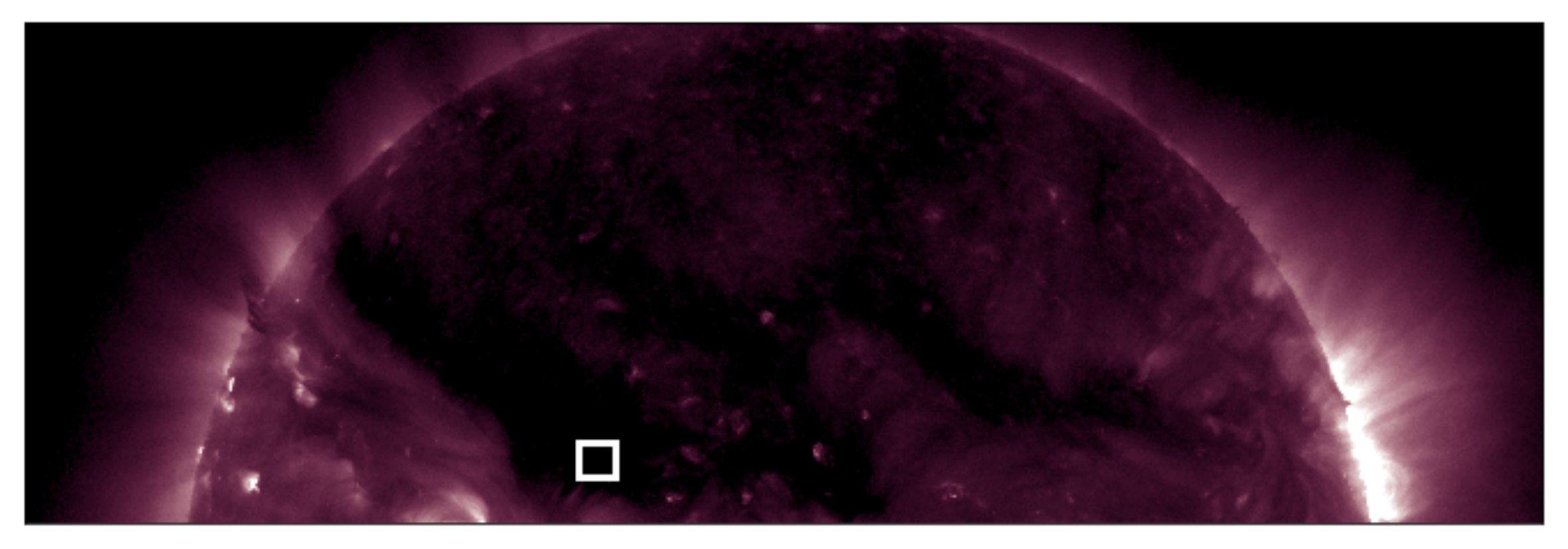}{0.92\textwidth}{(a)}}
\gridline{\fig{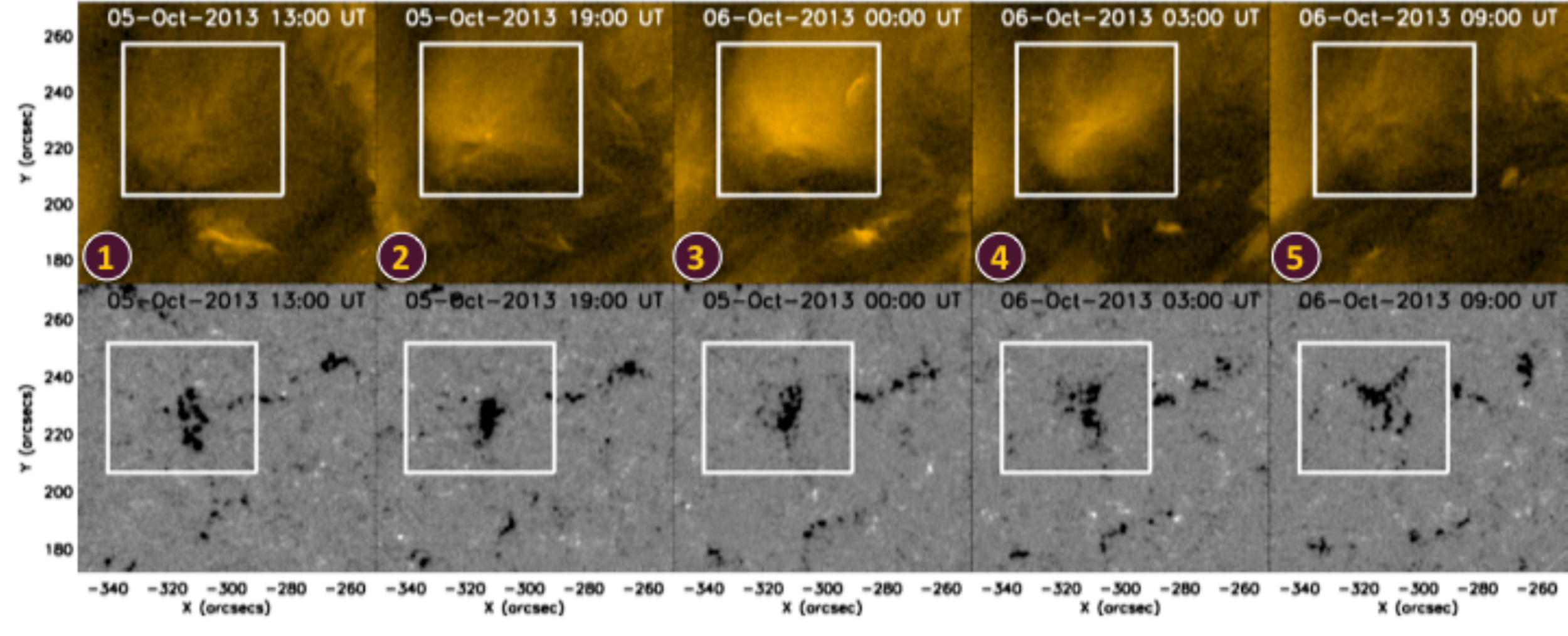}{0.99\textwidth}{(b)}}
\gridline{\fig{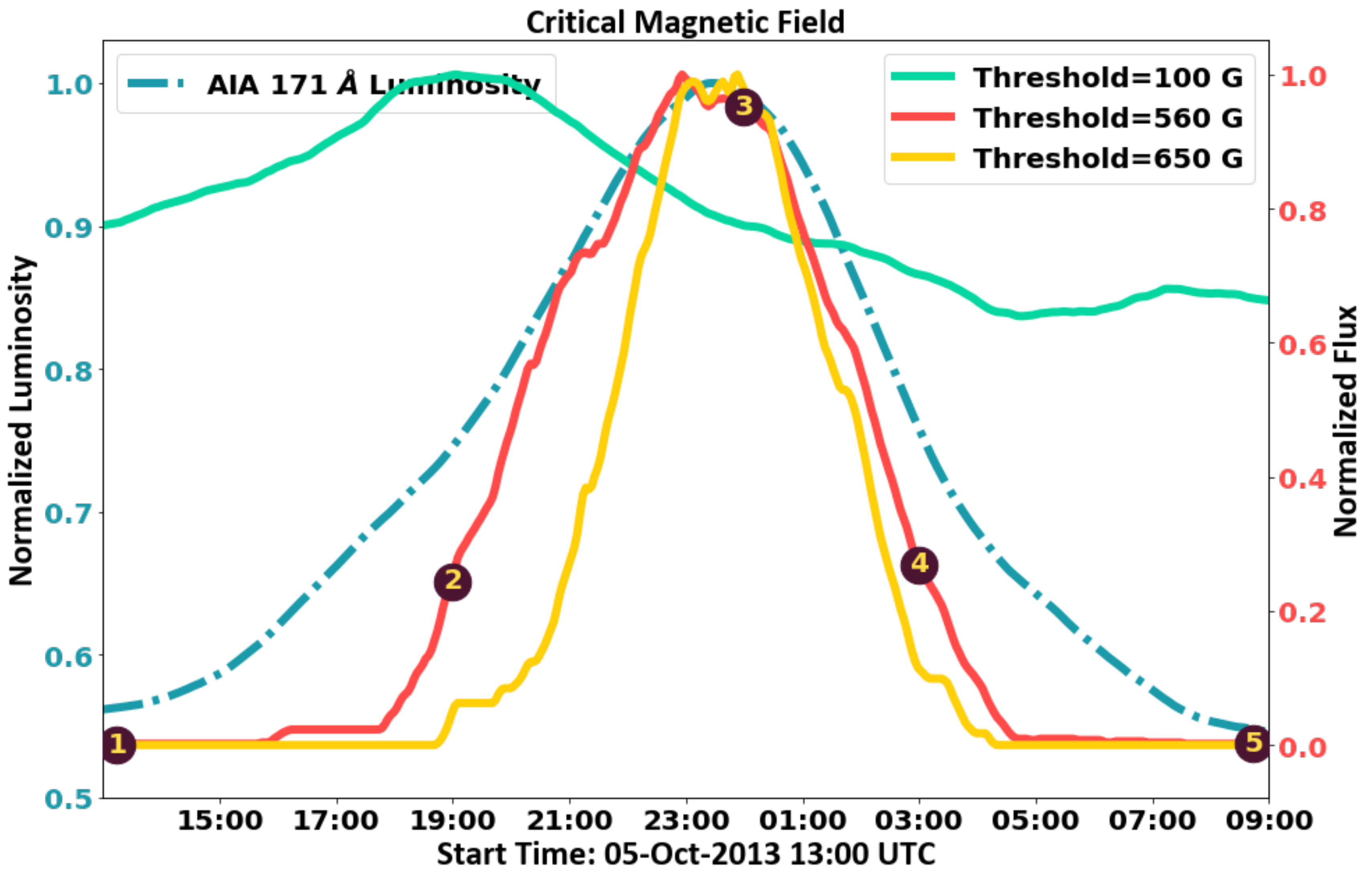}{0.49\textwidth}{(c)}
            \fig{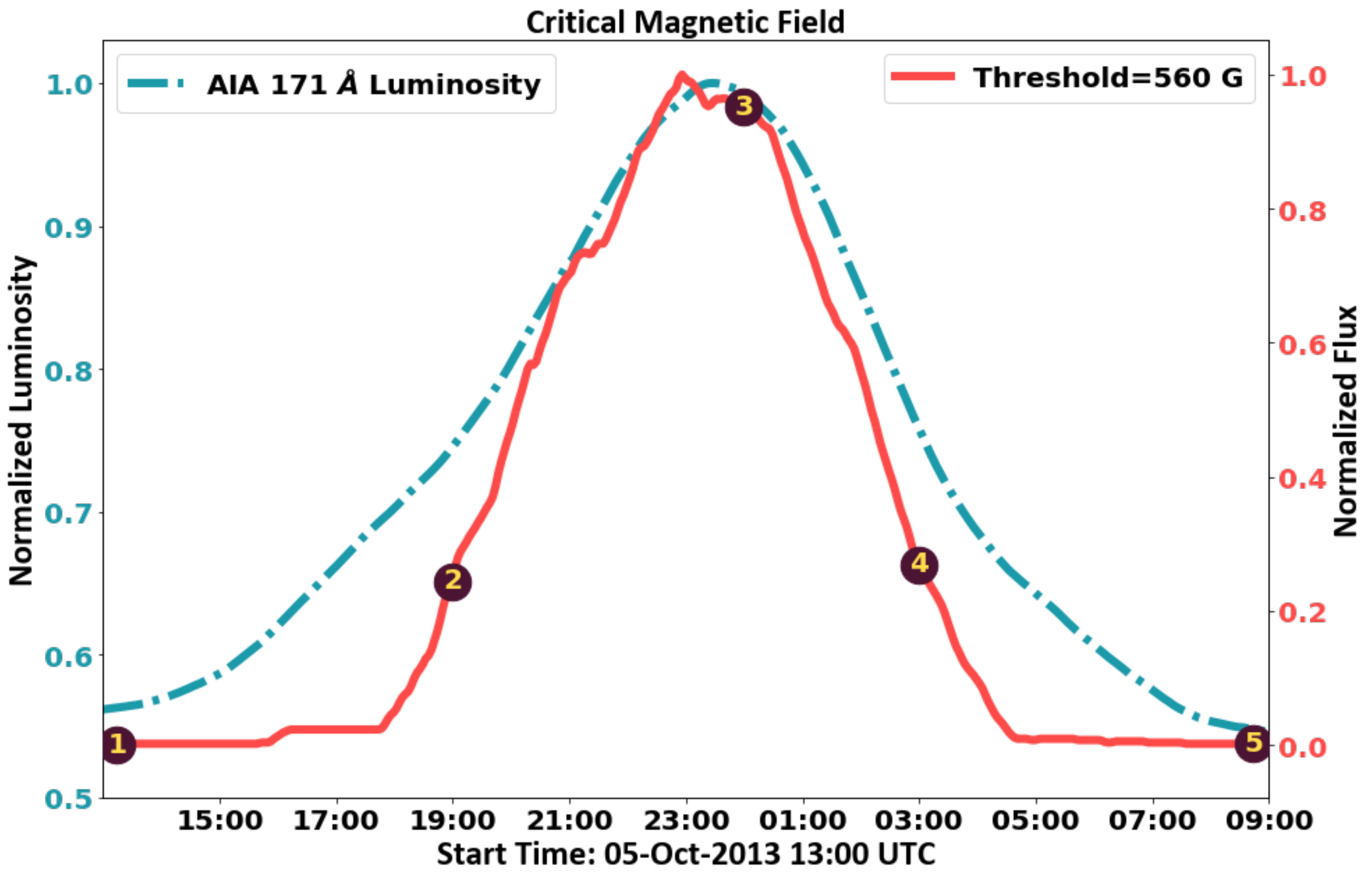}{0.49\textwidth}{(d)}}
\caption{\(P_{CH3}\), which occurred on October 5, 2013. (a) shows an AIA 211 \AA\, image of the Northern half the solar disk at 00:00 UTC on October 6, when the plume is at peak intensity (best seen in AIA 171 \AA, displayed in the upper row of panel (b)). The white box in (a) corresponds to the region in which AIA 171 \AA\, measurements were taken. We can see that \(P_{CH3}\) is in an ECH. (b) shows AIA 171 \AA\, images and HMI magnetograms of \(P_{CH3}\) throughout its lifetime. HMI magnetograms are saturated at \(\pm\)100 G. (c) shows this plume's normalized luminosity over its lifetime along with the normalized flux over its lifetime at three different thresholds: 100 G, 560 G, and 650 G. (d) shows the normalized luminosity and normalized flux over its lifetime at the critical magnetic field strength threshold, 560 G. The dots on (c) and (d) correspond to the frame locations in (b). The flux and luminosity data peak at approximately the same time, within one hour of each other. A movie of AIA 171 \AA\ images and HMI magnetograms, CH3{\_}movie.mp4, is available, beginning at 5-Oct-2013 13:00 UTC and ending at 6-Oct-2013 09:00 UTC with a 32 second duration.
\label{CH3}}
\end{figure}

\begin{figure}[ht!]
\gridline{\fig{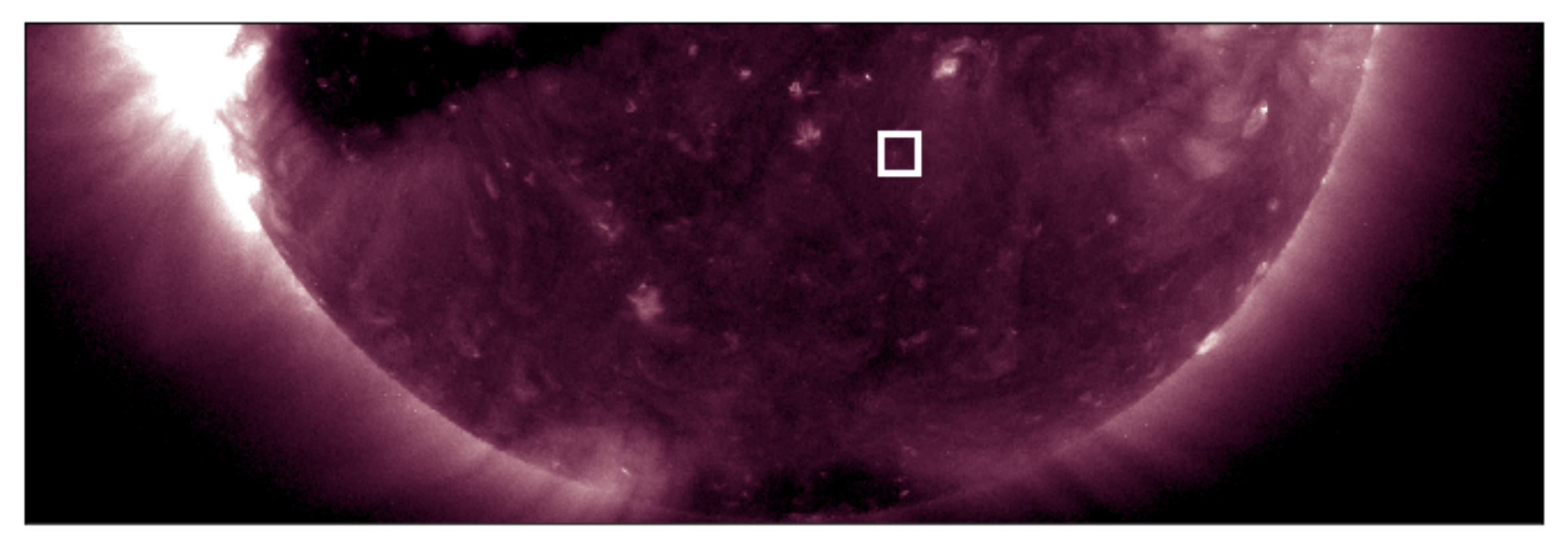}{0.92\textwidth}{(a)}}
\gridline{\fig{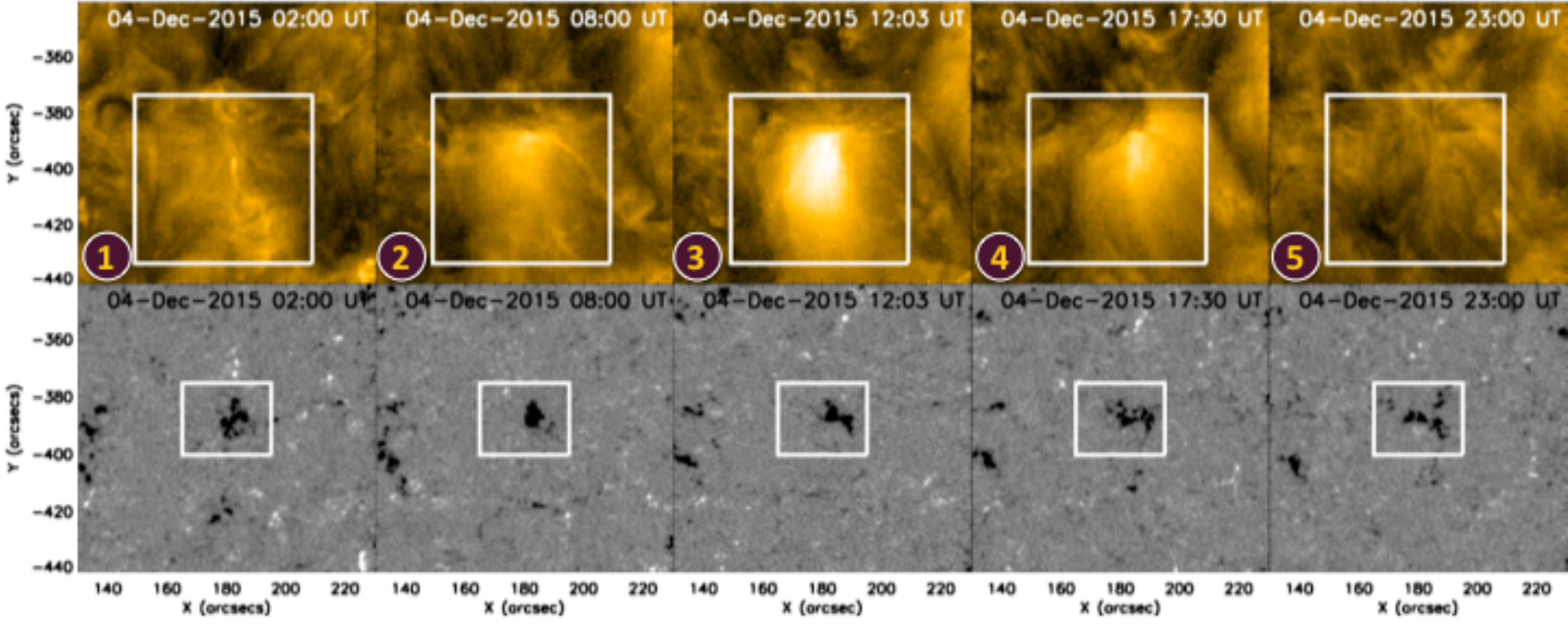}{0.99\textwidth}{(b)}}
\gridline{\fig{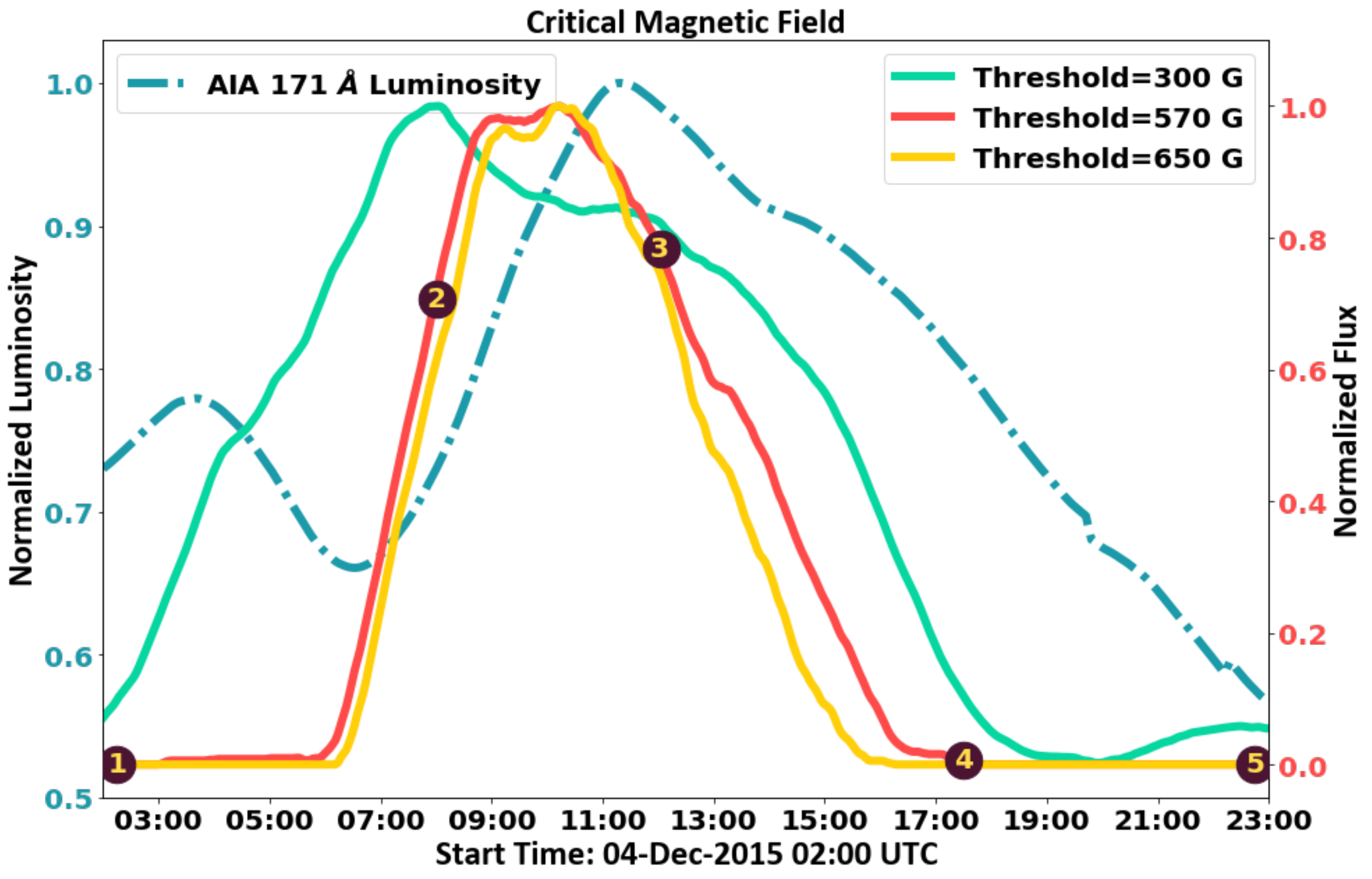}{0.49\textwidth}{(c)}
            \fig{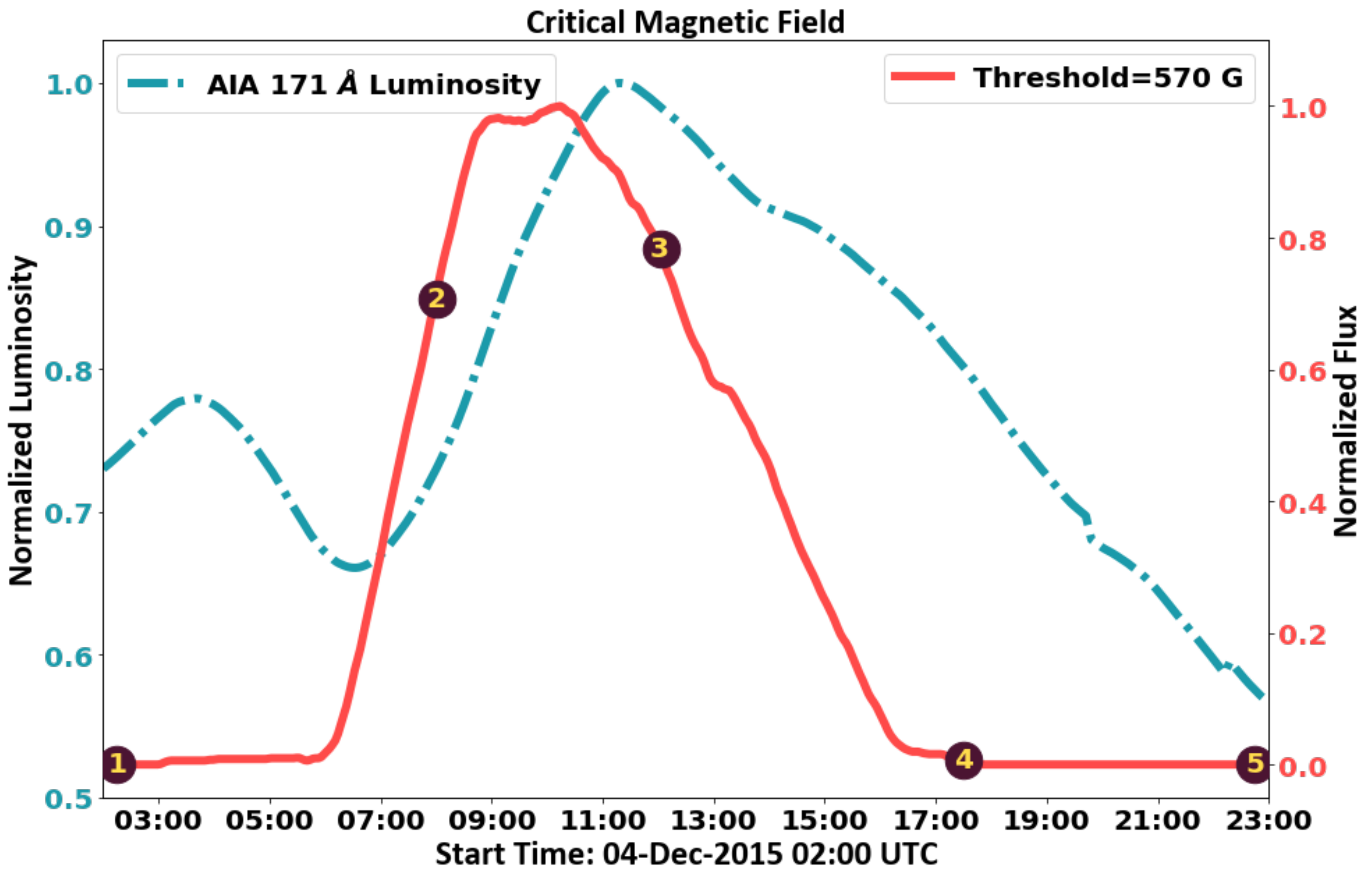}{0.49\textwidth}{(d)}}
\caption{\(P_{QR2}\), which occurred on December 4, 2015. (a) shows an AIA 211 \AA\, image of the Southern half the solar disk at 13:00 UTC, when the plume is at peak intensity (best seen in AIA 171 \AA, displayed in the upper row of panel (b)). The white box in (a) corresponds to the region in which AIA 171 \AA\, measurements were taken. We can see that \(P_{QR2}\) is in a non-CH QR. (b) shows AIA 171 \AA\, images and HMI magnetograms of \(P_{QR2}\) throughout its lifetime. HMI magnetograms are saturated at \(\pm\)100 G. (c) shows this plume's normalized luminosity over its lifetime along with the normalized flux over its lifetime at three different thresholds: 300 G, 570 G, and 650 G. (d) shows the normalized luminosity and normalized flux over its lifetime at the critical magnetic field strength threshold, 570 G. The dots on (c) and (d) correspond to the frame locations in (b). The flux and luminosity data peak within three hours of each other, with flux leading luminosity. A movie of AIA 171 \AA\ images and HMI magnetograms, QR2{\_}movie.mp4, is available, beginning at 4-Dec-2015 02:00 UTC and ending at 4-Dec-2015 23:00 UTC with a 34 second duration.
\label{QR2}}
\end{figure}

\begin{figure}[ht!]
\gridline{\fig{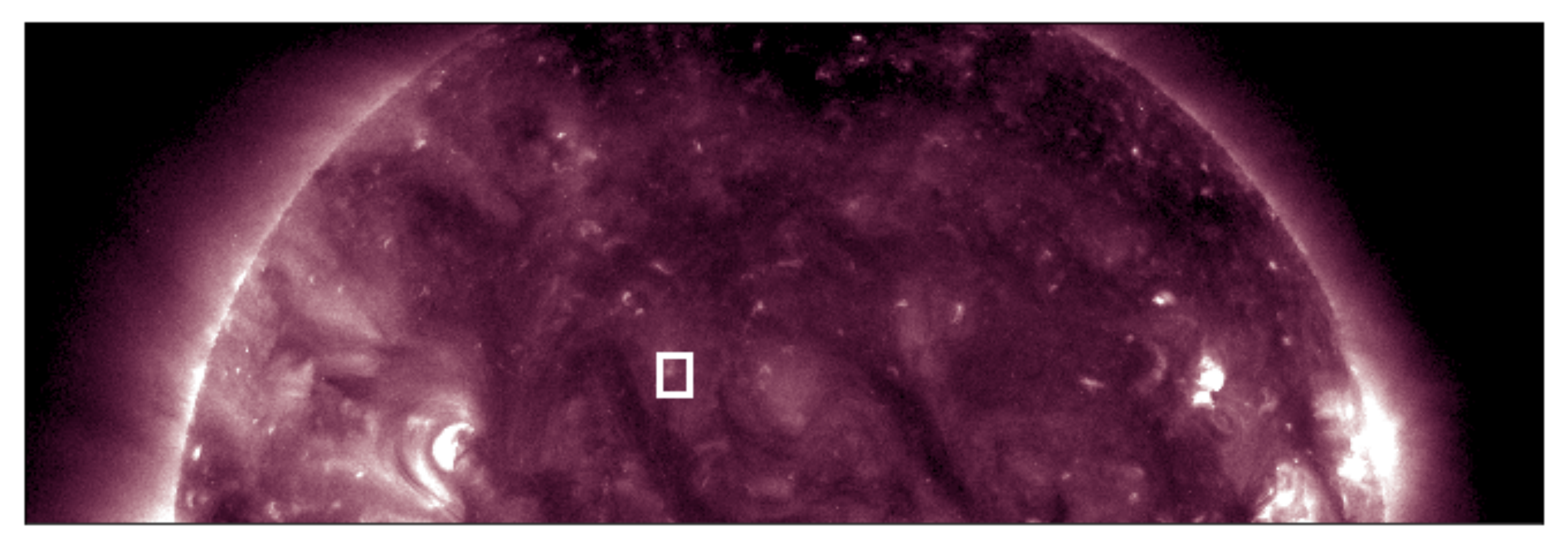}{0.92\textwidth}{(a)}}
\gridline{\fig{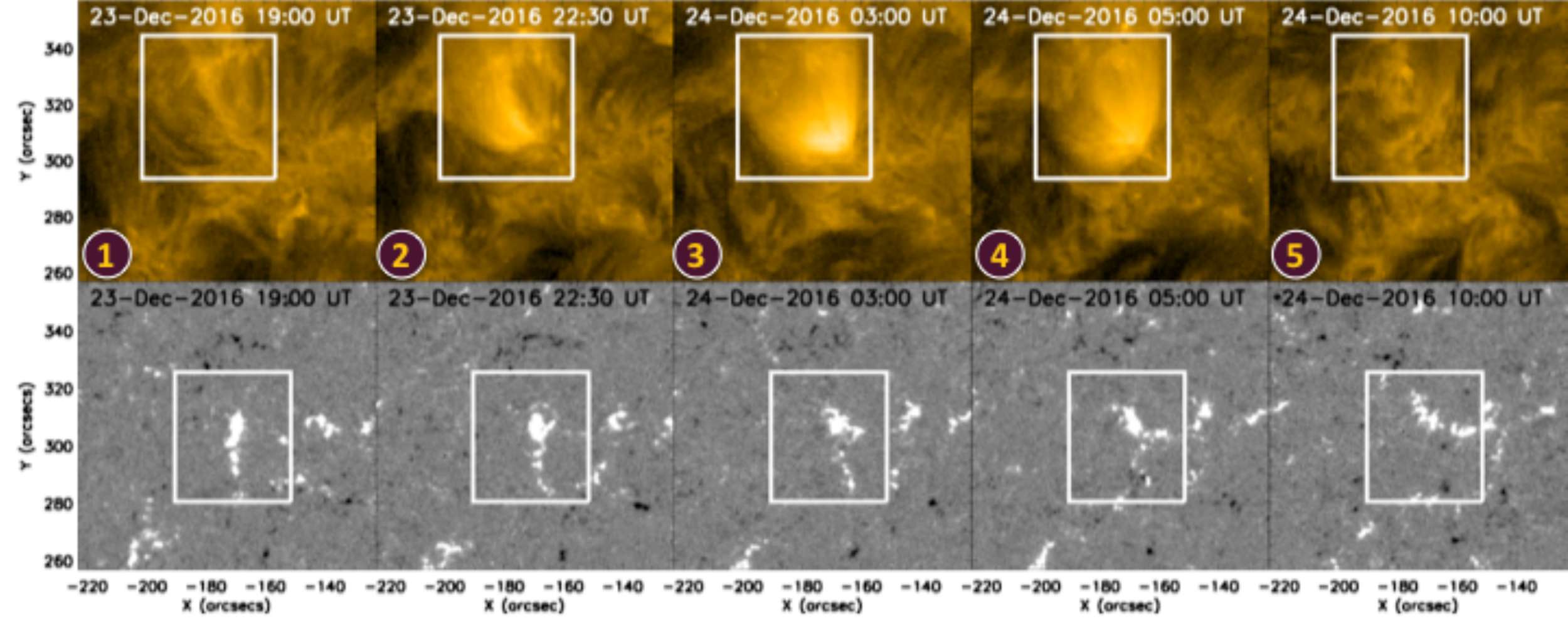}{0.99\textwidth}{(b)}}
\gridline{\fig{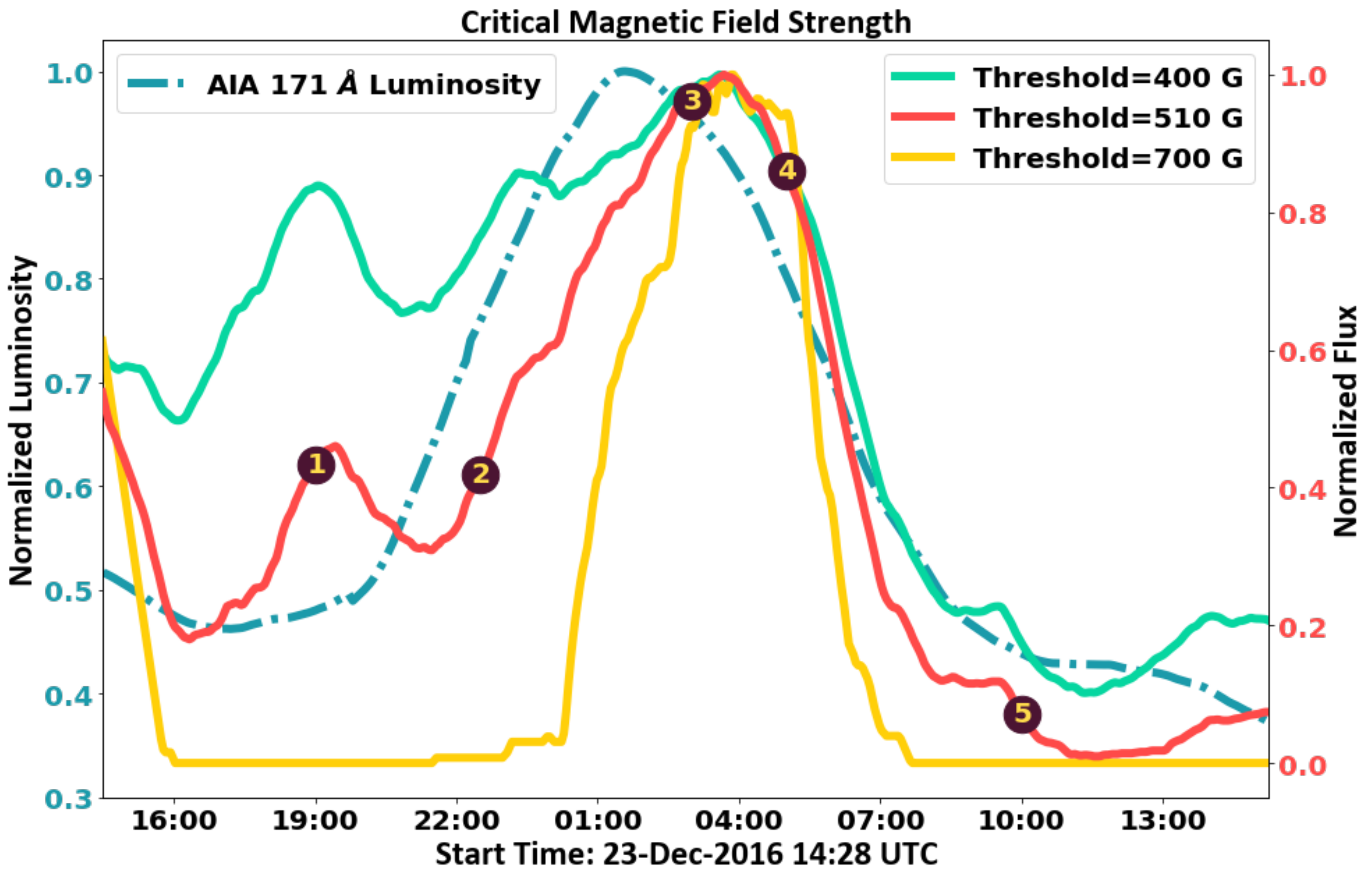}{0.49\textwidth}{(c)}
            \fig{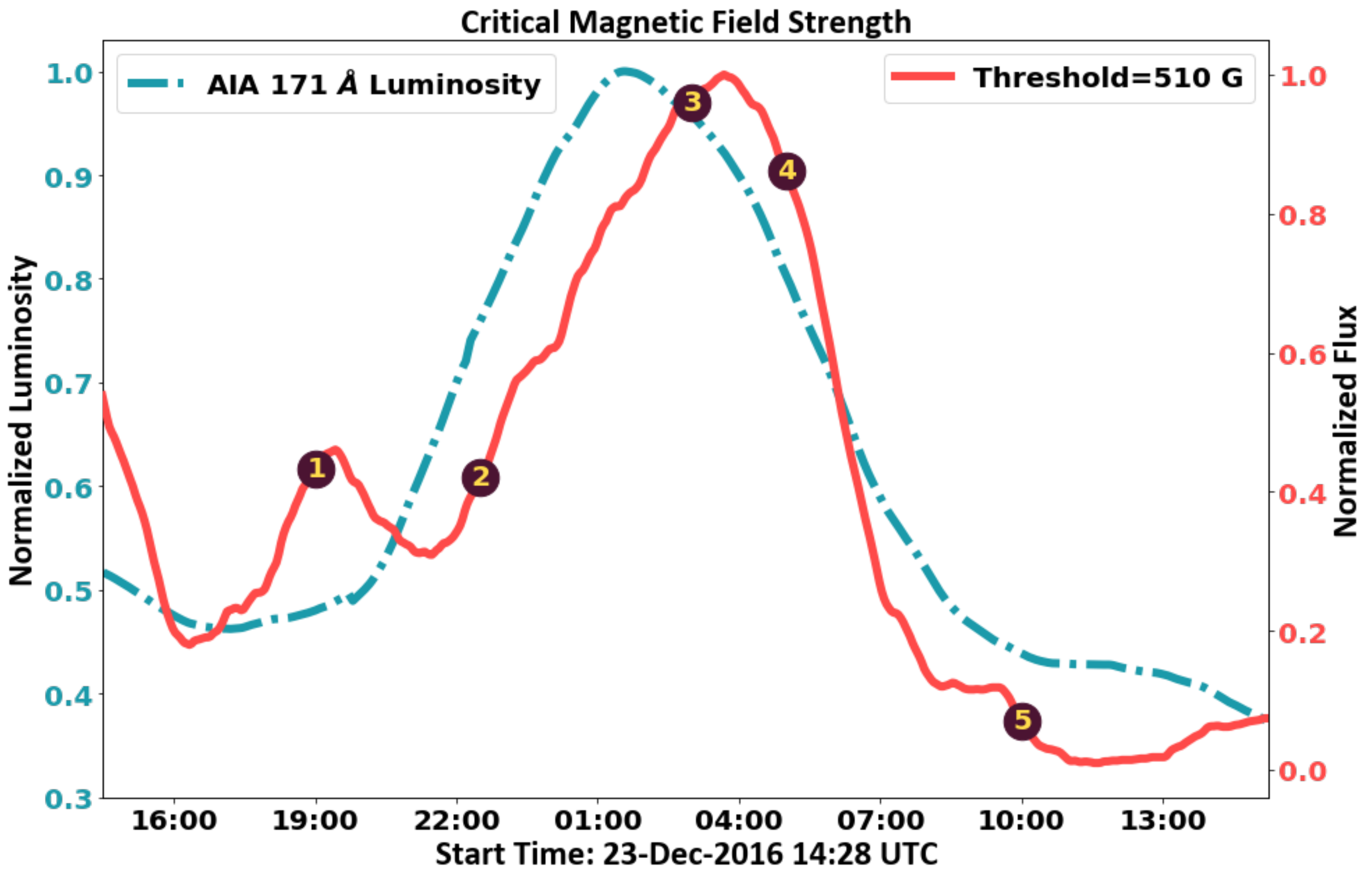}{0.49\textwidth}{(d)}}
\caption{\(P_{QR3}\), which occurred on December 23, 2016. (a) shows an AIA 211 \AA\, image of the Northern half the solar disk at 02:00 UTC on December 24, when the plume is at peak intensity (best seen in AIA 171 \AA, displayed in the upper row of panel (b)). The white box in (a) corresponds to the region in which AIA 171 \AA\, measurements were taken. We can see that \(P_{QR3}\) is in a non-CH QR. (b) shows AIA 171 \AA\, images and HMI magnetograms of \(P_{QR3}\) throughout its lifetime. HMI magnetograms are saturated at \(\pm\)100 G. (c) shows this plume's normalized luminosity over its lifetime along with the normalized flux over its lifetime at three different thresholds: 400 G, 510 G, and 700 G. (d) shows the normalized luminosity and normalized flux over its lifetime at the critical magnetic field strength threshold, 510 G. The dots on (c) and (d) correspond to the frame locations in (b). The flux and luminosity data peak within three hours of each other, with flux lagging luminosity. A movie of AIA 171 \AA\ images and HMI magnetograms, QR3{\_}movie.mp4, is available, beginning at 23-Dec-2016 14:30 UTC and ending at 24-Dec-2016 15:15 UTC with a 40 second duration.
\label{QR3}}
\end{figure}

\begin{figure}[ht!]
\gridline{\fig{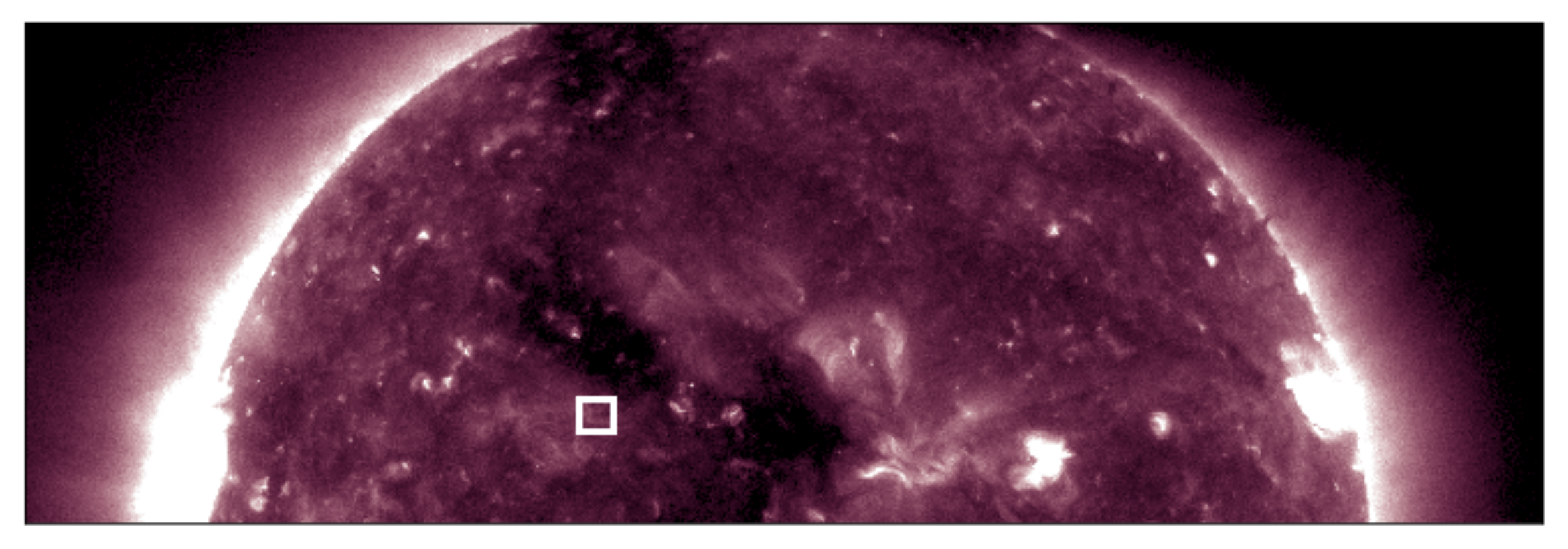}{0.92\textwidth}{(a)}}
\gridline{\fig{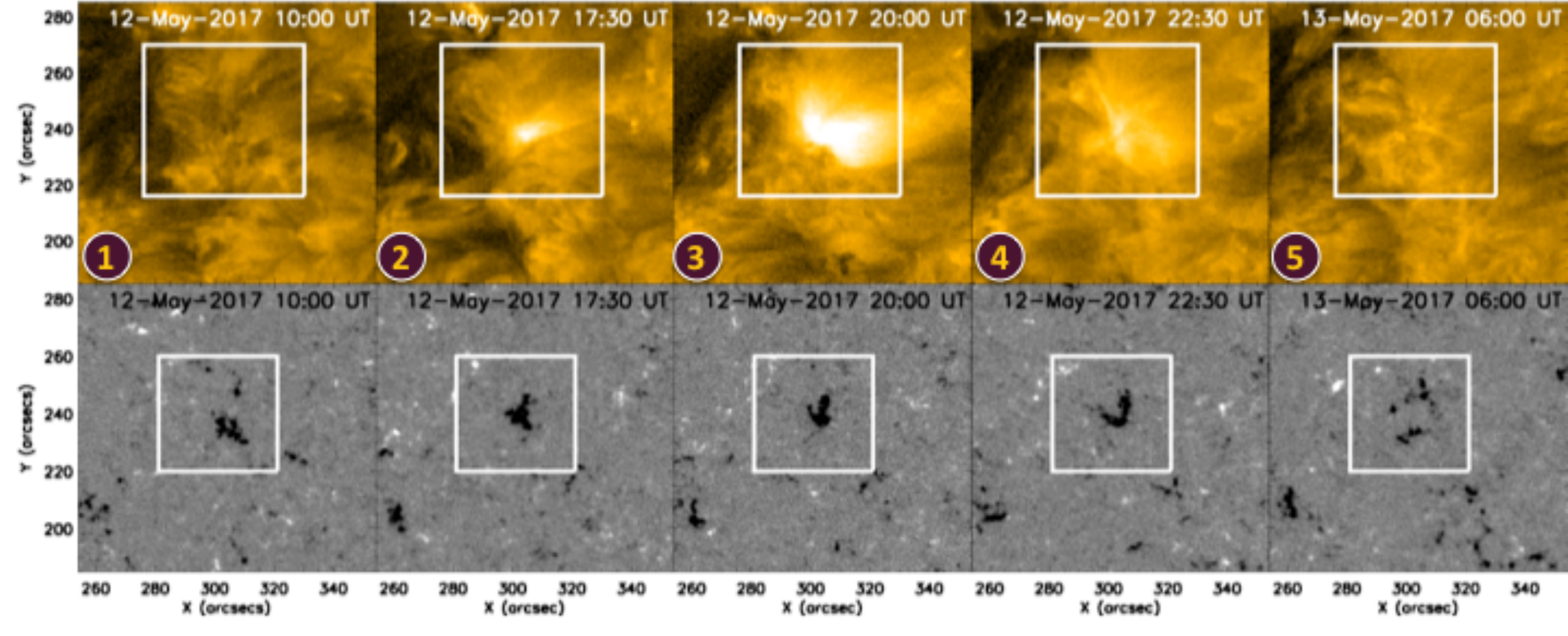}{0.99\textwidth}{(b)}}
\gridline{\fig{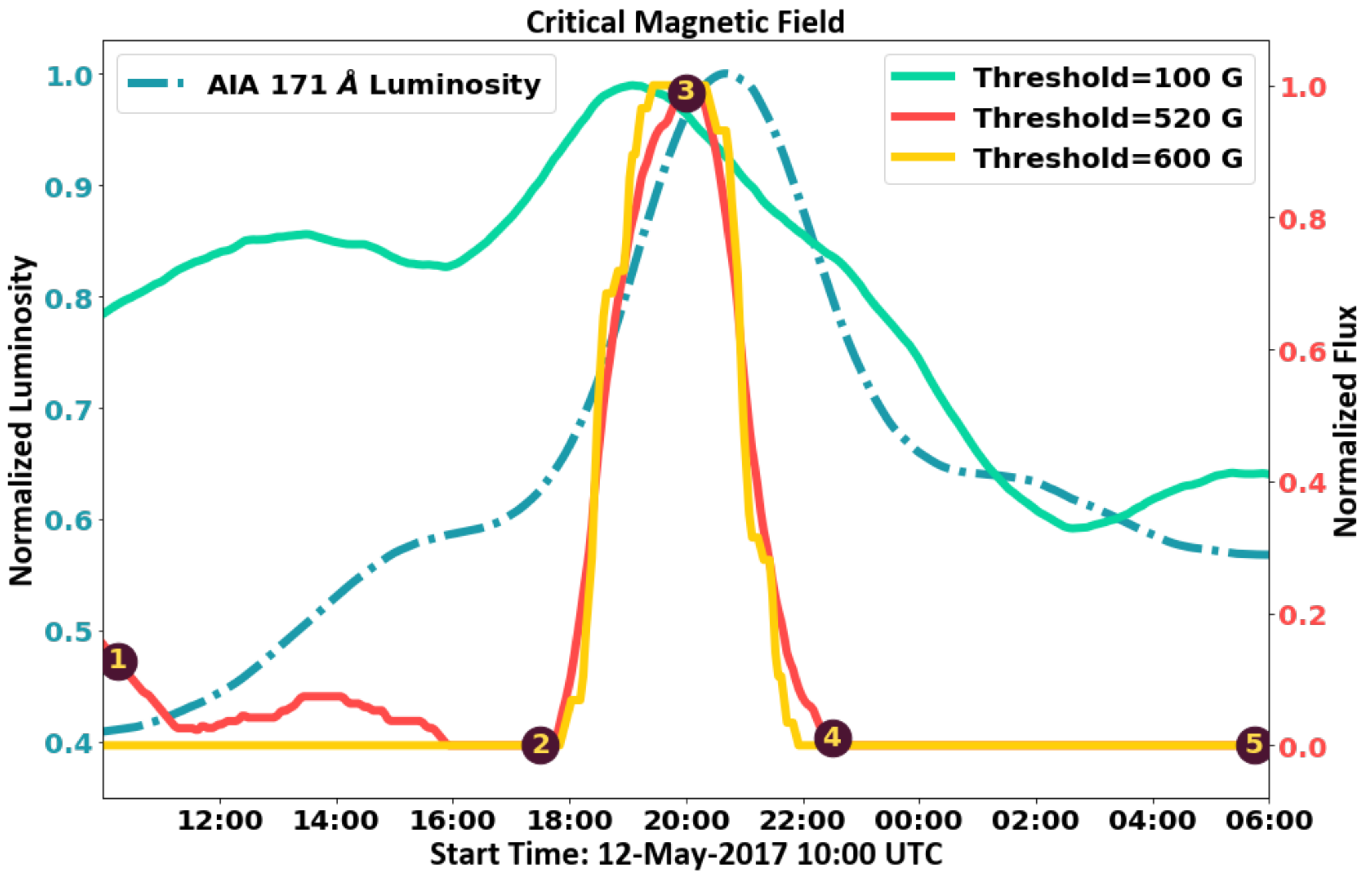}{0.49\textwidth}{(c)}
            \fig{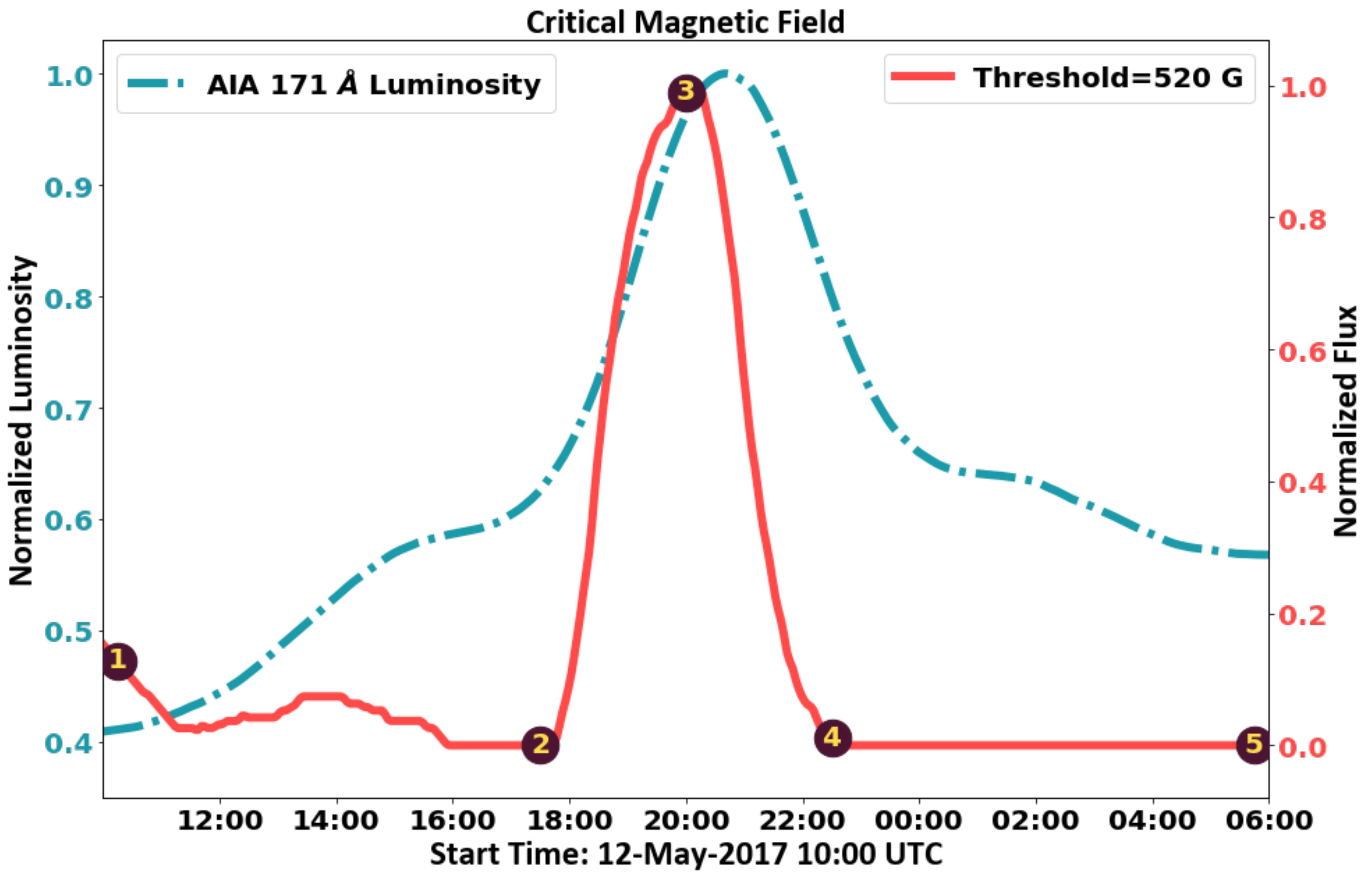}{0.49\textwidth}{(d)}}
\caption{\(P_{QR4}\), which occurred on May 12, 2017. (a) shows an AIA 211 \AA\, image of the Northern half the solar disk at 20:00 UTC, when the plume is at peak intensity (best seen in AIA 171 \AA, displayed in the upper row of panel (b)). The white box in (a) corresponds to the region in which AIA 171 \AA\, measurements were taken. We can see that \(P_{QR4}\) is in a non-CH QR. (b) shows AIA 171 \AA\, images and HMI magnetograms of \(P_{QR4}\) throughout its lifetime. HMI magnetograms are saturated at \(\pm\)100 G. (c) shows this plume's normalized luminosity over its lifetime along with the normalized flux over its lifetime at three different thresholds: 100 G, 520 G, and 600 G. (d) shows the normalized luminosity and normalized flux over its lifetime at the critical magnetic field strength threshold, 520 G. The dots on (c) and (d) correspond to the frame locations in (b). The flux and luminosity data peak at approximately the same time, within one hour of each other. A movie of AIA 171 \AA\ images and HMI magnetograms, QR4{\_}movie.mp4, is available, beginning at 12-May-2017 10:00 UTC and ending at 13-May-2017 06:00 UTC with a 32 second duration.
\label{QR4}}
\end{figure}

\begin{figure}[ht!]
\gridline{\fig{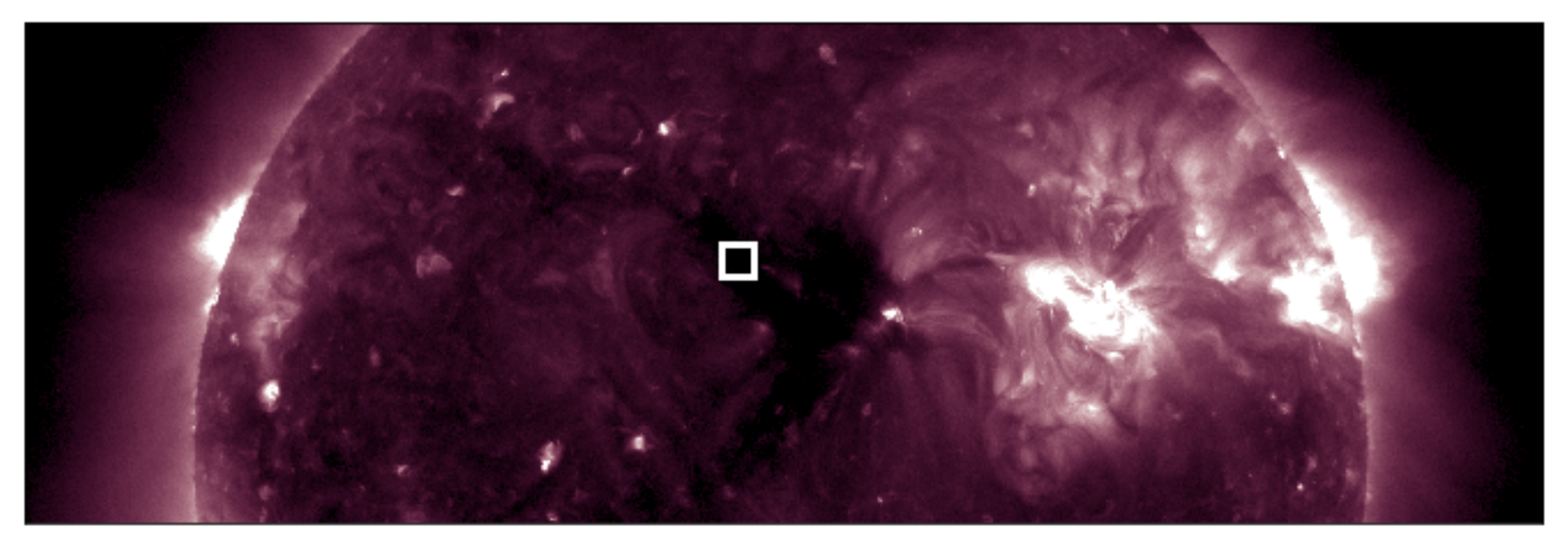}{0.92\textwidth}{(a)}}
\gridline{\fig{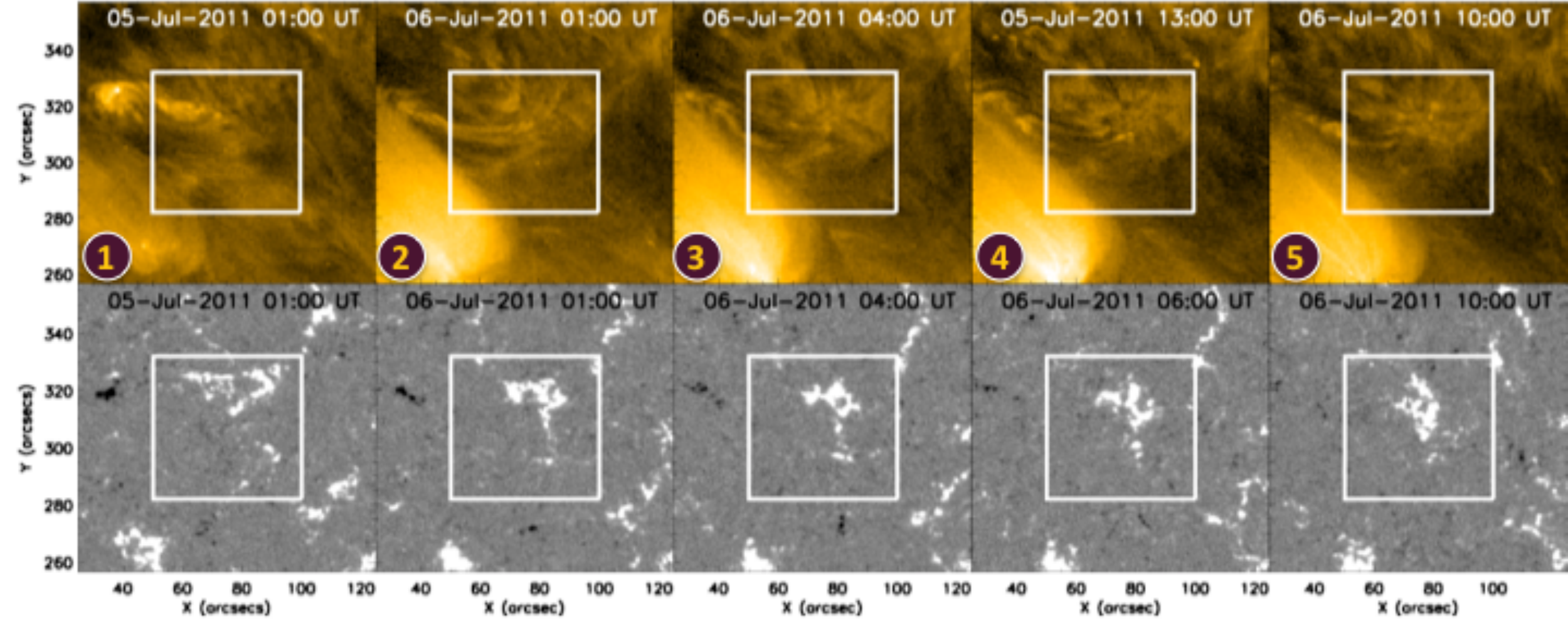}{0.99\textwidth}{(b)}}
\gridline{\fig{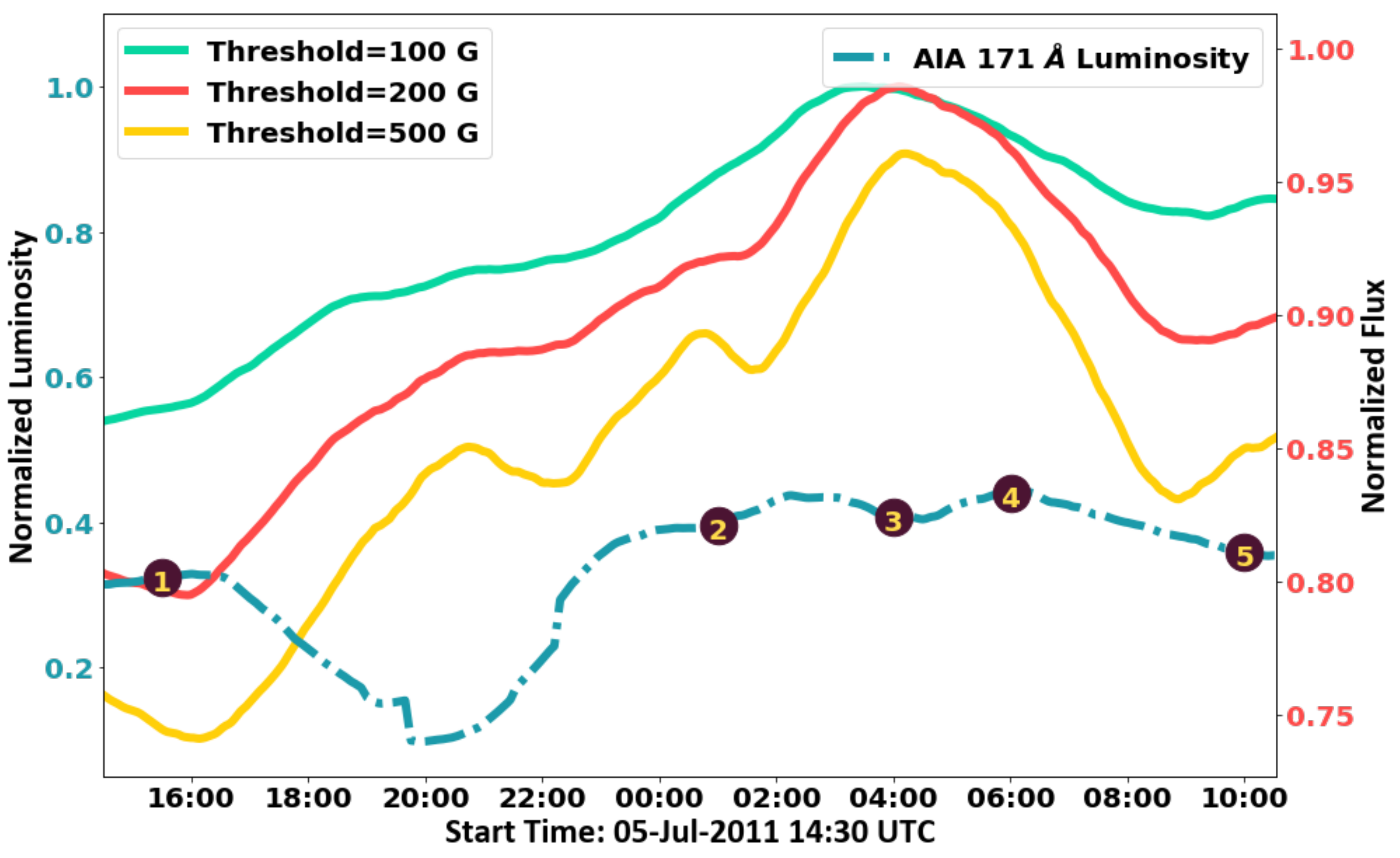}{0.49\textwidth}{(c)}
            \fig{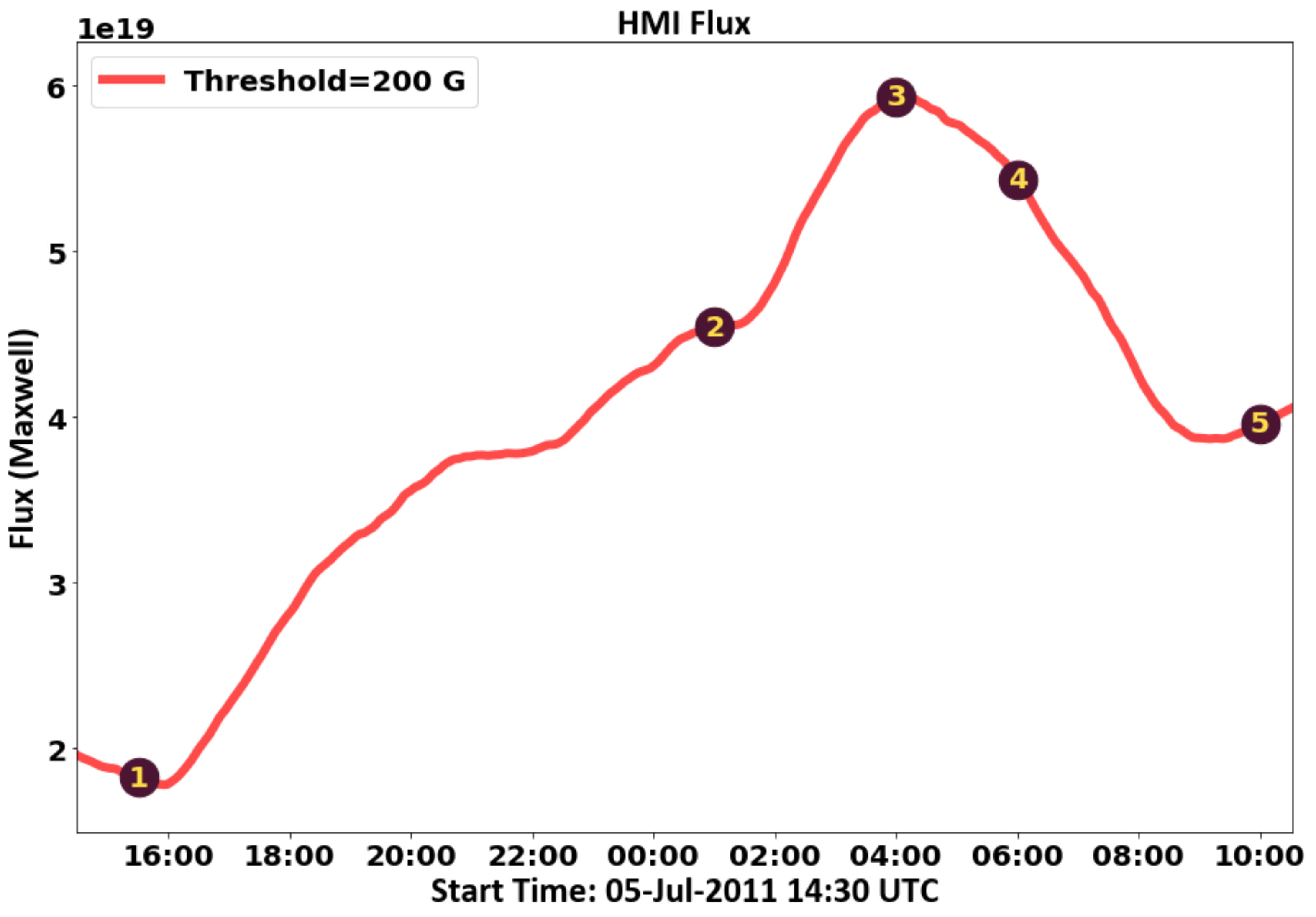}{0.44\textwidth}{(d)}}
\caption{\(P_{no1}\), a photospheric magnetic flux patch which does not form a plume that occurred on July 5, 2011. (a) shows an AIA 211 \AA\, image the solar disk at 04:00 UTC on July 6, when the flux patch is at peak convergence. The white box in (a) corresponds to the region where AIA 171 \AA\, measurements were taken. We see that \(P_{no1}\) is in an ECH. (b) shows AIA 171 \AA\, images and HMI magnetograms of \(P_{no1}\) throughout its lifetime. HMI magnetograms are saturated at \(\pm\)100 G. (c) shows the normalized luminosity of \(P_{no1}\) over its lifetime. We see that the flux over lifetime plot follows the same trend as the patches that form plumes, as shown by (d). However, the luminosity of the region in AIA 171 \AA, shown in (c) with flux plots at thresholds of 100 G, 200 G, and 300 G, does not follow the plume trend exhibited by the flux plots or get as bright as regions which form plumes. The dots on (c) and (d) correspond to the frame locations in (b). A movie of AIA 171 \AA\ images and HMI magnetograms, no1{\_}movie.mp4, is available, beginning at 5-Jul-2011 14:30 UTC and ending at 6-Jul-2011 23:54 UTC with a 53 second duration.
\label{no1}}
\end{figure}

\end{document}